\documentclass[final,11pt]{article}
\usepackage{multirow}
\usepackage[nottoc]{tocbibind}
\usepackage{geometry}
\usepackage{subcaption}
\usepackage{placeins}
\usepackage[affil-it, auth-sc]{authblk}
\usepackage[english]{babel}
\usepackage[utf8]{inputenc}
\usepackage{graphicx}
\usepackage{amssymb}
\usepackage{amsthm}
\usepackage{amsmath}
\usepackage{amsfonts}
\usepackage{verbatim}
\usepackage{lineno}
\usepackage{todonotes}
\presetkeys{todonotes}{inline, color=blue!30, size=\small}{}
\usepackage{slashed}
\usepackage{color, soul}
\definecolor{darkred}{rgb}{0.9, 0.0, 0.0}
\definecolor{darkgreen}{rgb}{0.0, 0.5, 0.0}
\usepackage[colorlinks,bookmarks,linkcolor=darkred, citecolor=darkgreen]{hyperref}
\usepackage{eso-pic}
\usepackage[sort&compress,numbers]{natbib}
\usepackage{doi}
\usepackage{paralist,epsfig}
\usepackage{relsize}
\usepackage{nicefrac,esint}
\usepackage{float}
\usepackage{cleveref}
\usepackage{marginnote}

\newcommand{\slashpi}{\protect{\slash\hspace{-0.5em}\pi}}

\begin{document}

\AddToShipoutPictureFG*{\AtPageUpperLeft{\put(-60,-60){\makebox[\paperwidth][r]{LA-UR-22-21034, INT-PUB-23-015}}}}

\title{\bf Effective field theory for radiative corrections \\
to charged-current processes I: Vector coupling}
\author[1]{Vincenzo~Cirigliano \thanks{cirigv@uw.edu}}
\affil[1]{Institute for Nuclear Theory, University of Washington, Seattle WA 98195, USA \vspace{1.2mm}}
\author[1]{Wouter~Dekens \thanks{wdekens@uw.edu}}
\author[2]{Emanuele~Mereghetti \thanks{emereghetti@lanl.gov}}
\affil[2]{Theoretical Division, Los Alamos National Laboratory, Los Alamos, NM 87545, USA \vspace{1.2mm}}
\author[2]{Oleksandr~Tomalak \thanks{tomalak@lanl.gov}}

\date{\today}

\maketitle

We study radiative corrections to low-energy charged-current processes involving nucleons, such as neutron beta decay and (anti)neutrino-nucleon scattering within a top-down effective-field-theory approach. We first match the Standard Model to the low-energy effective theory valid below the weak scale and, using renormalization group equations with anomalous dimensions of $\mathcal{O}(\alpha, \alpha \alpha_s, \alpha^2)$, evolve the resulting effective coupling down to the hadronic scale. Here, we first match to heavy-baryon chiral perturbation theory and subsequently, below the pion-mass scale, to a pionless effective theory, evolving the effective vector coupling with anomalous dimensions of $\mathcal{O}(\alpha, \alpha^2)$ all the way down to the scale of the electron mass, relevant for beta decays. We thus provide a new evaluation of the ``inner" radiative corrections to the vector coupling constant and to the neutron decay rate, discussing differences with the previous literature. Using our new result for the radiative corrections, we update the extraction of the Cabibbo-Kobayashi-Maskawa matrix element $V_{ud}$ from the neutron decay.

\newpage

\tableofcontents

\section{Introduction}
\label{sec1}

Low-energy processes mediated by the charged-current (CC) weak interaction provide promising ways to test the Standard Model (SM) and indirectly search for new physics, provided sufficiently high experimental and theoretical precision can be achieved. In recent years, there has been a resurgence of interest in  beta decays and CC neutrino scattering on nuclei. On the one hand, the study of beta decays at the sub-permille level provides a unique window into possible new physics at the multi-TeV scale. Recent 
analyses~\cite{Seng:2018yzq,Seng:2018qru,Czarnecki:2019mwq,Shiells:2020fqp,Hayen:2020cxh,Seng:2020wjq,Hardy:2020qwl,Cirigliano:2022yyo} have uncovered a 3$\sigma$ tension with the Standard Model interpretation of these processes in terms of the unitary Cabibbo-Kobayashi-Maskawa (CKM) quark mixing matrix \cite{Hardy:2020qwl,ParticleDataGroup:2020ssz}. Moreover, global analyses of beta decay observables~\cite{Falkowski:2020pma,Gonzalez-Alonso:2018omy}, including decay correlations, offer unique ways to probe non-standard CC interactions with Lorentz structures different from the SM ``$V-A$". On the other hand, the interest in the CC neutrino scattering process stems primarily from neutrino oscillation experiments~\cite{Nunokawa:2007qh,NOvA:2007rmc,T2K:2011qtm,KamLAND:2013rgu,JUNO:2015zny,Hyper-KamiokandeProto-:2015xww,RENO:2018dro,DayaBay:2018yms,DoubleChooz:2019qbj,T2K:2019bcf,NOvA:2019cyt,DUNE:2020ypp,JUNO:2021vlw}, as precise theoretical predictions are needed to calibrate the neutrino fluxes and reconstruct the neutrino energy~\cite{Mueller:2011nm,Huber:2011wv,MINERvA:2016iqn,NuSTEC:2017hzk,JUNO:2020ijm,CLAS:2021neh}. In what follows, we will focus on beta decays ($n \to p e \bar \nu_e$) but our results, based on a low-energy effective theory, apply to neutrino scattering  processes such as $\bar{\nu}_e  p \to e^+ n$ and $\nu_e  n \to e p$ at low energy as well.

One of the key ingredients to achieve high theoretical precision in beta decays (sub-permille, allowing one to probe physics up to 20 TeV) is the calculation of electromagnetic radiative corrections, controlled by an expansion in $\alpha/\pi$, where $\alpha \approx 1/137.036$ is the fine-structure constant. The analysis of radiative corrections to beta decays has a long history, predating the formulation of the Standard Model of electroweak and strong interactions. In the early work from the 1950s~\cite{Behrends:1955mb,Kinoshita:1958ru}, the nucleon was treated as point-like and the weak interaction was described in terms of the $(V-A) \times  (V-A)$ current-current contact operator. In the framework of the local  $(V-A) \times  (V-A)$ theory, two  developments from the 1960s have influenced all the subsequent literature. In Ref.~\cite{Sirlin:1967zza}, Sirlin identified a set of ultraviolet(UV)-finite and gauge-invariant corrections to the beta spectrum and decay rate that are independent of the details of the strong interaction, the so-called universal ``outer" corrections. Ref.~\cite{Sirlin:1967zza} also identified a set of ``inner" corrections that essentially shift the strength of the vector (Fermi) and axial-vector (Gamow-Teller) couplings at the single-nucleon level, pointing out that in principle these ``inner" corrections depend on the strong-interaction dynamics. Shortly afterwards, using current-algebra techniques, the authors of Ref.~\cite{Abers:1968zz}  showed that to $\mathcal{O} (\alpha)$ the contribution of the weak vector current $V$ to the Fermi transition ``inner" correction is calculable without knowledge of details about the strong interactions, leading to a universal UV-divergent correction. Ref.~\cite{Abers:1968zz} also showed that the contribution to the Fermi transition due to the weak axial-vector current $A$ does depend on the strong-interaction details. This class of ``inner" corrections was parameterized in terms of correlation functions of weak and electromagnetic hadronic currents in the nucleon state, and crudely estimated with available models of strong interactions.

The current-algebra formulation of radiative corrections was later embedded in the Standard Model by Sirlin ~\cite{Sirlin:1977sv,Sirlin:1981ie}, who also computed the leading logarithmic corrections to $\mathcal{O} (\alpha)$ and $\mathcal{O} (\alpha \alpha_s)$. Since then, the calculation of the terms that depend on the strong interactions has been performed in this framework with more sophisticated  hadronic models, culminating in the 2006 prediction~\cite{Marciano:2005ec} for the ``inner" correction to the vector amplitude. Large logarithms originating from both the UV (from $M_Z$ to the hadronic scale of the order of nucleon mass $m_N$) and IR (from $m_N$ to $m_e$) have been resummed in the leading logarithmic approximation (see  Ref.~\cite{Czarnecki:2004cw} and references therein). Ref.~\cite{Czarnecki:2004cw} also includes next-to-leading logarithms in $\alpha$ that are enhanced by the number of fermion species.

The next important development in the field has been the calculation of the non-perturbative input for the ``inner" corrections using dispersive methods,  pioneered by Seng et al.~\cite{Seng:2018yzq,Seng:2018qru}. This has led to a reduced uncertainty and an increase in the central value of the ``inner" correction to the Fermi coupling, later reproduced by Refs.~\cite{Czarnecki:2019mwq,Shiells:2020fqp,Hayen:2020cxh}. In this framework, lattice QCD has been used to supplement non-perturbative input in the meson sector~\cite{Seng:2020jtz,Feng:2020zdc,Ma:2021azh,Yoo:2022lmt} and efforts to do the same for nucleon decay are underway~\cite{Feng:2020zdc,Yoo:2023gln}.

All the results described above are rooted in the current-algebra framework developed by Sirlin~\cite{Sirlin:1977sv}. While this method is rigorous, it does not take full advantage of modern effective field theory (EFT) techniques, neither at the level of short-distance physics (the evolution of the interactions from the electroweak scale to the hadronic scale), nor at the level of strong interactions (chiral EFT for mesons, nucleons, and eventually nuclei). The use of EFT techniques is not a mere reformulation of the problem. EFT provides a rigorous way to connect scales and estimate uncertainties. Moreover, EFT methods can  bring new insights to the problem. In fact,  by providing a simple framework to analyze hadronic correlation functions,  the study of neutron decay to $\mathcal{O} (G_F \alpha)$ in Heavy Baryon Chiral Perturbation Theory (HBChPT)~\cite{Cirigliano:2022hob} has uncovered a new \%-level ``inner" correction to the ratio $g_A/g_V$ of axial-vector to vector nucleon couplings, missed in previous analyses based on current algebra~\cite{Hayen:2020cxh,Gorchtein:2021fce}.

In the HBChPT framework for single-nucleon weak CC processes, developed in Refs.~\cite{Ando:2004rk,Cirigliano:2022hob}, the active degrees of freedom are the light leptons, photons, pions, and nucleons. The effect of both electroweak- and other hadronic-scale physics is encoded in a number of low-energy constants (LECs). The goal of this paper is to develop a matching procedure  to express the relevant LECs in terms of perturbatively calculable Wilson coefficients and hadronic correlation functions that can then be estimated with  non-perturbative methods, such as dispersive methods or lattice QCD. Since there are multiple thresholds, the electroweak scale $\sim M_{W,Z}$, the chiral symmetry breaking scale $\Lambda_\chi \sim m_N\sim$~GeV, with $m_N$ the mass of nucleon, and the pion mass, we adopt a multi-step matching strategy. The first step connects the full Standard Model to the so-called low-energy effective theory (LEFT) below the weak scale, which coincides with the $V-A$ theory of weak interactions augmented by QED and QCD. This is a perturbative matching step. The second step connects the LEFT to HBChPT and involves non-perturbative physics. These first two steps are similar in spirit to the analysis of Ref.~\cite{Descotes-Genon:2005wrq} for the meson sector. The third step consists of integrating out the pions, by matching HBChPT onto a pionless EFT ($\slashpi$EFT) as detailed in Ref.~\cite{Cirigliano:2022hob}. The main novel aspects of our work are the following:
\begin{itemize}
    \item We evaluate the relevant LEFT Wilson coefficient to  next-to-leading logarithm accuracy in $\alpha$: we implement the matching condition at $\mu_{SM}  \sim M_Z$ at one loop and the running via the two-loop anomalous dimension of $\mathcal{O}(\alpha^2)$, for which we provide for the first time the full expression.  We also use the known two-loop anomalous dimension of $\mathcal{O}(\alpha \alpha_s)$ and present solutions of the Renormalization Group Equations (RGEs) summing leading and next-to-leading logarithms of the ratio $M_Z/m_N$.

    \item We set up the general formalism and provide explicit expressions for the HBChPT LECs that shift the vector coupling $g_V$. The relevant non-perturbative input can be obtained either from the existing dispersive analyses~\cite{Seng:2018qru} or lattice QCD in the future.

    \item We solve the RGEs for the  vector coupling $g_V (\mu_\chi)$, using one- and two-loop anomalous dimensions in $\slashpi$EFT. This allows us to sum the leading and next-to-leading logarithms involving the ratio $m_N/E_0$, where $E_0 \simeq 2.530~m_e$ is the electron energy endpoint, representing the infrared (IR) scale of the problem. The RGE evolution thus allows us to identify all terms in the amplitude proportional to $\alpha^2 \ln (m_N/E_0)$. Our treatment of these next-to-leading large logarithms differs from the one found in the literature, as discussed in Section~\ref{sec2}.

    \item Throughout, we use dimensional regularization with modified minimal subtraction ($\overline{\rm MS}$~\cite{Bardeen:1978yd}) in the LEFT, and the chiral version of it ($\overline{\rm MS}_\chi$~\cite{Gasser:1983yg}), specifying at every step the $\gamma_5$ and evanescent operator scheme. In this framework, the renormalization group (RG) equations have a very simple form, and the standard results on leading logarithm and next-to-leading logarithm  resummation can be applied. The residual sensitivity to the renormalization scale order by order in RG-improved perturbation theory gives us a rigorous way to estimate the perturbative  uncertainties. More generally, our results provide a new framework to analyze low-energy CC processes to $\mathcal{O} (G_F \alpha)$, largely independent from the current algebra formalism~\cite{Sirlin:1977sv}.

    \item As a first application of the new framework, using the dispersive input from Refs.~\cite{Seng:2018yzq,Seng:2018qru,Czarnecki:2019mwq,Shiells:2020fqp,Hayen:2020cxh,Seng:2020wjq} as compiled in Ref.~\cite{Cirigliano:2022yyo}, we evaluate the combination of LECs that determine the ``inner" corrections to the Fermi transition effective coupling $g_V$. We combine this with the  known $\mathcal{O}(\alpha)$ radiative corrections to the matrix element in HBChPT~\cite{Ando:2004rk,Cirigliano:2022hob}. We further resum the Coulomb-enhanced terms scaling as $(\pi \alpha/\beta)^n$ ($\beta \equiv p_e/E_e$) as well as subleading $\alpha/\pi (\pi \alpha/\beta)^n$ terms, in the nonrelativistic Fermi function, which is the natural quantity appearing in $\slashpi$EFT. In practice, this amounts to replacing the relativistic Fermi function (which contains large logarithms  $\sim \alpha^2 \ln (R_p p_e)$, with the proton radius $R_p \sim 1/\Lambda_\chi$) with its nonrelativistic  counterpart. Finally, we study the impact on the extraction of $V_{ud}$ from neutron decay. For the total corrections to the neutron decay rate, we find a result that is one $\sigma$ above the previous results, pointing to a correspondingly smaller value for $V_{ud}$.
\end{itemize}

The paper is organized as follows. In Section~\ref{sec2}, we provide a high-level summary of the results worked out in the rest of the paper, highlighting the connections to and differences from the previous literature. Following a top-down approach, we perform a multi-step matching to connect electroweak physics with neutron and nuclear decays. The first step, connecting the full Standard Model to the LEFT, is presented in Section~\ref{sec3}. The second step, connecting the LEFT to HBChPT, is presented in Section~\ref{sec4}. The resulting effective vector coupling $g_V (\mu_\chi \sim m_N)$ at the matching scale $\mu_\chi \sim m_N$ and its evolution to the scale of the decay, $\mu_\chi \sim E_0$, is discussed in Section~\ref{sec:subsec45}. In Section~\ref{sec6}, we discuss the implications for neutron decay and the determination of $V_{ud}$ and comment on the relation to superallowed $0^+ \to 0^+$ transitions. Conclusions and outlook are presented in Section~\ref{sec8}. Appendix~\ref{sect:alpha} contains details about electric charge renormalization and running in the LEFT and Chiral Perturbation Theory. Appendix~\ref{app:nrqed} discusses the factorization of the nonrelativistic Fermi function in nonrelativistic QED, while Appendix~\ref{sect:gammas} contains details on the extraction of the $\mathcal O(\alpha^2)$ anomalous dimension in LEFT and HBChPT/$\slashpi$EFT.

\section{Statement of the problem and results}
\label{sec2}

Neutron decay is a low-energy process characterized by the energy scales of the neutron-proton mass difference, $m_n - m_p \approx 1.3~\mathrm{MeV}$, and the electron mass $m_e \approx 511~\mathrm{keV}$. These scales, which we denote by $q_{\rm ext}$, are much smaller than the pion mass, $m_\pi \approx 137$ MeV,  the nucleon mass, $m_N \approx 939$ MeV, and the $W$ boson mass, $M_W \approx 80$ GeV. The existence of widely separated mass scales makes the process amenable to a description based on EFTs. In this work, we systematically implement EFT methods to study low-energy charged-current processes such as neutron decay. We first integrate out the heavy particles ($W$, $Z$, $h$, $t$) and match the full Standard Model onto the so-called LEFT. Subsequently, we  integrate out the scale of the nucleon mass, by matching the LEFT onto HBChPT~\cite{Jenkins:1990jv}. We finally integrate out physics at the scale of the pion mass, following~\cite{Cirigliano:2022hob}, by matching HBChPT onto $\slashpi$EFT. The neutron decay rate is thus organized in an expansion in several small parameters (besides $G_F q_{\rm ext}^2$, which sets the overall scale): the electromagnetic coupling constant $\alpha$, $\epsilon_{\rm recoil} = q_{\rm ext}/m_N$, which describes small kinematic corrections, $\epsilon_\slashpi = q_{\rm ext}/m_\pi$, which captures the radiative pion contributions, and the HBChPT expansion parameter $\epsilon_\chi = m_\pi/\Lambda_\chi$ with the scale $\Lambda_\chi = 4 \pi F_\pi \approx 1~\mathrm{GeV}$.

The neutron decay rate is most conveniently computed starting from the $\slashpi$EFT in which $\beta$ decays are described by the Lagrangian ~\cite{Ando:2004rk,Raha:2011aa,Falkowski:2021vdg}
\begin{equation}
\mathcal L_{\slashpi} = - \sqrt{2} G_F V_{ud} \, \bar{e} \gamma_\rho \mathrm{P}_\mathrm{L} \nu_e  \, \bar{N}_v \left(  g_V v^\rho - 2 g_A S^\rho \right) \tau^+  N_v   + \mathcal{O} \left(  \alpha, \epsilon_{\rm recoil} , \epsilon_{\slashpi}, \epsilon_\chi \right) + \mathrm{h.c.},\label{eq:Lagrangian_at_leading_order}
\end{equation}
where $N_v = \left( p,~n \right)^T$ denotes the heavy-nucleon field doublet, $v^\rho$ is the nucleon velocity, and  $S^\rho = \left( 0,  \vec\sigma/2 \right)$ denotes the nucleon spin, with the Pauli matrices $\sigma$, while $\tau$ denotes Pauli matrices in the isospin space, satisfying $ [\tau^a,\tau^b] = 2 i \varepsilon^{a b c} \tau^c$, $ \{\tau^a,\tau^b \} = 2 \delta^{a b}$, and $\tau^+ = \frac{1}{2} \left( \mathrm{\tau}^1 + i \mathrm{\tau}^2 \right)$. Higher-order terms in Eq.~\eqref{eq:Lagrangian_at_leading_order} include the contributions of weak magnetism, recoil corrections, and induced tensor coupling \cite{Cirigliano:2022hob}. The couplings $g_V$ and $g_A$ themselves have an expansion in $\alpha$, $\epsilon_{\slashpi}$, and $\epsilon_{\chi}$. At leading order, one has $g_V = 1$. At $\mathcal O(\alpha)$, $g_V$ does not receive any long-distance corrections from pion or photon loops and only picks up contributions from local $\mathcal O(e^2 p)$ operators in the HBChPT Lagrangian~\cite{Cirigliano:2022hob}:
\begin{align}
    g_V &= C_\beta^r\left[1 + \frac{\alpha}{2 \pi} \hat{C}_V \right],\label{eq:vector_coupling_CT}  \\
    \hat{C}_{V} &= 8 \pi^2 \left[ -\frac{X_6}{2} + 2 \left({V}_1 +{V}_2+{V}_3 + {V}_4 \right) - g_9 \right].\label{eq:vector_contribution_LECs_v1}
\end{align}
Here, $C_\beta^r = 1 + O(\alpha)$ is the Wilson coefficient of the Fermi operator in LEFT [see Eq.\ \eqref{eq:LLEFT}], which captures electroweak corrections from energy scales above $\Lambda_\chi$. The LECs $X_6, g_9, V_{1,2,3,4}$, and associated HBChPT operators will be defined below in Eqs.~\eqref{eq:LX6},~\eqref{eq:Lagrangian_electromagnetic}, and~\eqref{eq:operators_electroweak}. The couplings $g_{V,A} (\mu_\chi)$ depend on the renormalization scale of the $\slashpi$EFT (in a way that cancels in the ratio $\lambda = g_A/g_V$) and encode contributions from the weak scale all the way down to the pion mass scale.

In the following sections, we will detail the various steps needed to connect the low-energy coupling $g_V$ to the electroweak scale, following a top-down approach. Key new results of this work are as follows:
(i) The expression for $g_V (\mu_\chi \sim \Lambda_\chi)$ in terms of the Wilson coefficient $C_\beta^r$ computed with anomalous dimensions of $\mathcal{O}(\alpha, \alpha \alpha_s, \alpha^2)$ and a ``subtracted" hadronic function, related to the traditional non-perturbative $\gamma W$ box contribution evaluated in the recent literature~\cite{Seng:2018yzq,Seng:2018qru,Czarnecki:2019mwq,Shiells:2020fqp,Hayen:2020cxh} (see Eq.~\eqref{eq:gV2} and discussion surrounding it);
(ii) The use of two-loop anomalous dimensions in the RGE~\eqref{eq:RGE_LE} needed to evolve the vector coupling down to $g_V (\mu_\chi \sim m_e)$, resumming large next-to-leading logarithms of order $\alpha^2 \ln \left(m_N/m_e \right)$. The resulting $g_V (\mu_\chi \sim m_e)$ is directly relevant to the calculation of neutron decay and can be used as input for the one-body contribution to nuclear decays.

In this work, we have focused on the application to neutron decay. With $g_V (\mu_\chi \sim m_e)$ at hand, we combined both virtual and real photon corrections to the decay rate~\cite{Sirlin:1967zza,Ando:2004rk,Cirigliano:2022hob} to obtain the effective phase-space correction $\Delta_f$ and the radiative correction $\Delta_R$ to the neutron lifetime, see Section~\ref{sec6}, and the relation
\begin{equation} 
    |V_{ud}|^2  \, \tau_n \, \left( 1 + 3 \lambda^2 \right) \left(1 + \Delta_f  \right) \left(1 + \Delta_R  \right) = 5283.321(5)~\mathrm{s},\label{eq:Vudn}
\end{equation}
with $\Delta_f$ and $\Delta_R$ given in Eqs.~\eqref{eq:DeltaF1} and~\eqref{eq:DeltaR1}, respectively. Our definitions for $\Delta_f$ and $\Delta_R$ differ from the  traditional approach both conceptually and numerically. Technically, the bulk of this difference is in shifting all short-distance contributions from $\Delta_f$ to $\Delta_R$. $\Delta_f$ describes Coulomb-enhanced long-distance contributions and recoil corrections, while $\Delta_R$ includes all electroweak and HBChPT short-distance contributions along with the non-Coulomb radiative corrections in $\slashpi$EFT, as specified in Eqs.~\eqref{eq:gV2},~\eqref{eq:gVme},~and~\eqref{eq:DeltaRf}. Numerically, we find
\begin{align}
    \Delta_f &= 3.573(5) \times 10^{-2},\label{eq:DFnum} \\
    \Delta_R &= 4.044(24)_{\rm Had} (8)_{\alpha \alpha_s^2} (7)_{\alpha \epsilon_\chi^2} (5)_{\mu_\chi} [27]_\text{total} \times 10^{-2}.\label{eq:DRnum}
\end{align}
The uncertainty in $\Delta_f$ stems from an estimate of mixed  recoil times Coulomb corrections. The dominant sources of uncertainty to $\Delta_R$ are given by the following: the non-perturbative hadronic contributions, associated with the ``$\gamma W$ box" diagram in the standard approach~\cite{Seng:2018yzq,Seng:2018qru,Czarnecki:2019mwq,Shiells:2020fqp,Hayen:2020cxh,Seng:2020wjq}; contributions of $\mathcal{O}(\alpha \alpha_s^2)$ not included in our renormalization group analysis in the LEFT; chiral corrections of $\alpha \epsilon_\chi^2$; and residual dependence on the $\slashpi$EFT renormalization scale, varied between $m_e/\sqrt{2}$ and $\sqrt{2} m_e$, which is an indicator of the $\mathcal{O}(\alpha^2)$ corrections. A detailed discussion of uncertainties is presented in Sections~\ref{sec:numerics} (for $g_V$) and \ref{sect:rate} (for the remaining contributions to $\Delta_R$).

Our result for $\Delta_f$ in Eq.~\eqref{eq:DFnum} differs from the one found in the literature $\Delta_f = 3.608 \times 10^{-2}$~\cite{Czarnecki:2004cw} by $-0.035\%$. This is because in the phase-space integration we use the nonrelativistic  Fermi function, for the reasons discussed in Section~\ref{sect:mecorr}, and neglect  corrections induced by modeling the proton as a uniformly charged sphere of radius $R_p \simeq 1$~fm~\cite{Wilkinson:1982hu} (this effect is at the level of $0.005\%$).

Our result for $\Delta_R$ in Eq.~\eqref{eq:DRnum} exceeds the current value $\Delta_R =  3.983(27) \times 10^{-2}$, compiled in Ref.~\cite{Cirigliano:2022yyo} by combining the results of \cite{Seng:2018yzq,Seng:2018qru,Czarnecki:2019mwq,Seng:2020wjq,Hayen:2020cxh,Shiells:2020fqp}, by about twice the estimated uncertainties. The $+0.061\%$ shift in the central value is almost entirely due to the different treatment of the next-to-leading logarithmic terms at the hadronic level, i.e., the terms that scale as $\alpha^2 \ln \left(m_N/m_e \right)$. In both approaches, there is a contribution of this type coming from the cross term between the one-loop RGE correction to $g_V$, scaling as $\frac{\alpha}{\pi} \ln \left(m_N/m_e \right)$, and $\mathcal{O} \left(\pi \alpha / \beta \right)$ terms in the Fermi function. In our approach, additional $\alpha^2 \ln \left(m_N/m_e \right)$ large logarithmic corrections arise entirely from the two-loop anomalous dimension contribution to the RGE~\eqref{eq:RGE_LE} for the effective coupling $g_V(\mu_\chi)$ and produce a positive shift in $\Delta_R$ of $0.010\%$. In the EFT approach, there are no other sources of large logarithms of the ratio $\left(m_N/m_e \right)$ in the matrix element of the four-fermion operator \eqref{eq:Lagrangian_at_leading_order} to $\mathcal{O}(\alpha^2)$. In the literature, this class of effects is not associated with the running of $g_V$, but arises through the {\it negative} correction $\alpha/(2 \pi) \times \delta  = - 0.043\%$, introduced in Ref.~\cite{Czarnecki:2004cw} by adapting the results of Refs.~\cite{Sirlin:1986cc,Jaus:1970tah}.\footnote{In the standard non-EFT approach, additional terms scaling as $\alpha^2 \ln \left(m_N/m_e \right)$ (or $\alpha^2 \ln (R_p m_e)$ after including finite nucleon size effects) are included in the relativistic Fermi function (see discussion in Section~\ref{sect:mecorr}) and booked as effective phase-space corrections appearing in $\Delta_f$. It is worth noting that, for neutron decay, the $\alpha^2 \ln (R_p m_e)$ terms in the relativistic Fermi function cancel the corresponding terms in the correction $\alpha/(2 \pi) \times \delta$~\cite{Czarnecki:2004cw}.} The mismatch of the two approaches produces a $+0.053\%$ shift in our results. The remaining difference is due to a combination of the following, individually smaller, effects:
(i) we re-evaluate the ``elastic" hadronic contribution, as discussed in Section~\ref{sect:box}, which leads to a $-0.006\%$ shift to $\Delta_R$;
(ii) for the next-to-leading logarithmic corrections of $\mathcal{O}(\alpha^2 \ln (M_W/m_c))$, our result differs from the one in Ref.~\cite{Czarnecki:2004cw}, producing a negative shift of approximately $-0.011\%$;
(iii) we do not include $\mathcal{O}(\alpha \alpha_s^2)$ terms in the running of our Wilson coefficient (corresponding to the ``deep inelastic scattering" region of the $\gamma W$ box in the literature) that amounts to a net $+0.007\%$ in $\Delta_R$; and 
(iv) finally, different choices in the factorization between electroweak and $m_N/m_e$ logarithms compared to Refs.~\cite{Czarnecki:2004cw,Cirigliano:2022yyo} account for the remaining mismatch.

Using  $\Delta_{f,R}$ from Eqs.~\eqref{eq:DFnum} and \eqref{eq:DRnum}, respectively, in the master formula~\eqref{eq:Vudn}, we can extract $V_{ud}$. This requires experimental input for the neutron lifetime $\tau_n$ and the ratio $\lambda$ of axial-vector to vector couplings. Using the PDG~\cite{ParticleDataGroup:2022pth,Workman:2022ynf} averages for the experimental input, we obtain
\begin{equation}
    V_{ud}^\text{n, PDG}= 0.97430 (2)_{\Delta_f} (13)_{\Delta_R}(82)_\lambda(28)_{\tau_n}[88]_\text{total}.
\end{equation}
Both $\tau_n$ and $\lambda$ carry an inflated error due to scale factors. Following Ref.~\cite{Cirigliano:2022yyo}, if we instead use the most precise neutron lifetime measurement $\tau_n = 877.75(36)~\mathrm{s}$ from UCN$\tau$@LANL~\cite{UCNt:2021pcg} and the determination of $\lambda$ from the most precise measurement of the beta asymmetry in polarized neutron decay by PERKEO-III~\cite{Markisch:2018ndu,Dubbers:2018kgh}, we obtain a very competitive extraction of $V_{ud}$ from  neutron decay:
\begin{equation}
    V_{ud}^\text{n, best} = 0.97402 (2)_{\Delta_f} (13)_{\Delta_R} (35)_\lambda  (20)_{\tau_n}[42]_\text{total},
\end{equation}
with an uncertainty approaching the currently quoted error  $\delta V_{ud}= 31 \times 10^{-5}$ from $0^+ \to 0^+$ nuclear beta decays~\cite{Hardy:2020qwl}. Compared to the baseline correction of Refs.~\cite{Cirigliano:2022yyo,Seng:2018yzq,Seng:2018qru,Czarnecki:2019mwq,Seng:2020wjq,Hayen:2020cxh,Shiells:2020fqp}, the positive shift of $+0.061\%$ in $\Delta_R$ and the negative shift of $-0.035\%$ in $\Delta_f$ partially compensate, producing a smaller positive shift of $+0.026\%$ in the correction to the rate. This one, in turn, provides a negative shift in $V_{ud}$,  $\delta V_{ud} \simeq - 13 \times 10^{-5}$, compared to the results quoted in Ref.~\cite{Cirigliano:2022yyo}.

In the remainder of this paper, we provide details on the derivation of the results presented above.

\section{Step I: matching the Standard Model to LEFT}
\label{sec3}

In this Section, we perform the matching of the Standard Model to the LEFT and present the RGE that control the effective couplings in the LEFT between the electroweak and QCD scales. We then  introduce spurions and external sources in the LEFT to describe the electromagnetic and weak interactions of quarks~\cite{Knecht:1999ag,Descotes-Genon:2005wrq}, which is particularly useful in the matching of LEFT to chiral perturbation theory, to be described in subsequent sections. Throughout, we regulate the UV divergences in dimensional regularization, working in $d=4-2 \epsilon$ spacetime dimensions.

\subsection{Wilson coefficient and RGE} 
\label{sec:subsec31}

The part of the LEFT Lagrangian relevant for muon and $\beta$ decays just below the weak scale reads
\begin{equation}
    {\cal L}_{\rm LEFT}  =   - 2 \sqrt{2} G_F \ \bar{e}_L \gamma_\rho \mu_{L}  \,  \bar{\nu}_{\mu L} \gamma^\rho \nu_{eL} - 2 \sqrt{2} G_F  V_{ud} \ C^r_\beta (a,\mu)  \ \bar{e}_L \gamma_\rho \nu_{eL}  \,  \bar{u}_{L} \gamma^\rho d_L + \ {\rm h.c.} + ... ~.\label{eq:LLEFT}
\end{equation}
\noindent  Here $\mu$ denotes the $\overline{\rm MS}$ renormalization scale and
\begin{equation}
    G_F = \frac{\pi \alpha \left(\mu \right) g\left(\mu \right)}{\sqrt{2} M_W^2 \left(\mu \right) s^2_W\left(\mu \right)},
\end{equation}
is the scale-independent Fermi constant that is extracted from precise measurements of the muon lifetime~\cite{vanRitbergen:1999fi,FAST:2007rsc,Casella:2013bla,MuLan:2012sih}, expressed in terms of  $\overline{\rm MS}$ Standard Model parameters (with $s_W^2 = 1 - M_W^2/M_Z^2$). The function $g\left(\mu \right)$ can be found in Ref.~\cite{Hill:2019xqk} and reduces to $g \left(\mu \right) =1$ at tree level. The effective coupling multiplying the semileptonic operator that mediates $\beta$ decays involves the {\it same} $G_F$ as the pure-leptonic term in Eq.~(\ref{eq:LLEFT}), the CKM matrix element $V_{ud}$, and the $\overline{\rm MS}$-subtracted Wilson coefficient $C^r_\beta \left(a, \mu \right)$, which reads~\cite{Hill:2019xqk,Dekens:2019ept,Sirlin:1981ie}
\begin{align}
    C^r_\beta \left( a, \mu \right) &= 1 + \frac{\alpha}{\pi} \, \ln \frac{M_Z}{\mu} + \frac{\alpha}{\pi} B \left( a \right)  - \frac{\alpha \alpha_s}{4 \pi^2} \ln \frac{M_W}{\mu}+ \mathcal{O} (\alpha \alpha_s) + \mathcal{O} (\alpha^2), \\
    B \left( a \right) &= \frac{a}{6} - \frac{3}{4}.
\end{align}
The finite $\mathcal{O}(\alpha)$ matching coefficient depends on the scheme through $B(a)$. We have used the Naive Dimensional Regularization (NDR) scheme for $\gamma_5$ and kept track of the additional evanescent operator scheme dependence via the parameter $a$, defined by~\cite{Buras:1989xd,Dugan:1990df,Herrlich:1994kh}
\begin{equation}
    \gamma^\alpha \gamma^\rho \gamma^\beta \mathrm{P}_\mathrm{L} \otimes \gamma_\beta \gamma_\rho \gamma_\alpha \mathrm{P}_\mathrm{L} = 4 \left[ 1 + a \left( 4-d\right) \right] \gamma^\rho \mathrm{P}_\mathrm{L} \otimes \gamma_\rho \mathrm{P}_\mathrm{L} + \mathrm{E} \left( a \right),
\end{equation}
with an evanescent operator $\mathrm{E}(a)$ that has a vanishing matrix element in $d=4$. Current conservation protects $C_\beta$ from $\mathcal{O}(\alpha_s)$ corrections. Concerning the terms of $\mathcal{O}(\alpha \alpha_s)$, we only keep logarithmic contributions, as the finite matching coefficients and  the corresponding three-loop  anomalous dimensions are not known.

The renormalized Wilson coefficient $C^r_\beta \left(a, \mu \right)$ obeys the following RGE:
\begin{subequations} \label{eq:RGE}
\begin{eqnarray}
    \mu \frac{\mathrm{d} C^r_\beta \left(a, \mu \right)}{\mathrm{d} \mu} &=&  \gamma  (\alpha, \alpha_s)  \ C^r_\beta \left(a, \mu \right), \\
    \gamma (\alpha, \alpha_s) &=&  \gamma_0 \, \frac{\alpha}{\pi} + \gamma_1 \,   \left( \frac{\alpha}{\pi} \right)^2  + \gamma_{se}  \, \frac{\alpha}{\pi}  \, \frac{\alpha_s}{4 \pi}   \ + \  \cdots, \\
    \gamma_0 &=& -1~\text{\cite{Sirlin:1981ie}}, \\
    \gamma_1^{NDR} (a)  &=& \frac{\tilde n}{18} \, \left(2 a +1 \right), \qquad \qquad \tilde n = \sum_f n_f Q_f^2, \\
    \gamma_{se} &=& + 1~\text{\cite{Sirlin:1981ie,Erler:2002mv,Hill:2019xqk}}, 
\end{eqnarray}
\end{subequations}
where $\tilde n$ is the scale-dependent effective number of fermions, $\alpha\left(\mu \right)$ and $\alpha_s\left(\mu \right)$ are the electromagnetic and strong running coupling constants. We have obtained  $\gamma_1^{NDR}(a)$ by adapting the QCD calculation in~\cite{Buras:1989xd}. As far as we know, this is the first time the full two-loop anomalous dimension is worked out.\footnote{Ref.~\cite{Czarnecki:2004cw} quotes the $\tilde n$-enhanced component of $\gamma_1$. Taking into account the different normalization, Ref.~\cite{Czarnecki:2004cw} obtains  $\gamma_1^{NDR} (a=-1) = -  (1/16) \times (44/9)  \tilde n + \mathcal{O}(\tilde n^0)$, while we find $\gamma_1^{NDR} (a=-1) = - (1/16) \times (8/9) \tilde n$ for the total.}  With appropriate rescalings of the QCD diagrams of Ref.~\cite{Buras:1989xd}, we also reproduce $\gamma_{se}=1$. $\gamma_0$ and $\gamma_{se}$ are scheme-independent. The scheme independence of $\gamma_{se}$ follows from the general argument given in Ref.~\cite{Buras:1992zv}, combined with the fact that there is no finite matching term nor anomalous dimension to $\mathcal{O}(\alpha_s$) for the operator under study here. On the other hand, $\gamma_1$ depends on both the treatment of $\gamma_5$ in $d$ spacetime dimensions and on the scheme used for evanescent operators.

In our final result, we will use the numerical solution for $C_\beta^r(a,\mu)$. However, it is quite instructive to provide an approximate analytic solution,  based on the perturbative treatment of the next-to-leading logarithm (NLL) terms associated with the scheme-dependent two-loop anomalous dimension $\gamma_1 (a) = \mathcal{O} (\alpha^2)$ and the finite one-loop matching condition $B(a)$. First, setting $\gamma_1 (a) \to 0$ and consistently taking as an initial condition $C^r_\beta (a, \mu_{SM}) = 1$, generalizing the result of Ref.~\cite{Erler:2002mv} given at the $\tau$-mass scale, we obtain the solution
\begin{align}
    \tilde C_\beta^r (\mu) = \left( \frac{\alpha \left( m_c \right)}{\alpha \left( \mu \right)} \right)^{\frac{3}{8}} \left( \frac{\alpha \left( m_\tau \right)}{\alpha \left( m_c \right)} \right)^{\frac{9}{32}} \left( \frac{\alpha \left( m_b \right)}{\alpha \left( m_\tau \right)} \right)^{\frac{9}{38}} \left( \frac{\alpha \left( \mu_{SM} \right)}{\alpha \left( m_b \right)} \right)^{\frac{9}{40}} \nonumber \\
    \times \left( \frac{\alpha_s \left( m_c \right)}{\alpha_s \left( \mu \right)} \right)^{\frac{1}{18} \frac{\alpha \left( \mu \right)}{\pi}} \left( \frac{\alpha_s \left( m_b \right)}{\alpha_s \left( m_c \right)} \right)^{\frac{3}{50} \frac{\alpha \left( m_c \right)}{\pi}} \left( \frac{\alpha_s \left( \mu_{SM} \right)}{\alpha_s \left( m_b \right)} \right)^{\frac{3}{46} \frac{\alpha \left( m_b \right)}{\pi}},
\end{align}
where we have subsequently integrated out the $b$ quark,  $\tau$ lepton, and $c$ quark, and the strong and electromagnetic running couplings are obtained by solving the one-loop RGEs. This solution resums all the terms of $\mathcal{O}( \alpha^n \ln^n (\mu_{SM}/\mu))$ and $\mathcal{O}( \alpha \alpha_s^n \ln^n (\mu_{SM}/\mu))$. We can then perturbatively include the effects of $\mathcal{O}(\alpha^2 \ln (\mu_{SM}/\mu))$ due to $\gamma_1 (a)$  and $B(a)$, arriving at
\begin{equation}
    C_\beta^r (a, \mu) = \left(1 + \frac{\alpha (\mu)}{\pi} B (a) \right) \times  \tilde C_\beta^r (\mu) \times  \delta_{NLL} (\mu),  \label{eq:analytic_sol}
\end{equation}
where
\begin{align}
    \delta_{NLL} (\mu) &= 1 - \kappa  \Bigg( \tilde n (m_b) \, \left( \frac{\alpha (m_b)}{\pi} \right)^2 \, \ln \frac{\mu_{SM}}{m_b}  \ +  \  \tilde n (m_\tau) \, \left( \frac{\alpha (m_\tau)}{\pi} \right)^2 \, \ln \frac{m_b}{m_\tau}  \nonumber \\
    &   \quad + \   \tilde n (m_c) \, \left( \frac{\alpha (m_c)}{\pi} \right)^2 \, \ln \frac{m_\tau}{m_c}  \ + \   \tilde n (\mu) \, \left(\frac{\alpha (\mu)}{\pi} \right)^2  \, \ln \frac{m_c}{\mu}\Bigg) \nonumber \\
    & \approx  \ 1 - \kappa \,  \tilde n (m_b) \left( \frac{\alpha (\mu)}{\pi} \right)^2 \, \ln \frac{\mu_{SM}}{\mu},
\end{align}
and the scheme-independent combination $\kappa$ is given by\footnote{Ref.~\cite{Czarnecki:2004cw} finds $\kappa =  2/9$, more than a factor of two smaller compared to our result.}
\begin{equation}
    \kappa  = \frac{1}{\tilde{n}} \left( \gamma_1 (a) +  \frac{1}{2} \beta_0 B (a) \right)  = \frac{5}{9}.
\end{equation}
In the equation above,  $\beta_0 =-(4/3) \tilde n$ controls the one-loop $\beta$ function for $\alpha$ via $\mu d \alpha/ d \mu = - (\beta_0/(2 \pi)) \alpha^2$. The scale-dependent effective number of fermions takes the values $\tilde n (\mu < m_c)=4$, $\tilde n(m_c) = 16/3$, $\tilde n (m_\tau)= 19/3$, and $\tilde n (m_b) = 20/3$. Note that the scheme dependence of $C_\beta^r (a, \mu)$  in the solution \eqref{eq:analytic_sol} appears only through the initial factor involving $B(a)$. As we will show below, this term  explicitly cancels when one includes the $\mathcal{O}(\alpha)$ corrections to the matrix element of the semileptonic operator $\bar{u}_L \gamma_\alpha d_L \bar{e}_L \gamma^\alpha \nu_{eL}$.

We also provide an analytic solution to the RGE~\eqref{eq:RGE} in terms of the evolution operator $U(\mu, \mu_{SM})$ to NLL accuracy, formally written as  $C^r_\beta \left(a, \mu \right) = U(\mu, \mu_{SM})  C^r_\beta (a, \mu_{SM})$, with the initial condition $C^r_\beta (a, \mu_{SM}) = 1 + (\alpha/\pi) ( \ln (M_Z/\mu_{SM}) + B \left(a \right))$. Using the two-loop running coupling $\alpha (\mu)$ and the one-loop running $\alpha_s (\mu)$, we resum the series of leading logarithms ($n \geq 1$) $\mathcal{O}( \alpha^n \ln^n (\mu_{SM}/\mu))$, and sub-leading logarithms $\mathcal{O}( \alpha \alpha_s^n \ln^n (\mu_{SM}/\mu))$ and $\mathcal{O}(\alpha^{n+1} \ln^n (\mu_{SM}/\mu))$. The NLL solution for the evolution operator $U(\mu_1,\mu_2)$, valid  between two mass thresholds $\mu_1$ and $\mu_2$ takes the form~\cite{Buras:1990fn,Buchalla:1995vs,Neubert:2000ag}
\begin{equation}
    U(\mu_1,\mu_2) =\left( \frac{\alpha(\mu_1)}{\alpha(\mu_2)}\right)^{-\frac{2\gamma_0}{\beta_0}} \left( \frac{\alpha_s(\mu_1)}{\alpha_s(\mu_2)}\right)^{-\frac{2\gamma_{se}}{\beta_{0,s}}\frac{\alpha(\mu_1)}{4\pi}}\left[1-\frac{2\gamma_1}{\beta_0}\frac{\alpha(\mu_1)-\alpha(\mu_2)}{\pi}\right]\,,\label{eq:U}
\end{equation}
where we expanded $\alpha$ with respect to its two-loop beta function, $\beta_1$, after which the $\beta_1$ dependence cancels in Eq.\ \eqref{eq:U}. Therefore, both $\alpha$ and $\alpha_s$ in Eq.\ \eqref{eq:U} are evaluated using the one-loop RGEs, and the QCD beta function $\beta_{0,s}$ is expressed in terms of the number of active quarks $n_f$ as $\beta_{0,s} = \left(11 N_c-2n_f\right)/3$. Neglecting two-loop matching conditions, the evolution operator between the electroweak scale, $\mu_{SM}$, and the low-energy scale, $\mu$, can then be obtained by using Eq.\ \eqref{eq:U} between each particle threshold $U(\mu,\mu_{SM})= U(\mu,m_c) U(m_c,m_\tau)U(m_\tau,m_b)U(m_b,\mu_{SM})$.

\subsection{External sources and spurions}
\label{sec:subsec32}

The matching of LEFT to HBChPT is conveniently performed by introducing classical source fields $\bar l^\mu(x)$ and $\bar r^\mu (x)$ for the left- and right-handed chiral currents of quarks  as well as electromagnetic left ${\bf q}_L$ and right ${\bf q}_R$ spurions, and the weak spurion ${\bf q}_W$~\cite{Urech:1994hd,Moussallam:1997xx,Knecht:1999ag,Descotes-Genon:2005wrq}. These allow one to handle the explicit chiral symmetry breaking introduced by the electromagnetic and weak interactions at the quark level in a compact way. With this motivation in mind, we write the source term for currents plus the QED and weak CC interactions of the light quarks $q^T = (u,d)$ as
\begin{equation}
    {\cal L}_{\rm LEFT} = \bar{q}_L \slashed{\bar l} q_L + \bar{q}_R \slashed{\bar r} q_R - e \left( \bar{q}_L {\bf q}_L \slashed{A} q_L + \bar{q}_R {\bf q}_R \slashed{A} q_R \right)  +   \left( \bar{e}_L \gamma_\rho \nu_{eL} \, \bar{q}_L {\bf q}_W \gamma^\rho  q_L + \mathrm{h.c.}\right) + ...,\ \label{eq:LEFT2}
\end{equation}
where $A^\mu$ denotes the photon field. The Lagrangian in~\eqref{eq:LEFT2} is invariant under local $G=SU(2)_L \times SU(2)_R \times U(1)_V$ transformations
\begin{equation}
    q_L \to L (x)e^{i\alpha_V(x)}  q_L, \quad q_R \to R(x) e^{i\alpha_V(x)}  q_R,
\end{equation}
with $L,R \in SU(2)_{L,R}$, provided ${\bf q}_{L,R}$ and ${\bf q}_W$ transform as ``spurions"  under the chiral group, namely ${\bf q}_{L,W} \to L {\bf q}_{L,W} L^\dagger$ and ${\bf q}_R \to R  {\bf q}_R R^\dagger$, and that $\bar{l}_\mu$ and $\bar{r}_\mu$ transform as gauge fields under $G$. At the physical point,
\begin{equation}
    {\bf q}_L = {\bf q}_R = {\rm diag} (Q_u, Q_d),\qquad \qquad {\bf q}_W = -2 \sqrt{2}  \mathrm{G}_\mathrm{F} V_{ud} C_\beta^r\, \tau^+,
\end{equation}
with $\tau^+ = (\tau^1 + i \tau^2)/2$ in terms of the Pauli matrices $\tau^{a}$. Note that  we include  the LEFT Wilson coefficient $C_\beta^r$  in the definition of the spurion ${\bf q}_{W}$.  With this identification, Eq.~\eqref{eq:LEFT2} reproduces the semileptonic piece of Eq.~\eqref{eq:LLEFT}.

The $\mathcal{O}(e^2 )$ counterterms in the LEFT Lagrangian can be written in terms of spurions as~\cite{Descotes-Genon:2005wrq}
\begin{align}
    {\cal L}^{\rm CT}_{\rm LEFT} &=  -2 e^2 Q^2_e g_{00} \, \bar{e} \left( i \slashed{\partial} - e Q_e \slashed{A} -m_e\right) e - i g_{23} e^2 \left( \bar{q}_L \left[ {\bf q}_L, D^\rho {\bf q}_L \right] \gamma_\rho q_L + \bar{q}_R \left[ {\bf q}_R, D^\rho {\bf q}_R \right] \gamma_\rho q_R \right) \nonumber \\
    & +   e^2 Q_e \Big( \bar{e}_L \gamma_\rho  \nu_L \left( g_{02} \,  \bar{q}_L {\bf q}_W  {\bf q}_L \gamma^\rho q_L + g_{03}  \, \bar{q}_L {\bf q}_L  {\bf q}_W \gamma^\rho  q_L \right) + \mathrm{h.c.} \Big), \label{eq:CT}
\end{align}
where $g_{00}$ is the counterterm related to the electron wavefunction renormalization, $g_{02}$ and $g_{03}$ come from the counterterm of $C_\beta$, while $g_{23}$ includes contributions from both the counterterm of $C_\beta$ as well as divergences related to the quark wavefunction renormalization. Furthermore,
\begin{align} 
    D^\rho {\bf q}_L  &\equiv  \partial^\rho {\bf q}_L - i \left[ l^\rho, {\bf q}_L \right], \label{eq:charge_covariant_derivative_left} \\
    D^\rho {\bf q}_R &  \equiv  \partial^\rho {\bf q}_R - i \left[ r^\rho, {\bf q}_R \right], \label{eq:charge_covariant_derivative_right}
\end{align}
are chiral covariant derivatives, expressed in terms of the fields $l^\mu (x)$ and $r^\mu (x)$ that combine the classical sources, the photon, the leptons, and the spurions:
\begin{align}
    l_\mu &= \bar l_\mu - e {\bf q}_L A_\mu  + {\bf q}_W \, \bar{e}_L \gamma_\mu \nu_{eL} + {\bf q}_W^\dagger \bar{\nu}_{eL} \gamma_\mu e_L, \\
    r_\mu &= \bar r_\mu - e {\bf q}_R A_\mu.\label{eq:sources}
\end{align}
In the $\overline{\rm MS}$ scheme, the $g_{ij}$ couplings appearing in Eq.~\eqref{eq:CT} are determined by the $1/\varepsilon$ divergences and can be written as
\begin{equation}
    g_{ij} = \frac{h_{ij}}{\left(4 \pi \right)^2} \left( \frac{1}{\varepsilon} - \gamma_E + \ln \left( 4 \pi \right) \right)  , \label{eq:gij}
\end{equation}
with $h_{00} = 1/2$, $h_{23}= (1/2) (1 - \alpha_s/\pi) $, $h_{02} = -1 -  \alpha_s/\pi$, and $h_{03} = 4 -2 \alpha_s/\pi$.

\section{Step II: matching LEFT to HBChPT}
\label{sec4}

The goal of this section is to find a representation for the LECs appearing in $\hat{C}_V$, see \eqref{eq:vector_contribution_LECs_v1}, in terms of the LEFT counterterms $g_{ij}$ and quark correlation functions, which can then be modeled or computed via non-perturbative techniques such as lattice QCD.

\subsection{The Chiral Lagrangian}
\label{sec:subsec41}

The chiral representation of Eq.~\eqref{eq:CT} can be built using standard spurion techniques. As in Eq.~\eqref{eq:CT}, we will need purely leptonic operators, purely electromagnetic operators, and operators with charged leptons and neutrinos. The corresponding chiral Lagrangians were built in Refs. \cite{Meissner:1997ii,Muller:1999ww,Knecht:1999ag,Gasser:2002am,Cirigliano:2022hob}. Here we extend the bases of \cite{Gasser:2002am,Cirigliano:2022hob} in order to avoid assumptions regarding $\mathbf{q}_L$ and $\mathbf{q}_R$, allowing us to keep the spurions $\mathbf{q}_{L,R}$ completely general. Moreover, we do not use the equations of motions to reduce the operator set in order to avoid hadronic contributions to purely leptonic LECs~\cite{Descotes-Genon:2005wrq}.

As we will see below, to perform the matching between LEFT and HBChPT it is convenient to introduce vector and axial-vector charge spurions and sources, which we define as
\begin{equation}
    {\bf{q}}_V = {\bf{q}}_L + {\bf{q}}_R, \qquad {\bf{q}}_A = {\bf{q}}_L - {\bf{q}}_R, \qquad \mathrm{v}_\rho = l_\rho + r_\rho, \qquad \mathrm{a}_\rho = l_\rho - r_\rho.
\end{equation}
It is also convenient to decompose the electromagnetic charge spurions in  isovector and isoscalar components
\begin{equation}
    {\bf{q}}^\mathrm{baryon}_J = {\bf q}^0_J + {\bf q}^a_J \tau^a, \qquad {\bf{q}}_J^\mathrm{quark} =  \frac{ {\bf q}_J^0}{3} + {\bf q}_J^a \tau^a,\label{eq:nucleon_and_quark_spurions}
\end{equation}
with  $J \in \{ L, R, V, A\}$. The physical values are ${\bf q}_{L,R}^0 = {\bf q}_{L, R}^3 = \frac{1}{2}$ for the left and right spurions, ${\bf q}_V^0 = {\bf q}_V^3 = 1$ for the vector spurion, and ${\bf q}_A^0 = {\bf q}_A^3 = 0$ for the axial-vector case.\footnote{In what follows, we will omit the superscripts in the charge spurions: whenever ${\bf q}_{L,R,V,A}$ appears in the HBChPT Lagrangian, it is understood to be ${\bf q}_{L,R,V,A}^\mathrm{baryon}$.}

The chiral Lagrangians are built using the chiral covariant functions of the charges and of the corresponding covariant derivatives in Eqs.~\eqref{eq:charge_covariant_derivative_left} and~\eqref{eq:charge_covariant_derivative_right}:
\begin{equation}
    {{\cal{Q}}}^W_L = u {\bf q}_W u^\dagger, \quad {{\cal{Q}}}_L = u {\bf{q}}_L u^\dagger, \quad  {{\cal{Q}}}_R = u^\dagger {\bf{q}}_R u, \quad {{\cal{Q}}}_\pm = \frac{ {{\cal{Q}}}_L \pm {{\cal{Q}}}_R}{2}, \quad c_\rho^\pm =  \frac{1}{2} \left( u \left(D_\rho  {\bf{q}}_L\right)  u^\dagger \pm u^\dagger \left(D_\rho {\bf{q}}_R\right)  u  \right),
\end{equation}
with $u^2 = U = \exp (i \boldsymbol{\pi} \cdot \boldsymbol\tau/F_\pi)$ and $F_\pi \approx 92$ MeV.

At lowest order, the HBChPT Lagrangian is given by
\begin{equation}
    \mathcal L^{p^2}_\pi + \mathcal L^{e^2}_\pi + \mathcal L^{p}_{\pi N} = \frac{F_\pi^2}{4} \langle u_\mu u^\mu + \chi_+  \rangle  + e^2 Z_\pi F_\pi^4 \langle {\cal Q}_L {\cal Q}_R  \rangle +  \bar N_v i v \cdot \nabla N_v + g^{(0)}_A \bar N_v S \cdot u N_v,\label{eq:LO}
\end{equation}
where $F_\pi$ and $g_A^{(0)}$ denote the pion decay constant and the nucleon axial-vector coupling in the chiral limit. $u_\mu$ and $\chi_+$ are given by
\begin{equation}
    u_\mu =  i \left[ u^\dagger (\partial_\mu - i r_\mu) u  - u (\partial_\mu - i l_\mu) u^\dagger\right], \qquad \chi_\pm = u^\dagger \chi u^\dagger \pm u \chi^\dagger u, \qquad \chi  =   B_0  (m_q + \bar{s} + i \bar{p}),
\end{equation}
with the light quark masses $m_q$ and a LEC of dimension of mass $B_0$, which is related to the quark condensate. We further introduced the nucleon chiral covariant derivative
\begin{eqnarray}
    \nabla_\mu N &\equiv & \left(\partial_\mu + \Gamma_\mu  \right) N,  \qquad \Gamma_\mu = \frac{1}{2} \left[ u (\partial_\mu - i l^{\rm baryon}_\mu) u^\dagger + u^\dagger (\partial_\mu - i r^{\rm baryon}_\mu) u \right],
\end{eqnarray}
where by the superscript $\textit{baryon}$ we indicate that the photon couples to the nucleon via the charge $\textbf{q}_V^{\rm baryon}$ in Eq.~\eqref{eq:nucleon_and_quark_spurions}. In addition to the weak and electromagnetic interactions arising from chiral covariant derivatives, Eq.~\eqref{eq:LO} contains electromagnetic effects mediated by high-momentum photons via the coupling $Z_\pi$, which is related to the pion-mass splitting.

The chiral Lagrangian needed at $\mathcal O(G_F \alpha)$ is given by
\begin{equation}
    \mathcal L = \mathcal L^{CT}_{\rm lept} + \mathcal L^{e^2 p}_{\pi N} + \mathcal L^{e^2 p}_{\pi N \ell}.
\end{equation}
$\mathcal L^{CT}_{\rm lept}$ is a purely leptonic counterterm Lagrangian
\begin{equation}
    \mathcal L^{CT}_{\rm lept} = e^2 X_6 \bar e \left( i \slashed{\partial} + e \slashed{A}\right) e.\label{eq:LX6}
\end{equation}
The coupling $X_6$ is determined by computing the electron propagator in LEFT and chiral perturbation theory, obtaining
\begin{equation}
    X^r_6 \left( \mu_\chi ,\mu\right)=  \frac{\xi}{\left( 4 \pi \right)^2} \left( 1 - \ln \frac{\mu_\chi^2}{\mu^2} \right),
\end{equation}
in arbitrary $R_\xi$ gauge, where $\mu$ and $\mu_\chi$
are the LEFT and HBChPT renormalization scales, respectively. 
$X_6^r(\mu_\chi,\mu)$ denotes the renormalized coupling, after subtraction of the $1/\varepsilon$ pole in the $\overline{\rm MS}_\chi$ scheme. Note that, following standard practice~\cite{Gasser:1983yg}, in the $\overline{\rm MS}_\chi$ scheme, we subtract
\begin{equation}
    \frac{1}{\varepsilon} - \gamma_E + \ln \left( 4 \pi \right) + 1,
\end{equation}
instead of the conventional $\overline{\rm MS}$ subtraction used in LEFT:
\begin{equation}
    \frac{1}{\varepsilon} - \gamma_E + \ln \left( 4 \pi \right).
\end{equation} 
For the electromagnetic Lagrangian $\mathcal L^{e^2 p}_{\pi N}$, we use the construction of Ref. \cite{Gasser:2002am}. Only one operator is required to describe Fermi transitions,
\begin{equation}
    \mathcal L^{e^2 p}_{\pi N} = e^2 g_9 \bar{N}_v \left( \frac{i}{2} \left[ {\cal{Q}}_+, v \cdot c^+ \right]  + \mathrm{h.c.} \right) N_v.\label{eq:Lagrangian_electromagnetic}
\end{equation}
For the electroweak sector with charged leptons and neutrinos, we provide the most general weak-interaction Lagrangian in the heavy-baryon sector with one charge and one weak spurion, where we only assumed the constraint $\langle{\bf{q}}_W\rangle = 0$~\cite{Tomalak:2023xgm}:
\begin{equation}
    \mathcal L^{e^2 p}_{\pi N \ell} = e^2 \sum \limits_{i=1}^{6} \bar{e}_L \gamma_\rho  \nu_{e L} \bar{N}_v \, \left(  {V}_i v^\rho - 2 {A}_i g^{(0)}_A S^\rho \right) {{\mathrm{O}_i }}  N_v  + \mathrm{h.c.}, \label{eq:Lagrangian_electroweak}
\end{equation}
where $g_A^{(0)}$ denotes the nucleon axial-vector coupling in the chiral limit and
\begin{align}
    { \mathrm{O}_1 } &= [{{\cal{Q}}}_{L}, {\cal{Q}}^W_L], \qquad ~~ { \mathrm{O}_2 } = [{{\cal{Q}}}_{R}, {\cal{Q}}^W_L], \nonumber \\
    { \mathrm{O}_3 } &= \{ {{\cal{Q}}}_{L}, {\cal{Q}}^W_L \}, \qquad ~{ \mathrm{O}_4 } = \{ {{\cal{Q}}}_{R}, {\cal{Q}}^W_L \}, \nonumber \\
    { \mathrm{O}_5 } &= \langle {{\cal{Q}}}_{L} {\cal{Q}}^W_L \rangle, \qquad \quad { \mathrm{O}_6 } = \langle {{\cal{Q}}}_{R} {\cal{Q}}^W_L \rangle.\label{eq:operators_electroweak}
\end{align}
The dimensionless low-energy coupling constants in Ref.~\cite{Cirigliano:2022hob} are related to the couplings in Eq.~(\ref{eq:Lagrangian_electroweak}) by the relations 
$\tilde{X}_1/C_\beta^r = V_1 + V_3 + V_4 - A_1 - A_3 - A_4$, $\tilde{X}_2/C_\beta^r = -V_2$, $\tilde{X}_3/C_\beta^r = 2 A_2 g^{(0)}_A$, $\tilde{X}_4/C_\beta^r = V_4 + V_6$, and $\tilde{X}_5/C_\beta^r = - 2 (A_4 + A_6) g^{(0)}_A$ when the spurions take physical values. In Ref.~\cite{Cirigliano:2022hob}, the authors have used the equations of motion to eliminate some $S^\rho$-dependent operators. In addition, they reduced operators that are bi-linear in spurions to linear expressions by exploiting the relations ${\bf q}_L {\bf q}_W = (2/3) {\bf q}_W$ and  ${\bf q}_W {\bf q}_L = -(1/3) {\bf q}_W$, valid for physical values of the spurions.

As realized in the mesonic sector in Refs.~\cite{Moussallam:1997xx,Descotes-Genon:2005wrq}, we can interpret amplitudes in LEFT and in HBChPT as functionals of the same charges $q^{0,a} (x)$, promoted to be spacetime-dependent external fields. The matching between the LEFT and HBChPT can then be obtained by equating functional derivatives of the effective action with respect to $q^{0,a} (x)$ in both theories. As we will see, this allows us to derive an explicit representation for the LECs and to keep track of unphysical scale and scheme dependence appearing at intermediate steps of the calculation.

\subsection{Electromagnetic coupling constant}
\label{sec:subsec42}

We start from the electromagnetic coupling $g_9$. Expanding the charge covariant derivative in Eq.~\eqref{eq:Lagrangian_electromagnetic}, we obtain
\begin{equation}
    \mathcal L^{e^2 p}_{\pi N} = e^2 g_9 \frac{i}{4} \bar{N}_v v^\rho \left( \left[{\bf q}_V, \partial_\rho {\bf q}_V\right] - \frac{i}{2} \left[{\bf q}_V, \left[ \mathrm{v}_\rho, {\bf q}_V \right]\right]- \frac{i}{2} \left[{\bf q}_V, \left[ \mathrm{a}_\rho, {\bf q}_A \right]\right] \right) N_v.
\end{equation}
$g_9$ can then be evaluated by taking functional derivatives with respect to two isovector charges, provided that the charge carries non-zero momentum, or by taking three derivatives, two with respect to the charges and one  with respect to a vector or axial-vector source. The first representation is more convenient, since, as we will see, it allows one to automatically obtain cancellations between electromagnetic and weak couplings.

More precisely, we will consider the following object:
\begin{equation}
    \Gamma_{VV} = \frac{\varepsilon^{a b c} \tau^c_{i j} \, \delta^{\sigma^\prime \sigma}}{12} \frac{i}{2} v_\rho \frac{\partial}{\partial r_\rho} \left(  \int \mathrm{d}^d x e^{i r \cdot x} \langle  N (k^\prime, \sigma^\prime, j) | \frac{\delta^2 W \left(  {\bf{q}}_V, {\bf{q}}_A  \right)}{\delta {\bf{q}}^b_{V} \left( x \right) \delta {\bf{q}}^a_{V} \left( 0 \right)} \Bigg|_{{\bf{q}}=0} | N (k, \sigma, i) \rangle   \right)  \Bigg|_{r_\rho=0},
\end{equation}
where $k$ and $k^\prime$ are the nucleon momenta, $\sigma$ and $\sigma^\prime$ denote the nucleon spins, and $i$, $j$ are the nucleon isospins. We take the nucleon to be at rest, $k = k^\prime = m_N v$ and use the nonrelativistic  normalization for heavy-particle states $ \langle N \left(k^\prime, \sigma^\prime, j \right) |N \left(k, \sigma, i \right) \rangle =  (2 \pi)^3  \delta^{(3)} ({\bf k} - {\bf k}^\prime) \delta^{ij} \delta^{\sigma \sigma^\prime}$. $W = - i \ln Z$ denotes the generating functional of the connected diagrams.

\begin{figure}
\center
\includegraphics[width=0.9\textwidth]{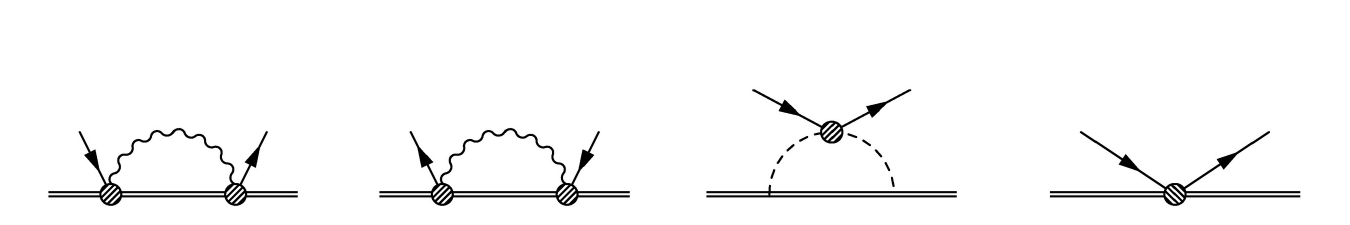}
\caption{Diagrams that contribute to $\Gamma_{VV}$ in HBChPT are shown. Double, wiggly, and dashed lines denote nucleons, photons, and pions, respectively. Dashed circles denote insertions of the sources ${\bf q}^{a, b}_V$. The arrows denote the flow of the momentum $r$ inserted by the sources. The first three diagrams originate from the leading-order $\pi$ and $\pi$N Lagrangians, $\mathcal L^{p^2}_\pi$, $\mathcal L^{e^2}_\pi$, and $\mathcal L^p_{\pi N}$~\cite{Gasser:2002am,Cirigliano:2022hob,Bernard:1995dp}, which are presented in Eq.~\eqref{eq:LO}. The last diagram denotes contributions from $\mathcal L^{e^2 p}_{\pi N}$ and is proportional to $g_9$.}\label{fig:g9_chipt}
\end{figure}

$\Gamma_{VV}$ needs to be computed in both HBChPT and LEFT, and, in both theories, it receives tree-level and loop contributions. The contributions to $\Gamma_{VV}$ in HBChPT are illustrated in Fig.~\ref{fig:g9_chipt}. The short-range contributions are determined by LECs in the $\mathcal L^{e^2  p}_{\pi N}$ Lagrangian. $g_9$ provides the only contribution to $\Gamma_{VV}$. The loops are determined by couplings in the leading-order (LO) pion and pion-nucleon Lagrangians. In particular, the diagram with pion-mass splitting $Z_\pi$ is symmetric in isospin, and vanishes once contracted with the Levi-Civita tensor, so that the loop corrections are purely determined by the minimal coupling of the photon to the nucleon. In arbitrary $R_\xi$ gauge, we introduce the photon mass $\lambda_\gamma$ as an infrared regulator and obtain 
\begin{align}
    \left. \Gamma_{VV}\right|^{\rm HB\chi PT } &= e^2 \left( g_9 + \int \frac{i \mathrm{d}^d q}{\left( 2 \pi \right)^d} \frac{1}{ \left( q^2 - \lambda_\gamma^2 \right)^2 } + \frac{1 - \xi}{2} \int \frac{i \mathrm{d}^d q}{\left( 2 \pi \right)^d} \frac{1}{\left( q^2 - \lambda_\gamma^2 \right) \left( q^2 -\xi \lambda_\gamma^2 \right) }\right) \nonumber \\
    & = \frac{e^2}{(4\pi)^2}  \left( (4\pi)^2 g^r_9(\mu_\chi,\mu)  -  \left(1 + \frac{1- \xi}{2}\right) \ln\frac{\mu^2_\chi}{\lambda_\gamma^2} + 1 - \frac{\xi}{2} \ln \xi \right).\label{eq:g9_chi_PT}
\end{align}
$g_9^r(\mu_\chi,\mu)$ in the second line denotes the renormalized coupling, after subtraction of the $1/\varepsilon$ pole in the $\overline{\rm MS}_\chi$ scheme. For $\xi=1$, the anomalous dimension of $g_9^r(\mu_\chi,\mu)$ agrees with the result of Ref. \cite{Gasser:2002am}, so that Eq. \eqref{eq:g9_chi_PT} is independent of the scale $\mu_\chi$.

In the LEFT, the same matrix element is given by
\begin{equation}
    \left. \Gamma_{VV}\right|^{\rm LEFT} = e^2  \left( - g_{23} + \int \frac{\mathrm{d}^d q}{\left( 2 \pi \right)^d} \frac{v \cdot q\, g_{\mu \nu}  T_{VV}^{\mu \nu} \left( q, v \right)}{\left( q^2 - \lambda_\gamma^2 \right)^2} + \frac{1 - \xi}{2} \int \frac{i \mathrm{d}^d q}{\left( 2 \pi \right)^d} \frac{1}{\left( q^2 - \lambda_\gamma^2 \right) \left( q^2 -\xi \lambda_\gamma^2 \right) } \right).\label{eq:g9_four_fermion}
\end{equation}
Eq. \eqref{eq:g9_four_fermion} contains a tree-level term, proportional to the counterterm $g_{23}$ that cancels the divergences generated by loop diagrams. The loop contribution contains the hadronic tensor ${T}^{\mu \nu}_{VV} \left( q,v\right)$, which can be expressed in terms of the two-point correlation function of quark currents. Here, we use the definition~\cite{Abers:1968zz}
\begin{equation}
    {T}^{\mu \nu}_{VV \left( A \right)} \left( q, v\right) = \frac{ \varepsilon^{a b c} \tau^c_{i j} \delta^{\sigma^\prime \sigma}}{12}  \frac{i}{4} \int \mathrm{d}^d x \,  e^{i q \cdot x}  \langle  N (k, \sigma^\prime, j)   | T \left[ \overline{q}  \gamma^\mu \tau^b q \left( x \right) \overline{q}  \gamma^\nu \left( \gamma_5 \right) \tau^a q  (0)  \right]  | N (k, \sigma, i)   \rangle.\label{eq:hadron_tensor_definition}
\end{equation}
The gauge-dependent term in Eq. \eqref{eq:g9_four_fermion} is obtained using
\begin{equation}
    q_\mu T^{\mu \nu}_{VV}(q,v)  = q_\mu T^{\nu \mu }_{VV}(q,v)  = i v^\nu,
\end{equation}
which follows from the conservation of the vector current. 

To highlight the UV structure of Eq. \eqref{eq:hadron_tensor_definition}, we add and subtract the high-energy limit of the hadronic tensor provided by the operator product expansion (OPE)
\begin{equation}
    \left. g_{\mu \nu} T^{\mu \nu}_{VV}(q,v) \right|_{\rm OPE} = \frac{i v \cdot  q}{q^2 - \mu_0^2}  \,  \left( 2 - d  + 2 \frac{\alpha_s}{\pi} \right),\label{eq:OPE1}
\end{equation}
where for the OPE of the relevant currents we use results from Refs.~\cite{Adler:1969ei,Adler:1970hu}, adapted to include the appropriate color factors~\cite{Sirlin:1977sv}. Since our calculation is only accurate at the leading logarithm in $\mathcal O(\alpha \alpha_s)$, the $\mathcal O(\alpha_s)$ correction to the OPE is computed in $d=4$. Note that in Eq.~\eqref{eq:OPE1} we have introduced an arbitrary scale $\mu_0$ to regulate  infrared divergences that appear when evaluating the  convolution integrals with $T_{\rm OPE}$. Performing the relevant integrations, we obtain 
\begin{align}
    \left. \Gamma_{VV}\right|^{\rm LEFT}& = \frac{e^2}{(4\pi)^2} \left(  \frac{1}{2} \left(1 - \frac{\alpha_s}{\pi} \right) \ln \frac{\mu^2}{\mu_0^2} + \frac{1}{4} - \frac{1-\xi}{2} \left(\ln \frac{\mu^2}{\lambda_\gamma^2} + 1\right)  - \frac{\xi}{2} \ln \xi \right. \nonumber \\ 
    & \left. + (4\pi)^2 \int \frac{\mathrm{d}^4 q}{\left( 2 \pi \right)^4} \frac{v \cdot q \,  g_{\mu \nu}  \overline{T}_{VV}^{\mu \nu}\left( q, v \right)}{\left( q^2 - \lambda_\gamma^2 \right)^2} \right),\label{eq:g9_four_fermion_3}
\end{align}
where $\overline{T}$ denotes the subtracted hadronic tensor, $\overline{T} = T - T_{\rm OPE}$. $\overline T$  depends on $\mu_0$ in such a way that the final results are $\mu_0$-independent. Finally, note that we are dropping terms of $\mathcal{O}(\alpha \alpha_s)$ that appear without logarithmic enhancements, because they are beyond the accuracy of our calculation.

Equating Eqs.~\eqref{eq:g9_chi_PT}~and~\eqref{eq:g9_four_fermion}, we obtain a representation for $g_9$:
\begin{align}
    g^r_9(\mu_\chi,\mu) &=   \int \hspace{-0.1cm} \frac{\mathrm{d}^4 q}{\left( 2 \pi \right)^4} \frac{v \cdot q \,  g_{\mu \nu}  \overline{T}^{\mu \nu}_{VV}\left( q, v \right)}{\left( q^2 - \lambda_\gamma^2 \right)^2} \nonumber \\
    & + \frac{1}{\left(4 \pi \right)^2} \left[ \ln\frac{\mu^2_\chi}{\lambda_\gamma^2}  +\frac{1}{2} \left(1 - \frac{\alpha_s}{\pi} \right) \ln \frac{\mu^2}{\mu_0^2} + \frac{1-\xi}{2}\ln\frac{\mu^2_\chi}{\mu^2} - \frac{5}{4} + \frac{\xi}{2} \right].
\end{align}
Alternatively, to control the infrared region and see a cancellation of the infrared divergences, we can introduce the combination $\tilde{T} = T - T_{\rm IR}$, where $T_{\rm IR}$ is the leading infrared contribution $g_{\mu \nu} T^{\mu \nu}_{\rm IR} = i/ \left( v \cdot q \right) $, and obtain
\begin{equation}
    g^r_9(\mu_\chi,\mu) =  \int \hspace{-0.1cm} \frac{\mathrm{d}^d q}{\left( 2 \pi \right)^d} \frac{v \cdot q \,  g_{\mu \nu}  \tilde{T}^{\mu \nu}_{VV}\left( q, v \right)}{\left( q^2 \right)^2}  +  \frac{1}{\left(4 \pi \right)^2} \left[ \left( 1 + \frac{1-\xi}{2} \right) \ln\frac{\mu^2_\chi}{\mu^2} - \frac{3}{2} + \frac{\xi}{2} \right].
\end{equation}

\subsection{Electroweak coupling constants}
\label{sec:subsec43}

\begin{figure}
\center
\includegraphics[width=0.9\textwidth]{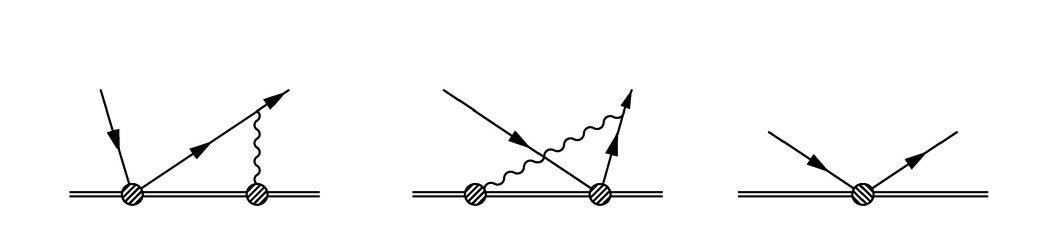}
\caption{Diagrams that contribute to $\Gamma_{VW}$ in HBChPT are shown. Single lines denote electrons and neutrinos. The remaining notations are the same as in Fig. \ref{fig:g9_chipt}. In this case, the sources inject zero momentum. The first two diagrams originate from the LO $\pi$N Lagrangian $\mathcal L^{p}_{\pi N}$, and the last diagram denotes contributions from $\mathcal L^{e^2 p}_{\pi N \ell}$. Diagrams with the sources coupling to pions do not contribute at this order.}\label{fig:AV_chipt}
\end{figure}

We follow the same strategy for the determination of the electroweak coupling constants. In this case, the operators $V_1$ and $V_2$ receive contributions from the isovector component of the electromagnetic charges, while $V_3$ and $V_4$ receive from the isoscalar component. We thus define two matrix elements
\begin{align}
    \bar{e}  \slashed{v} \mathrm{P}_\mathrm{L} \nu_e  \, \Gamma^{(1)}_{VW} &= \frac{ \varepsilon^{a b c} \tau^c_{i j} \delta^{\sigma^\prime \sigma}}{12}   \frac{i}{2} \int \mathrm{d}^d x \langle e^- \bar{\nu}_e N(k, \sigma^\prime, j ) |  \frac{\delta^2 W  \left( {\bf{q}}_V, {\bf{q}}_A, {\bf{q}}_W \right)}{ \delta {\bf{q}}^b_{V} \left( x \right) \delta {\bf{q}}^a_{W} \left( 0 \right)}   \Bigg|_{{\bf{q}}=0} | N(k, \sigma, i) \rangle,\\
    \bar{e}  \slashed{v} \mathrm{P}_\mathrm{L} \nu_e  \, \Gamma^{(0)}_{VW} &= \frac{ \tau^a_{i j} \delta^{\sigma^\prime \sigma}}{12}  \int \mathrm{d}^d x \langle e^- \bar{\nu}_e N(k, \sigma^\prime, j ) | \frac{\delta^2 W  \left( {\bf{q}}_V, {\bf{q}}_A, {\bf{q}}_W \right)}{ \delta {\bf{q}}^0_{V} \left( x \right) \delta {\bf{q}}^a_{W} \left( 0 \right)}   \Bigg|_{{\bf{q}}=0} | N(k, \sigma, i) \rangle.
\end{align}
At the order we are working, the electron and neutrinos can be taken to be massless and to carry zero momentum.

The HBChPT diagrams contributing to $\Gamma^{(0,1)}_{V W}$ are shown in Fig.~\ref{fig:AV_chipt}. The loop diagrams cancel for isoscalar electromagnetic couplings, so that we obtain
\begin{align}
    \left. \Gamma^{(1)}_{VW} \right|^{\rm HB\chi PT} &= e^2 \left( {V}_1 +{V}_2 - \frac{1}{2} \int \frac{i \mathrm{d}^d q}{\left( 2 \pi \right)^d} \frac{1}{q^2 \left( q^2 - \lambda_\gamma^2 \right)} + \frac{1 - \xi}{2} \int \frac{i \mathrm{d}^d q}{\left( 2 \pi \right)^d} \frac{1}{\left( q^2 - \lambda_\gamma^2 \right) \left( q^2 -\xi \lambda_\gamma^2 \right)} \right),\\ 
    \left. \Gamma^{(0)}_{VW} \right|^{\rm HB\chi PT} &=   e^2  \left( {V}_3 +{V}_4 \right).
\end{align}

In the LEFT, the isovector and isoscalar components are given by
\begin{align}
  \bar{e}  \slashed{v} \mathrm{P}_\mathrm{L} \nu_e \,  \left. \Gamma^{(1)}_{VW} \right|^{\rm LEFT} &=  e^2   \bar{e}  \slashed{v} \mathrm{P}_\mathrm{L} \nu_e \left( \frac{g_{02} - g_{03}}{4} + \frac{1 - \xi}{2} \int \frac{i \mathrm{d}^d q}{\left( 2 \pi \right)^d} \frac{1}{\left( q^2 - \lambda_\gamma^2 \right) \left( q^2 -\xi \lambda_\gamma^2 \right)}\right) \nonumber\\
  &   -   \frac{e^2}{2}\int  \frac{ \mathrm{d}^d q}{\left( 2 \pi \right)^d} \frac{1}{q^2 \left( q^2 - \lambda_\gamma^2 \right)} \bar e \gamma_\mu \slashed{q} \gamma_\nu  P_L  \nu_e \,  \left(  T^{\mu \nu}_{VV}(q,v) - T^{\mu \nu}_{VA}(q,v) \right),  \\
 \bar{e}  \slashed{v} \mathrm{P}_\mathrm{L} \nu_e \, \left. \Gamma^{(0)}_{VW} \right|^{\rm LEFT} &= -  e^2   \bar{e}  \slashed{v} \mathrm{P}_\mathrm{L} \nu_e  \frac{g_{02} + g_{03}}{12} \nonumber \\ 
  & +   \frac{e^2}{2}\int  \frac{i \mathrm{d}^d q}{\left( 2 \pi \right)^d} \frac{1}{q^2 \left( q^2 - \lambda_\gamma^2 \right)} \bar e \gamma_\mu \slashed{q} \gamma_\nu  P_L  \nu_e \,  \left(  T^{\mu \nu}_{VV,\, 0}(q,v) - T^{\mu \nu}_{VA, \, 0}(q,v) \right).
\end{align}

The hadronic tensors with two isovector currents are defined in Eq. \eqref{eq:hadron_tensor_definition}, while we define the hadronic tensor with one isoscalar vector current as
\begin{equation}
 {T}^{\mu \nu}_{VV \left( A \right),\, 0} \left( q, v\right) = \frac{  \tau^a_{i j} \delta^{\sigma^\prime \sigma}}{12}  \frac{i}{6} \int \mathrm{d}^d x \,  e^{i q \cdot x}  \langle  N (k, \sigma^\prime, j)   | T \left[ \bar{q}  \gamma^\mu  q \left( x \right) \bar{q}  \gamma^\nu \left( \gamma_5 \right) \tau^a q  (0)  \right]  | N (k, \sigma, i) \rangle.
\end{equation}
As in Sec.~\ref{sec:subsec42}, the UV divergences in the LEFT are determined by the operator product expansion. In NDR, the leading-order OPEs of $T^{\mu \nu}_{VV} - T^{\mu \nu}_{VA}$ and $T^{\mu \nu}_{VV, 0} - T^{\mu \nu}_{VA, 0}$ are proportional to the symmetric and antisymmetric combinations of Dirac matrices $(\gamma^{\mu} \slashed{q} \gamma^\nu \pm \gamma^{\nu} \slashed{q} \gamma^\mu) P_L$, respectively. The symmetric combination does not depend on the scheme, while the antisymmetric piece depends on the definition of the evanescent operators, in such a way as to compensate the dependence of the couplings in the LEFT. Using the OPE, we obtain
\begin{align}
    \bar e \gamma_\mu \slashed{q} \gamma_\nu P_L \nu_e  \left(\left.  T^{\mu \nu}_{VV} - T^{\mu \nu}_{VA}  \right)\right|_{\rm OPE} & = i \left[ \frac{  3 d - 2 }{d} \, - \frac{1}{2} \frac{\alpha_s}{\pi}  \right] \frac{ q^2}{q^2 -\mu_0^2} \, \bar e \slashed{v}  P_L \nu_e, \\  
    \bar e \gamma_\mu \slashed{q} \gamma_\nu P_L \nu_e \left( \left.  T^{\mu \nu}_{VV,\, 0} - T^{\mu \nu}_{VA, 0} \right) \right|_{\rm OPE} &= \left[ \frac{1}{d} \left( \left( 4-d \right) \left( 1 + \frac{4a}{3} \right) - 2 \right) + \frac{1}{2} \frac{\alpha_s}{\pi} \right] \frac{q^2}{q^2 - \mu_0^2} \, \, \bar e \slashed{v}  P_L \nu_e.
\end{align}
The integrals of the subtracted hadronic tensors $\bar{T}$ are convergent, so that we can perform the Dirac algebra on the leptonic leg in the $d=4$ dimension. Putting everything together, we arrive at the matching equations
\begin{align}
    2 (V_1 + V_2)(\mu_\chi,\mu) &=  \int  \frac{ \mathrm{d}^4 q}{\left( 2 \pi \right)^4} \frac{1}{q^2 \left( q^2 - \lambda_\gamma^2 \right)} \, \left( v \cdot q \, g_{\mu \nu} \overline{T}^{\mu  \nu}_{VV}(q,v) + i \varepsilon_{\mu \rho \nu\sigma} q^\rho v^{\sigma} {\overline T}^{\mu \nu}_{VA} (q, v)  \right)  \nonumber \\ 
    &+  \frac{1}{(4\pi)^2}\left[ 2 \ln \frac{\mu^2}{\lambda_\gamma^2} + \frac{1}{2}   \left(1 - \frac{\alpha_s}{\pi} \right) \ln \frac{\mu^2}{\mu_0^2 } - \ln \frac{\mu_\chi^2}{\lambda_\gamma^2}  + \frac{9}{4} + (1-\xi) \left(\ln \frac{\mu_\chi^2}{\mu^2} - 1\right) \right] ,\label{eq:v1v2_match}  \\
    2 (V_3 + V_4)(a,\mu_\chi,\mu) &=  - \int \frac{i \mathrm{d}^4 q}{(2\pi)^4} \frac{1}{q^2 \left( q^2 - \lambda_\gamma^2 \right)} \left(  v \cdot q \, g_{\mu  \nu} \overline{T}^{\mu  \nu}_{VV,\, 0}(q,v) + i \varepsilon_{\mu \rho \nu\sigma} q^\rho v^{\sigma} {\overline T}^{\mu \nu}_{VA,\, 0} (q, v) \right)\nonumber  \\ 
    &+ \frac{1}{(4\pi)^2}\left[ \frac{1}{2}  \left(1 - \frac{\alpha_s}{\pi} \right)   \ln \frac{\mu^2}{\mu_0^2} + \frac{3 - 8 a}{12} \right] .\label{eq:v3v4_match}
\end{align}
To obtain the second line of Eqs.~\eqref{eq:v1v2_match} and~\eqref{eq:v3v4_match}, we used the Ward identities on the subtracted tensors,
\begin{equation}
    q_\mu \overline{T}^{\mu \nu}_{VV}(q,v)  =  i v^\nu \left(1 - \frac{q^2}{q^2 - \mu_0^2} \right), \qquad q_\mu \overline{T}^{\mu \nu}_{VV,\, 0}(q,v)  =  0,
\end{equation}
the symmetry (antisymmetry) of unpolarized hadronic tensors $T^{\mu \nu}_{VV (0)}$ $(T^{\mu \nu}_{VA (0)})$ under $\mu \leftrightarrow \nu$, and, in the contractions with the Levi-Civita tensor, we replaced 
\begin{equation}
    \bar e \gamma^\sigma P_L \nu_e \rightarrow  \bar e \slashed{v} v^{\sigma} P_L \nu_e  +  \bar e \left(\gamma^\sigma - \slashed{v} v^\sigma \right) P_L \nu_e.
\end{equation}

The non-perturbative QCD input in the LECs is encoded in the subtracted hadronic tensors $\overline{T}_{VV}$, $\overline{T}_{VA}$, $\overline{T}_{VV,\, 0}$, and $\overline{T}_{VA,\, 0}$. Using time reversal and crossing symmetry \cite{Sirlin:1967zza,Seng:2018qru}, we can show that the scalar functions in the matching equations~\eqref{eq:v1v2_match} and~\eqref{eq:v3v4_match} are odd or even under $q \rightarrow -q$. Explicitly we have
\begin{align}
    g_{\mu  \nu} \overline{T}^{\mu  \nu}_{VV}(q^2,v \cdot q) &= - g_{\mu  \nu} \overline{T}^{\mu  \nu}_{VV}(q^2,-v \cdot q), \quad i \varepsilon_{\mu \rho \nu\sigma} q^\rho v^{\sigma} {\overline T}^{\mu \nu}_{VA} (q^2, v \cdot q) = - i \varepsilon_{\mu \rho \nu\sigma} q^\rho v^{\sigma} {\overline T}^{\mu \nu}_{VA} (q^2, -v \cdot q),\label{eq:crossing1} \\
    g_{\mu  \nu} \overline{T}^{\mu  \nu}_{VV,\, 0}(q^2,v \cdot q) &=  g_{\mu  \nu} \overline{T}^{\mu  \nu}_{VV,\,0}(q^2,-v \cdot q), \quad i \varepsilon_{\mu \rho \nu\sigma} q^\rho v^{\sigma} {\overline T}^{\mu \nu}_{VA,\, 0} (q^2, v \cdot q) =  i \varepsilon_{\mu \rho \nu\sigma} q^\rho v^{\sigma} {\overline T}^{\mu \nu}_{VA,\,0 } (q^2, -v \cdot q),\label{eq:crossing2}
\end{align}
where we indicated that the functions depend only on the invariants 
\begin{equation}
    Q^2 = - q^2, \qquad \nu = v\cdot q.
\end{equation}
As a consequence of Eqs. \eqref{eq:crossing1} and~\eqref{eq:crossing2}, $T_{VA}$ and $T_{VV,\,0}$ do not contribute to the matching, and the final expressions for the combinations of LECs $V_1+V_2$ and $V_3 + V_4$ are 
\begin{align}
    2 (V_1 + V_2)(\mu_\chi,\mu) &=  \int  \frac{ \mathrm{d}^4 q}{\left( 2 \pi \right)^4} \frac{1}{q^2 \left( q^2 - \lambda_\gamma^2 \right)} \,  v \cdot q \, g_{\mu \nu} \overline{T}^{\mu  \nu}_{VV}(q,v) \nonumber \\
    &+  \frac{1}{(4\pi)^2}\left[ 2 \ln \frac{\mu^2}{\lambda_\gamma^2} + \frac{1}{2}  \left(1 - \frac{\alpha_s}{\pi} \right) \ln \frac{\mu^2}{\mu_0^2 } - \ln \frac{\mu_\chi^2}{\lambda_\gamma^2}  + \frac{9}{4} + (1-\xi) \left(\ln \frac{\mu_\chi^2}{\mu^2} - 1\right) \right] , \\
    2 (V_3 + V_4)(a,\mu_\chi,\mu) &= - \int \frac{i \mathrm{d}^4 q}{(2\pi)^4} \frac{1}{q^2 \left( q^2 - \lambda_\gamma^2 \right)}  i \varepsilon_{\mu \rho \nu\sigma} q^\rho v^{\sigma} {\overline T}^{\mu \nu}_{VA,\, 0} (q, v) \nonumber  \\ 
    &  + \frac{1}{(4\pi)^2}\left[ \frac{1}{2}  \left(1 - \frac{\alpha_s}{\pi} \right) \ln \frac{\mu^2}{\mu_0^2} + \frac{3 - 8 a}{12} \right].
\end{align}
Note that in this framework, the LECs depend not only on the chiral renormalization scale ($\mu_\chi$) but also on the LEFT renormalization scale ($\mu$) and the schemes adopted for $\gamma_5$ and the evanescent operators.

\section{Corrections to $g_V$}
\label{sec:subsec45}

In this Section, we combine the coupling constants of the heavy-baryon chiral perturbation theory into the counterterm of $g_V$ in $\slashed{\pi}\mathrm{EFT}$. We subsequently evaluate the non-perturbative inputs to the vector coupling constant, resum logarithms between the chiral and electron-mass scales, and provide numerical results for $g_V$.

\subsection{Matching at the baryon-mass scale}

Having determined the electroweak coupling constants $V_1$-$V_4$ and the electromagnetic coupling constant $g_9$, we can evaluate the  $\mathcal{O}(\alpha)$ contribution to $g_V$ in the low-energy effective theory, cf. Eqs.~(\ref{eq:vector_coupling_CT}) and \eqref{eq:vector_contribution_LECs_v1}. These corrections are known in the literature as ``inner" radiative corrections. 

Before getting to the final result, we can combine the LECs that depend on the $VV$ hadronic tensor, $g_9$ and $V_1+V_2$, and the lepton wavefunction renormalization $X_6$, obtaining 
\begin{align}
   \left(  - \frac{X_6}{2} +   2 \left( V_1 + V_2 \right) - g_9 \right)(\mu_\chi,\mu)  &= \frac{1}{\left( 4 \pi \right)^2} \left[ 1 + \frac{3}{2} \left( 1 - \ln \frac{\mu^2_\chi}{\mu^2} \right) \right]  \nonumber \\ 
   & - \int  \frac{ \mathrm{d}^4 q}{\left( 2 \pi \right)^4} \frac{\lambda_\gamma^2}{q^2 \left( q^2 - \lambda_\gamma^2 \right)^2} \,  v \cdot q \, g_{\mu \nu} \overline{T}^{\mu  \nu}_{VV}(q,v),   
\end{align}
which is independent of the gauge parameter $\xi$. $T_{VV}$ enters this combination of LECs multiplied by the IR regulator $\lambda_\gamma^2$. The only contribution to the integral can thus come from the infrared limit of $T_{VV}$, where the hadronic tensor is well approximated by the elastic piece. The integral over the hadronic tensor then only leaves behind a finite piece, yielding 
\begin{equation}
    \left( - \frac{X_6}{2} +  2 \left( V_1 + V_2 \right) - g_9 \right)(\mu_\chi,\mu) = \frac{1}{(4\pi)^2}  \frac{3}{2} \left( 1 - \ln \frac{\mu^2_\chi}{\mu^2} \right).\label{eq:mm1}
\end{equation}
Thus, the only contributions to $ - \frac{X_6}{2} + 2 \left( V_1 + V_2 \right) - g_9$ are due to the different renormalization scales, $\mu$ vs $\mu_\chi$, and the different subtraction schemes commonly used in HBChPT, $\overline{\rm MS}_\chi$ vs $\overline{\rm MS}$. 

The other combination of LECs $V_3 + V_4$ is conveniently expressed in terms of the  scalar amplitude $T_3 \left( \nu, Q^2 \right)$ as
\begin{equation}
    2 (V_3 + V_4)(a,\mu_\chi,\mu) =  \frac{1}{(4\pi)^2}\left[ \frac{1}{2}  \left(1 - \frac{\alpha_s}{\pi} \right) \ln \frac{\mu^2}{\mu_0^2} + \frac{3 - 8 a}{12} \right] - \int \frac{i \mathrm{d}^4 q}{\left( 2 \pi \right)^4} \frac{ \nu^2 + Q^2}{ Q^4 }  \frac{{\overline T}_3 (\nu, Q^2)}{2 m_N \nu},\label{eq:mm2}
\end{equation}
where we defined the amplitude $T_3$ from the tensor decomposition of the hadronic tensor as~\cite{Ji:1993ey,Blumlein:1996tp,Maul:1996dx,Blumlein:2012bf,Drell:1966kk,Bjorken:1966jh}\footnote{Note that $T_3$ defined in this paper is equal to $i$ times the  $T_3$ defined in \cite{Gorchtein:2021fce}, which in turn is twice as large as the $T_3$ defined in \cite{Seng:2018yzq}.}
\begin{equation}
    T_{VA, 0}^{\mu \nu} = i \varepsilon^{\mu \nu \sigma \rho } q_\rho v_{\sigma} \frac{T_3}{4 m_N \nu} +  \   \cdots,
\end{equation}
with the OPE-subtracted expression
\begin{equation}
    {\overline T}_3 (\nu, Q^2) = T_3 (\nu, Q^2) -\frac{4}{3} \frac{m_N \nu}{Q^2 + \mu^2_0} \, \left(1 - \frac{\alpha_s} {\pi} \right).
\end{equation}
In the OPE, we have retained the $\mathcal{O}(\alpha_s)$ correction, which is needed to cancel the $\mu$-dependent term proportional to $\alpha \alpha_s \ln (M_W/\mu)$ in $C_\beta^r$. To the order we are working, we can use $\alpha_s (\mu)$ at any $\mu$ where QCD is perturbative. We will use $\alpha_s (\mu_0)$ in what follows.

Combining the HBChPT coupling constants into the $\slashed{\pi}\mathrm{EFT}$ counterterm $\hat{C}_V$ according to Eqs.~\eqref{eq:vector_coupling_CT},~\eqref{eq:vector_contribution_LECs_v1},~\eqref{eq:mm1},~and~\eqref{eq:mm2}, we achieve the matching condition
\begin{align}
  g_V  \left( \mu_\chi \right) &=  C^r_\beta \left( a, \mu \right) \Bigg[ 1 - \frac{\alpha\left( \mu_\chi \right)}{2 \pi} \left( 2 B(a) + \frac{5}{8} + \frac{3}{4} \ln \frac{\mu_\chi^2}{\mu_0^2} + \left( 1 - \frac{\alpha_s}{4 \pi} \right)  \ln \frac{\mu_0^2}{\mu^2}  \right) \nonumber \\
  & - e^2 \int \frac{i \mathrm{d}^4 q}{\left( 2 \pi \right)^4} \frac{ \nu^2 + Q^2}{Q^4} \frac{{\overline T}_3 (\nu, Q^2)}{2 m_N \nu} \Bigg],\label{eq:vector_isovector_finite_integrals} 
\end{align}
where we resummed logarithms in the Wilson coefficient  $C^r_\beta \left( a, \mu \right)$, as it is described in Section~\ref{sec:subsec31}. This expression does not contain electroweak-scale parameters or artificial hadronic scales, besides the dependence contained in the coupling constant $C^r_\beta \left( a, \mu \right)$. 
The vector coupling $g_V \left( \mu_\chi \right)$ does not depend on the scale and scheme used in the LEFT at the one-loop level.

We can further simplify the expression for $g_V (\mu_\chi)$ and connect it to the previous literature. First, we eliminate the evanescent scheme dependence by defining the scheme-independent NLO Wilson coefficient~\cite{Buras:1989xd}
\begin{equation}
    \overline{C}^r_\beta (\mu) = \frac{C^r_\beta (a, \mu)}{1 + \frac{\alpha (\mu)}{\pi} B(a)},
\end{equation}
which can be immediately read off from Eq.~\eqref{eq:analytic_sol}. We then have
\begin{equation}
    g_V  \left( \mu_\chi \right) =  \overline{C}^r_\beta \left(\mu \right) \left[ 1 + \overline{\Box}_\mathrm{Had}^V (\mu_0) - \frac{\alpha\left( \mu_\chi \right)}{2 \pi} \left( \frac{5}{8} + \frac{3}{4} \ln \frac{\mu_\chi^2}{\mu_0^2} + \left( 1 - \frac{\alpha_s}{4 \pi} \right) \ln \frac{\mu_0^2}{\mu^2}  \right) \right], \label{eq:gV2}
\end{equation}
where the non-perturbative input is in the ``subtracted" hadronic contribution $\overline{\Box}_\mathrm{Had}^V (\mu_0)$, which is closely related to the standard $\Box_{\gamma W}^V$ of Refs.~\cite{Seng:2018qru,Seng:2018yzq,Seng:2020jtz}
\begin{align}
    \overline{\Box}_\mathrm{Had}^V (\mu_0)  &= -e^2 \int \frac{i \mathrm{d}^4 q}{\left( 2 \pi \right)^4} \frac{ \nu^2 + Q^2}{Q^4} \left[ \frac{{ T}_3 (\nu, Q^2)}{2 m_N \nu} -\frac{2}{3} \frac{1}{Q^2 + \mu^2_0}\, \left(1 - \frac{\alpha_s (\mu_0^2)}{\pi}   \right) \right],\label{eq:boxbar} \\
    \Box_{\gamma W}^V &= - e^2 \int \frac{i \mathrm{d}^4 q}{\left( 2 \pi \right)^4}  \frac{M_W^2}{Q^2 + M_W^2} \frac{ \nu^2 + Q^2}{Q^4} \frac{{ T}_3 (\nu, Q^2)}{2 m_N \nu}.
\end{align}
We will evaluate the non-perturbative input in Eq.~\eqref{eq:boxbar} in Sec.~\ref{sect:box}.

Eq.~\eqref{eq:gV2} encodes the so-called ``inner" radiative corrections to the Fermi transitions in the EFT language in the form of a $\mu_\chi$-dependent coupling $g_V (\mu_\chi)$, which appears in the effective Lagrangian of  Eq.~\eqref{eq:Lagrangian_at_leading_order}. Once all large electroweak logarithms are resummed via the RGE in $\overline C_\beta (\mu)$, Eq.~\eqref{eq:gV2} does not contain additional large logarithms when the scales $\mu_\chi$, $\mu$, and $\mu_0$ are similar and of order $\Lambda_\chi \sim 1$~GeV. As shown below, the $\mu_\chi$-scale dependence in $g_V (\mu_\chi)$ is canceled in physical amplitudes by the $\mu_\chi$ dependence of the virtual photon corrections computed in the pionless theory.  Since the only scale of these loops is ${\cal O}(m_e)$, we will evolve $g_V (\mu_\chi)$ down to the scale $\mu_\chi \sim m_e$ in order to avoid large logarithms, see Sec.~\ref{sec:RGE2}.

\subsection{Evaluation of the non-perturbative input} 
\label{sect:box}

As shown in Refs.~\cite{Seng:2018qru,Seng:2018yzq}, the box function can be represented as a one-dimensional integral over the $Q^2>0$ variable
\begin{equation}
    \Box_{\gamma W}^V = \frac{\alpha}{8 \pi} \int_0^\infty \mathrm{d} Q^2 \frac{M_W^2}{M_W^2 + Q^2} F (Q^2),
\end{equation}
where $F (Q^2) = (12/Q^2) M_3^{(0)} (1, Q^2)$ and $M_3^{(0)} (1, Q^2)$ is the first Nachtmann moment of the structure function defined in terms of the imaginary part of $T_3 (\nu, Q^2)$. Following Refs.~\cite{Seng:2018qru,Seng:2018yzq}, it is useful to isolate the well-defined elastic contribution to $F(Q^2)$, which we denote by $F_{\rm el} (Q^2)$, known in terms of the nucleon isoscalar magnetic vector and axial-vector form factors, and define 
\begin{equation}
    F (Q^2) = F_{\rm el} (Q^2) + \overline F (Q^2),
\end{equation}
where $\overline F (Q^2)$ includes inelastic contributions. For $Q^2 \leq Q_0^2 = 2~{\rm GeV}^2$,\footnote{The value of $Q_0$ is somewhat arbitrary, and here we follow the choice of Refs.~\cite{Seng:2018qru,Seng:2018yzq}.} $\overline F(Q^2)$ contains contributions from the resonance region and the so-called Regge region. Current knowledge is based on modeling~\cite{Seng:2018qru,Seng:2018yzq,Czarnecki:2019mwq,Shiells:2020fqp,Hayen:2020cxh} and  lattice QCD input~\cite{Seng:2020wjq}. For $Q^2 \geq Q_0^2 = 2~{\rm GeV}^2$, one enters the deep inelastic scattering region (DIS), controlled by the OPE with Wilson coefficients computed in perturbative QCD (pQCD). The OPE representation of $\overline F (Q^2)$ is known to leading order in $1/Q^2$, with coefficients known to $\mathcal{O}(\alpha_s^4)$~\cite{Czarnecki:2019mwq,Larin:1990zw,Larin:1991tj,Baikov:2010je}: 
\begin{equation}
    \overline F_{\rm DIS} (Q^2) = \frac{1}{Q^2} \left( 1 - \Delta (Q^2) \right), \qquad \qquad \Delta (Q^2) = \sum_{n=1}^4 \tilde{c}_n \left(\frac{\alpha_s (Q)}{\pi}\right)^n.
\end{equation}
In practice, we will use only the $n=1$ term (with coefficient $\tilde c_1=1$) in $\Delta (Q^2)$, as higher-order terms are beyond the accuracy of our NLL LEFT analysis. Moreover, for consistency with the OPE terms that we subtract in the matching procedure, we will use $\Delta (Q^2) \to \Delta (\mu_0^2)$ in $\overline F_{\rm DIS} (Q^2)$.

In terms of the quantities defined above, the subtracted hadronic contribution reads 
\begin{equation}
    \overline \Box_\mathrm{Had}^V (\mu_0) = \frac{\alpha}{8 \pi} \int_0^\infty \mathrm{d}Q^2     \left[ F_{\rm el} (Q^2) + \overline F (Q^2) - \frac{1}{Q^2 + \mu_0^2} \left(1 - \Delta (\mu_0^2) \right) \right].
\end{equation}
Isolating the elastic contribution and separating the integration in the regions below and above $Q_0^2 = (\sqrt{2}~{\rm GeV})^2$, we find
\begin{align}
    \Box_{\gamma W}^V &=      \Box_{\gamma W}^V \Big \vert_{\rm el} +      \frac{\alpha}{8 \pi}      \int_0^{Q_0^2} \mathrm{d}Q^2  \  \overline F (Q^2)  \ + \ \frac{\alpha}{8 \pi} \Big(1 - \Delta (\mu_0^2) \Big)  \ln \frac{M_W^2}{Q_0^2}  + \mathcal{O}\left(\frac{Q_0^2}{M_W^2} \right), \label{eq:box1} \\
    \overline \Box_\mathrm{Had}^V (\mu_0) &= \ \Box_{\gamma W}^V \Big \vert_{\rm el} + \frac{\alpha}{8 \pi} \int_0^{Q_0^2} \mathrm{d}Q^2  \  \overline F (Q^2)  \ + \   \frac{\alpha}{8 \pi} \Big(1 - \Delta (\mu_0^2) \Big)  \ln \frac{\mu_0^2}{Q_0^2},
\end{align}
Numerically, for the non-perturbative contributions we find
\begin{subequations}
\begin{align}
   \Box_{\gamma W}^V \Big \vert_{\rm el}  &=  1.030(48) \times 10^{-3}, \\
    \int_0^{Q_0^2} \mathrm{d} Q^2  \  \overline F (Q^2) &  \qquad \quad \longrightarrow \qquad \qquad \delta \overline \Box_\mathrm{Had}^V\Big \vert_{\rm Regge + Res.} = (0.49(11)  + 0.04(1)) \times 10^{-3}.\label{eq:RR}
\end{align}
\end{subequations}

We  evaluated the elastic contribution with the isoscalar magnetic vector form factor, which is extracted from experimental $ep$ and $en$ scattering data, measurements of the neutron scattering length, and $\mu\mathrm{H}$ spectroscopy~\cite{Borah:2020gte}. For the axial-vector form factor, we use the fit to the experimental $\nu_\mu\mathrm{D}$ scattering data from Ref.~\cite{Meyer:2016oeg}. Our result is in reasonable agreement with previous evaluations of the elastic contribution to $\Box_{\gamma W}^V$, giving $\left(1.05\pm0.04\right)\times10^{-3}$~\cite{Shiells:2020fqp}, $\left(1.06\pm0.06\right)\times10^{-3}$~\cite{Seng:2018qru,Seng:2018yzq,Gorchtein:2021fce}, $\left(1.06\pm0.06\right)\times10^{-3}$~\cite{Hayen:2020cxh}, and $ \left(0.99\pm0.10\right)\times10^{-3}$~\cite{Czarnecki:2019mwq}, but contains an improved uncertainty estimate since our errors are directly propagated from the experimental data.

Up to negligible contributions of $\mathcal{O}(Q_0^2/M_W^2)$, the integral of $\overline F (Q^2)$ between $0$ and $Q_0^2$ in Eq.~\eqref{eq:RR} coincides with the $Q^2 \leq Q_0^2$ inelastic  piece of the ``box diagram", recently considered in the literature~\cite{Seng:2018yzq,Seng:2018qru,Czarnecki:2019mwq,Seng:2020wjq,Hayen:2020cxh,Shiells:2020fqp}. The result is usually written as the sum of the ``Regge" plus ``Resonance" contributions. The various evaluations in the literature have recently been combined by Ref.~\cite{Cirigliano:2022yyo}, leading to the numbers used in Eq.~\eqref{eq:RR}. This part of our result is fully correlated with previous work and carries the dominant contribution to the error budget for the radiative corrections.

\subsection{RG evolution of $g_V$ below the baryon scale}\label{sec:RGE2}

To account for higher-order perturbative logarithms, which are needed for precise predictions of $\beta$-decay rates and (anti)neutrino-nucleon scattering, we evolve the low-energy coupling constant $g_V (\mu_\chi)$ from the matching scale $ \mu_\chi \sim \Lambda_\chi$ to the physical scale $\mu_\chi \sim m_e$ using the one- and two-loop anomalous dimensions. The vector coupling constant evolves according to
\begin{subequations} \label{eq:RGE_LE}
\begin{align}
    \mu_\chi \frac{\mathrm{d} g_V\left( \mu_\chi \right)}{\mathrm{d} \mu_\chi} &=  \gamma (\alpha) \  g_V \left( \mu_\chi \right),  \\
    \gamma (\alpha) &= \tilde{\gamma}_0 \, \frac{\alpha}{\pi} + \tilde{\gamma}_1 \left( \frac{\alpha}{\pi} \right)^2 \  +  \   \cdots, \\
    \tilde{\gamma}_0 &= -\frac{3}{4},  \\
    \tilde{\gamma}_1 &=  \frac{5 \tilde{n}}{24} + \frac{5}{32} - \frac{\pi^2}{6},
\end{align}
\end{subequations}
with the effective number of particles $\tilde n$, as it is described in Appendix~\ref{appa2}. The appropriate one-loop anomalous dimension $\tilde \gamma_0$ has been identified in Refs.~\cite{Ando:2004rk,Fukugita:2004cq,Czarnecki:2004cw,Raha:2011aa,Cirigliano:2022hob,Tomalak:2022xup}.\footnote{Note that the one-loop anomalous dimension in the theory with relativistic nucleons is a factor of $2$ larger than $\tilde{\gamma}_0$ in Eqs.~\eqref{eq:RGE_LE}, and, therefore, our coupling constant can be used for the calculation of radiative corrections only in the theory with heavy nucleons.} It can also be extracted from  calculations of the ``heavy-light" current QCD anomalous dimension in the context of heavy quark physics, as for example in Refs.~\cite{Voloshin:1986dir,Politzer:1988wp}. As discussed in Appendix~\ref{sect:gammas}, we can exploit this analogy to extract the QED two-loop anomalous dimension $\tilde \gamma_1$, by adapting the results from Ref.~\cite{Gimenez:1991bf} (see also Refs.~\cite{Ji:1991pr,Broadhurst:1991fy,Broadhurst:1991fz}). The above expression for $\tilde \gamma_1$ only includes two-loop diagrams involving two virtual photons in the pionless theory. Possible contributions arising from diagrams involving pions and photons are not included. Note that the term in $\tilde \gamma_1$ proportional to $\pi^2$ can lead to contributions to the decay rate that scale as $\alpha^2 \ln \left(m_N/m_e \right)$, larger than a typical two-loop contribution.

Using the expression for the evolution operator in Eq.~\eqref{eq:U}, we solve the RGE in \eqref{eq:RGE_LE} and resum the leading and subleading logarithms between particle thresholds according to
\begin{align}
    g_V \left( \mu_\chi \right) &= \tilde U(\mu_\chi,m_\mu)\tilde U(m_\mu,m_\pi)\tilde U(m_\pi,\Lambda_\chi) g_V(\Lambda_\chi)\,,\nonumber\\
    \tilde U(\mu_1,\mu_2) &=  \left( \frac{\alpha\left( \mu_1 \right)}{\alpha\left( \mu_2 \right)} \right)^{-2\tilde \gamma_0/\tilde\beta_0}\left[1-\frac{2\tilde \gamma_1(\mu_1)}{\tilde\beta_0(\mu_1)}\frac{\alpha(\mu_1)-\alpha(\mu_2)}{\pi}\right]\,.\label{eq:gVme}
\end{align}
Below the baryon scale, we determine $\alpha$ from its value in the Thomson limit by evolving it up in scale with the electron, muon, and charged pion as active degrees of freedom, which leads to
\begin{equation}
    \tilde \beta_0 = -\frac{4}{3}\sum_{\ell=e,\mu}Q_\ell^2\theta(\mu-m_\ell)-\frac{1}{3} Q_\pi^2\theta(\mu-m_\pi)\,.
\end{equation}
See Appendix~\ref{sect:alpha} for details on the definition of the fine-structure constant in both LEFT and $\chi$PT. 

In Eq.~\eqref{eq:gVme}, $g_V (\Lambda_\chi)$ is obtained by evaluating Eq.~\eqref{eq:gV2} at $\mu_\chi= \Lambda_\chi \sim m_N$. Note that both $\tilde \gamma_0$ and $\tilde \gamma_1$ are negative, implying $g_V(m_e)/g_V (\Lambda_\chi) >1$.

\subsection{Numerical results and uncertainty estimates} 
\label{sec:numerics}

We next present numerical results for the vector coupling $g_V(m_e)$ and discuss the various sources of uncertainty. We start by providing some intermediate results that illustrate the impact of corrections at various orders in our RGE analysis. 

For the semileptonic Wilson coefficient  $C^r_\beta$, we include $\alpha,~\alpha \alpha_s$, and $\alpha^2$ contributions to the running, as described in Section~\ref{sec:subsec31}.\footnote{We perform the one-loop running for $\alpha \left( \mu \right)$ and $\alpha_s \left( \mu \right)$ in LEFT, consistently with the order of our calculation. We have checked that using the higher-order couplings as in Ref.~\cite{Hill:2019xqk} modifies our final results at the level of 0.001\%.} To illustrate the effect of running from the electroweak to GeV scales, we provide results for the fixed order (LO) $C_\beta (m_c) = 1 + (\alpha (m_c)/\pi) \ln (M_Z/m_c)$, leading logarithms (LL), next-to-leading logarithms NLL1, which includes the anomalous dimensions up to order $\alpha \alpha_s$, and next-to-leading logarithms NLL2, including the anomalous dimensions up to orders $\alpha \alpha_s$ and $\alpha^2$. For the initial conditions, we specify
\begin{align}
    C^\mathrm{LL,NLL1}_\beta (M_W) &= 1 + \frac{\alpha (M_W)}{\pi} \ln \frac{M_Z}{M_W}, \\
    C^\mathrm{NLL2}_\beta (M_W) &= 1 + \frac{\alpha (M_W)}{\pi} \ln \frac{M_Z}{M_W}  + \frac{\alpha (M_W)}{\pi} \, B(a=-1).
\end{align}
After numerically solving the RGEs, we obtain the following values for the effective couplings at $\mu = m_c$:
\begin{subequations}
    \begin{align}
    C_\beta^\mathrm{LO} (m_c) &= 1.01014,   \\
    C_\beta^\mathrm{LL} (m_c) &= 1.01043,    \\
    C_\beta^\mathrm{NLL1} (m_c) &= 1.01027,   \\
    \overline C_\beta^\mathrm{NLL2} (m_c) &= 1.01018.
\end{align}
\end{subequations} 
The effects of NLL1 and NLL2 resummations combine to essentially ``undo" the effect of LL resummation. The final result is very close to the perturbative one. The numerical solution of the RGEs agrees with the analytic solutions provided in Section~\ref{sec:subsec31}. Our result for the NLL1 correction is consistent with the finding of Ref.~\cite{Erler:2002mv}. The impact of NLL2 corrections in our result is more than a factor of 2 larger than in Ref.~\cite{Czarnecki:2004cw}, reflecting the difference discussed in Section~\ref{sec:subsec31}.

For the running of the vector coupling constant $g_V$, we include the ${\cal O}(\alpha)$ and ${\cal O}(\alpha^2)$ anomalous dimensions, as described in Section~\ref{sec:RGE2}.\footnote{We match LEFT at the scale $\mu = m_c$ to the $\mathrm{HB}\chi\mathrm{PT}$ at the scale $\mu_\chi = m_p$, below which we perform the running of $\alpha$ with the one-loop anomalous dimension for leptons and pions~\cite{Cirigliano:2022yyo,Hill:2019xqk}.}
We provide the relative running contributions for the one-loop logarithm (LO), namely $g_V (m_e)/g_V (m_p) \vert_\mathrm{LO} = 1 + (3/4) (\alpha/\pi) \ln (m_p/m_e)$, the LL resummation, where we include only $\tilde \gamma_0$ in the RGE, and NLL resummation, where we also include $\tilde \gamma_1$ in the RGE:
\begin{subequations}
\begin{align}
    \frac{g_V (m_e)}{g_V (m_p)} \Bigg \vert_\mathrm{LO} &= 1.01308, \\
    \frac{g_V (m_e)}{g_V (m_p)} \Bigg \vert_\mathrm{LL} &=  1.01325, \\
    \frac{g_V (m_e)}{g_V (m_p)} \Bigg \vert_\mathrm{NLL} &=  1.01330.
\end{align}
\end{subequations}
At the level of decay rate, our NLL correction implies an increase of $1.0 \times 10^{-4}$ (roughly half of the final uncertainty of the radiative corrections).

Putting together all the results obtained so far, we evaluate the vector coupling constant $g_V \left( \mu_\chi \right)$ in the $\overline{\mathrm{MS}}$ renormalization scheme of $\chi\mathrm{PT}$ at the scale $\mu_\chi = m_e$, where $\mathcal{O}(\alpha^n)$ loop corrections to the matrix elements of the Lagrangian~\eqref{eq:Lagrangian_at_leading_order} do not contain large logarithms:
\begin{equation}
    g_V \left( \mu_\chi = m_e \right) - 1 = \left(2.499\pm0.013 \right)\%.
\end{equation}
In contrast, the vector coupling at fixed order $g_V ^{1-\mathrm{loop}}$ (i.e., without resummation, without $\alpha \alpha_s$ corrections, and with taking the value for the electromagnetic coupling constant in the Thomson limit) takes the value
\begin{equation}
    g_V ^{1-\mathrm{loop}} \left( \mu_\chi = m_e \right) - 1 = \left(2.430\pm0.012 \right)\%.
\end{equation} 
In the RGE evolution, the electromagnetic  and $\alpha \alpha_s$ effects contribute with opposite signs, resulting in a net increase of $g_V$ at the level of 0.07\%.

For the uncertainty estimate, we add the following dominant sources in quadrature:
\begin{itemize}
    \item $0.012\%$: the hadronic error for Regge, resonance, and $\pi N$ contributions from Ref.~\cite{Gorchtein:2021fce} is added in quadrature to the uncertainty propagated from the lepton-nucleon experimental data for the elastic contribution.
    
    \item $0.004\%$: the higher-order $\alpha \alpha_s^2$ uncertainty is estimated by including the known terms of $\mathcal{O}(\alpha_s^2)$~\cite{Czarnecki:2004cw,Czarnecki:2019mwq,Seng:2018qru,Seng:2018yzq} in the pQCD correction $\Delta (\mu_0^2)$ that controls the DIS region of $\Box_{\gamma W}^V$ in Eq.~\eqref{eq:box1}. In our approach, this DIS contribution maps onto the $\alpha \alpha_s^2$ anomalous dimension for the Wilson coefficient $C_\beta (\mu)$ in LEFT.

    \item $0.003\%$: the higher-order $\chi\mathrm{PT}$ uncertainty is estimated by assuming the natural size for unaccounted corrections, i.e., $\frac{\alpha}{\pi} \frac{m^2_\pi}{16 \pi^2 F_\pi^2}$.
\end{itemize}
All other perturbative and parametric sources of uncertainties are at the level $0.001\%$ or even below.

We conclude this section by noting that the effective coupling $g_V (\mu_\chi \approx m_e)$ captures the ``inner" corrections to one-body weak transitions through  NLL, i.e., up to and including terms of order $\alpha^2 L^2$ and $\alpha^2 L$ (where $L$ indicates large logarithms of $M_Z/m_N$ and $m_N/m_e$), with residual uncertainty at $\mathcal{O}(\alpha^2)$ due to finite terms in two-loop diagrams. Importantly, $g_V$ controls both neutron decay and the one-body contribution to nuclear $\beta$ decays, in combination  with appropriate $n \to p e \bar \nu_e$ and $(N,Z) \to (N-1,Z+1) e \bar \nu_e$ matrix elements computed to the same accuracy. For applications in neutrino and nuclear physics, in Table~\ref{tab:table_gV} we provide the coupling constant $g_V$ for a few  values of the renormalization scale up to $50$~MeV.
\begin{table}[htb]
\centering 
 \begin{tabular}{|c|c|c|c|c|c|c|} \hline  
    $\mu_\chi$, MeV & $1$ & $5$  & $10$ & $20$ & $30$ & $50$  \\ \hline
    $g_V -1,~\%$ & 2.379 & 2.090 & 1.966 & 1.842 & 1.770 & 1.678 \\ \hline
 \end{tabular}
 \caption{The coupling constant $g_V$ is presented for a few values of the renormalization scale $\mu_\chi$. \label{tab:table_gV}}
\end{table}

\section{Corrections to neutron decay and impact on $V_{ud}$}\label{sec6}

We can now use the $\slashpi$EFT Lagrangian in Eq.~\eqref{eq:Lagrangian_at_leading_order} with $g_V(\mu_\chi=m_e)$ from Eq.~\eqref{eq:gVme} to compute the neutron decay rate including radiative corrections. The final ingredient is the square modulus of the $n \to p e \bar \nu_e$ and $n \to p e \bar \nu_e \gamma$ matrix elements in HBChPT, evaluated at $\mu_\chi \sim m_e$. To match the accuracy achieved in $g_V (m_e)$, since $\ln (\mu_\chi/m_e) \sim \mathcal{O}(1)$, we will need the matrix elements to $\mathcal{O}(\alpha)$ and will ignore terms of $\mathcal{O}(\alpha^2)$ and higher. The only exceptions are ``Coulomb"-enhanced terms scaling as $(\pi \alpha/\beta)^n$ and $\alpha/\pi (\pi \alpha/\beta)^n$, where $\beta \equiv p_e/E_e$, which are parametrically large, diverge for $\beta \to 0$, and can be resummed in the nonrelativistic Fermi function.

\subsection{``Long distance" electromagnetic corrections and differential decay rate}
\label{sect:mecorr}

After including the contributions from both virtual and real photons~\cite{Ando:2004rk,Cirigliano:2022hob} as well as recoil corrections~\cite{Ando:2004rk,Wilkinson:1982hu}, the differential decay rate $\mathrm{d}\Gamma_n$ for unpolarized neutrons takes the form~\cite{Sirlin:1967zza,Wilkinson:1982hu}
\begin{equation}  
    \frac{\mathrm{d}\Gamma_n}{\mathrm{d}E_e} = \frac{G_F^2  \,  |V_{ud}|^2 }{(2\pi )^5}   \left(  1 + 3  \lambda^2 \right) \  p_e E_e (E_0 - E_e)^2  \left[ g_V (\mu_\chi) \right]^2 \ \Big( 1  +  \tilde \delta_{\rm RC} (E_e,\mu_\chi) \Big) \Big( 1+  \delta_{\rm recoil} (E_e)\Big),
\end{equation}
where $E_0 =  (m_n^2 - m_p^2 + m_e^2)/(2m_n)$  is the electron endpoint energy and $\lambda \equiv g_A/g_V$ is the ratio of effective axial-vector and vector couplings in the low-energy Lagrangian \eqref{eq:Lagrangian_at_leading_order}. The ratio $\lambda = \lambda^{\rm QCD} (1 + {\delta}^{(\lambda)}_{\rm RC})$ is affected by a $\mu_\chi$-independent electromagnetic correction ${\delta}^{(\lambda)}_{\rm RC}$ parameterized in terms of calculable pion loops and certain chiral LECs (see Ref.~\cite{Cirigliano:2022hob}). $\lambda$ itself can be extracted from beta decay correlation experiments, so that we do not need to know $\delta^{(\lambda)}_{\rm RC}$ for the purpose of studying total decay rates and the extraction of $V_{ud}$ . $\delta_{\rm recoil} (E_e)$ collects recoil corrections that can be found in Ref.~\cite{Ando:2004rk}. They are usually factorized since the impact of the product of radiative times recoil corrections is estimated to be well below $10^{-4}$. Finally, $\tilde{\delta}_{\rm RC} (E_e)$ represents the  electromagnetic corrections arising from the matrix element squared. To $\mathcal{O}(\alpha)$, one finds
\begin{equation}
     \tilde \delta^{}_{\rm RC} (E_e,\mu_\chi) = \frac{\alpha \left( \mu_\chi \right)}{2\pi} \left( \frac{2 \pi^2}{\beta} + \frac{3}{2} \ln \frac{\mu_\chi^2}{m_e^2} + \frac{5}{4} + \hat g \left( E_e, E_0 \right)   \right), \label{eq:msq1}
\end{equation}
where $\hat g (E_e, E_0)$  is a ``subtracted" Sirlin function
\begin{equation}
    \hat g \left( E_e, E_0 \right) = g \left( E_e, E_0 \right) - \frac{3}{2} \ln \frac{m_N^2}{m_e^2},
\end{equation}
defined in terms of the Sirlin function $g \left( E_e, E_0 \right)$ of Ref.~\cite{Sirlin:1967zza}. $\hat g \left( E_e, E_0 \right)$ arises naturally in  the EFT calculation and does not contain any large logarithm of $m_N/m_e$.

The corrections proportional to $\pi \alpha/\beta$ in Eq.~\eqref{eq:msq1} are enhanced by a factor of $\pi^2$ compared to the naive  scaling of loop corrections, and are numerically dominant even for $\beta \sim \mathcal{O}(1)$. The leading terms in the series in $\pi \alpha/\beta$ arise from the momentum regions of loop integrals in which the photon momentum has potential scaling, $k_0 \sim m_e \beta^2 \ll |\vec k| \sim m_e \beta$, and they can be identified with nonrelativistic  EFT methods~\cite{Hoang:1997sj,Hoang:1997ui,Czarnecki:1997vz,Beneke:1999qg}. Their resummation leads to the nonrelativistic  Fermi function $F_{NR} (\beta)$~\cite{Sommerfeld:1931qaf,Fermi:1934hr,Konopinski:1935zz,Morita:1963zz,Wilson:1968pwx,Halpern:1970it,Halpern:1968zz,Hoferichter:2009gn,Matsuzaki:2012qb,Matsuzaki:2013twa,Heller:2019dyv},
\begin{equation}
    F_{NR} \left( \beta \right) = \frac{2 \pi \alpha}{\beta} \frac{1}{1 - e^{- \frac{2 \pi \alpha}{\beta}}} \approx 1 + \frac{\pi \alpha}{\beta}  + \frac{\pi^2 \alpha^2}{3 \beta^2} - \frac{\pi^4 \alpha^4}{45 \beta^4} + ...,
\end{equation}
which we include in the matrix element squared as
\begin{align}
    1+ \tilde \delta_{\rm RC} (E_e,\mu_\chi) &= F_{NR} (\beta) + \frac{\alpha \left( \mu_\chi \right)}{2\pi} \left(\frac{3}{2} \ln \frac{\mu_\chi^2}{m_e^2} + \frac{5}{4} + \hat g \left( E_e, E_0 \right)   \right)  \nonumber \\ 
    & \longrightarrow F_{NR} (\beta) \Big( 1+  \delta_{\rm RC} (E_e,\mu_\chi) \Big) + \mathcal{O}\left( \alpha^2 \right),\label{eq:factorization}
\end{align}
where
\begin{equation}
     \delta_{\rm RC} (E_e,\mu_\chi) = \frac{\alpha \left( \mu_\chi \right)}{2\pi} \left( \frac{3}{2} \ln \frac{\mu_\chi^2}{m_e^2} + \frac{5}{4} + \hat g \left( E_e, E_0 \right)   \right).
\end{equation}
As we discuss in Appendix \ref{app:nrqed}, the factorization ansatz in Eq.~\eqref{eq:factorization} captures all numerically-enhanced leading and subleading terms in $1/\beta$, and reproduces similar results for the production of two heavy quarks at threshold, derived with nonrelativistic QCD and potential nonrelativistic QCD~\cite{Beneke:1997jm,Hoang:1997sj,Hoang:1997ui,Czarnecki:1997vz,Beneke:1999qg,Beneke:2009ye,Kawamura:2012cr,Piclum:2018ndt}. At $\mathcal O(\alpha^2)$, Eq. \eqref{eq:factorization} gives
\begin{equation}
    F_{NR} (\beta) \Big( 1+  \delta_{\rm RC} (E_e,\mu_\chi) \Big) = F_{NR} (\beta) - \frac{11}{4} \frac{\alpha^2}{\beta} + \frac{\left( E_0 - m_e \right)^2}{12 m^2_e} \frac{\alpha^2}{\beta} +   \delta_{\rm RC} (E_e,\mu_\chi) + \mathcal{O}\left( \alpha^2 \right).
\end{equation}
Indeed, the first cross term  $-(11/4) \alpha^2/\beta$ corresponds to the matching coefficient of heavy-light to heavy-heavy current~\cite{Grozin:2004yc} in the $\overline{\mathrm{MS}}_\chi$ renormalization scheme. The second cross term $ (\alpha^2/\beta) \left( E_0 - m_e \right)^2/ \left(12 m^2_e \right)$ comes from the product of the Fermi function with real radiation. These terms are beyond the accuracy of our calculation and can be booked as $\mathcal O(\alpha^2 \beta^3)$ in the nonrelativistic limit. In the case of neutron decay,  this term provides a negligible shift  of $1.6\times10^{-5}$ to the decay rate.
\begin{figure}
\centering
\includegraphics[width=0.9\textwidth]{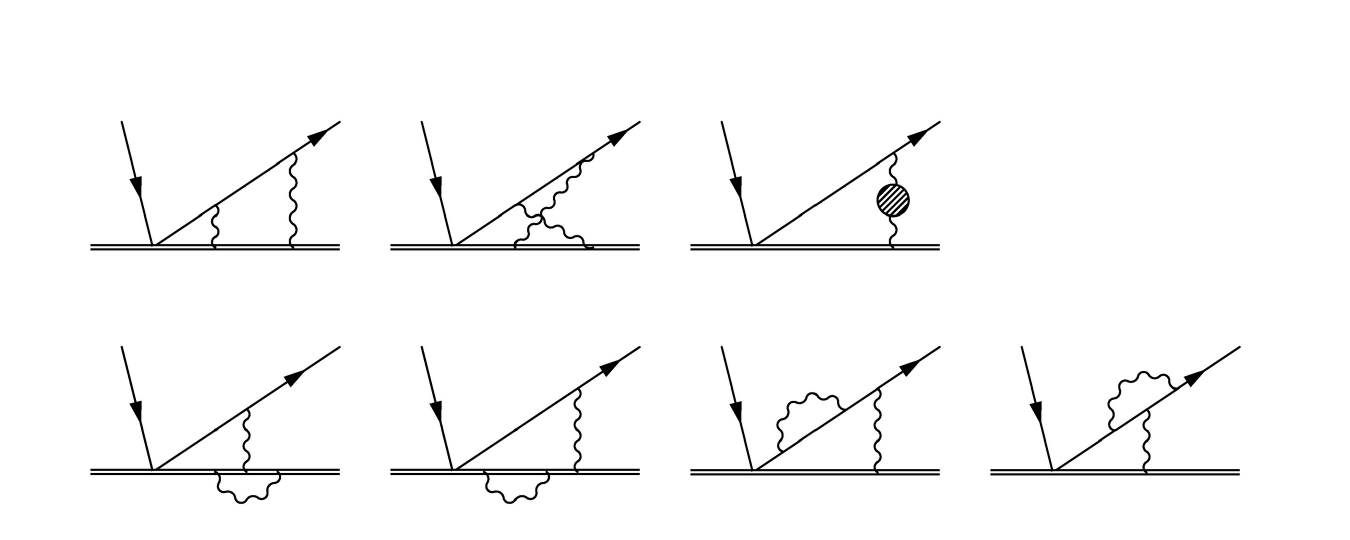}
\caption{HBChPT diagrams contributing to the anomalous dimension of $g_V$ and to $\tilde{\delta}_\mathrm{RC}$ at two loop. Only the first two diagrams give rise to terms in the $\tilde{\gamma}_1$ enhanced by $\pi^2$ \cite{Gimenez:1991bf}. These diagrams also give rise to the leading $\pi^2 \alpha^2/\beta^2$ behavior captured by the nonrelativistic  Fermi function.}\label{fig:HBChPT_digarams}
\end{figure}

We thus arrive to our final form for the differential  decay rate:
\begin{equation} \label{eqn:jtw3}  
    \frac{\mathrm{d}\Gamma_n}{\mathrm{d}E_e} = \frac{G_F^2  \,  |V_{ud}|^2 }{(2\pi )^5}   \left(  1 + 3  \lambda^2 \right) \  p_e E_e (E_0 - E_e)^2  \    \left[ g_V (\mu_\chi) \right]^2 \ F_{NR} (\beta) \bigg(  1  +   \delta_{\rm RC} (E_e, \mu_\chi) \bigg) \bigg(1+  \delta_{\rm recoil} (E_e)\bigg).
\end{equation}
Compared to state-of-the-art analyses of neutron decay in the literature (see e.g. Ref.~\cite{Czarnecki:2004cw}), our result~\eqref{eqn:jtw3} amounts to replacing the relativistic Fermi function~\cite{Fermi:1934hr,Konopinski:1935zz,Morita:1963zz,Behrens:1969,Behrens:1970ncy,Behrens:1971rbq,Behrens:1972zya,Wilkinson:1982hu} with the nonrelativistic one, $F_0 \to F_{NR}$. While we arrived at this result by constructing the relevant terms of the amplitude in the EFT framework, one could also argue for this replacement along the following lines. First, recall that the leading corrections to the phase space coming from the distortion of the electron wavefunction in the Coulomb field of the proton is usually captured by the function~\cite{Wilkinson:1982hu}
\begin{equation}
    F_0(\beta) = \frac{2}{1+\gamma} F(\beta) = 4 (2 E_e \beta R)^{2(\gamma-1)} e^{\pi y} \frac{\left|\Gamma(\gamma + i y )\right|^2}{(\Gamma(1+2\gamma))^2}, \quad y = \frac{\alpha}{\beta},\quad \gamma = \sqrt{1- \alpha^2}. 
\end{equation} 
This form is obtained by solving the Dirac equation for an electron moving in the charge distribution of a uniformly charged sphere of radius $R$ \cite{Wilkinson:1982hu}, but corresponds to a rescaling of the solution of the Dirac equation for a point-like proton, $F(\beta)$, evaluated not at the origin, where the wavefunction diverges logarithmically, but at the ``nucleon radius" $R$. $R$ corresponds to a mass scale much larger than $m_e$ and effectively acts as a UV regulator. So we see that while $F_0(\beta)$ coincides with $F_{NR}(\beta)$ at the one-loop level, $F_0$ includes a dependence on the UV regulator via the logarithms of $R$ that first appear at $\mathcal O(\alpha^2)$. Expanding $F_0$ in series of $\alpha$, one obtains
\begin{equation}
    F_0(\beta) = F_{NR} \left( \beta \right) \left[ 1 - \alpha^2  \left( \gamma_E - 3 + \ln (2 E_e R \beta) \right) + \mathcal O(\alpha^4) \right]. \label{eq:F0vsFNR}
\end{equation}
The dependence on the UV regulator $R \sim 1/\mu_\chi$ does not match the $\mu_\chi$-dependence of $g_V (\mu_\chi)$ in the $\overline{\rm MS}_\chi$ scheme presented so far. In dimensional regularization, indeed, the $\ln R$ term in Eq.~\eqref{eq:F0vsFNR} corresponds to a UV singularity that appears in the first two diagrams in Fig.~\ref{fig:HBChPT_digarams}, when we consider only the contribution arising from picking the two nucleon poles. This is only one piece of the full anomalous dimension $\tilde\gamma_1$. In order not to double-count large logarithms, one should set the logarithmic term in $F_0$ to zero when using the RGEs to evaluate the large logarithms as we do here. The remaining $\mathcal{O}(\alpha^2)$ terms in Eq.~\eqref{eq:F0vsFNR} are incomplete and beyond the accuracy of our calculation, which allows us to drop them and replace the relativistic Fermi function $F_0$ by its nonrelativistic  counterpart $F_{NR}$.

\subsection{Total decay rate and  extraction of $V_{ud}$} 
\label{sect:rate}

Upon performing the integration over $E_e$ in Eq.~\eqref{eqn:jtw3}, the decay rate can be written as 
\begin{equation}
    \Gamma_n   =  \frac{G_F^2  |V_{ud}|^2  m_e^5}{2 \pi^3}  \left(1 + 3 {\lambda}^2\right) \cdot f_0 \cdot \big( 1 + \Delta_{f} \big) \cdot \big(1 + \Delta_R  \big), \label{eq:taun}
\end{equation}
where the phase-space integral is given by 
\begin{align}
    f_0 &= \int_{1}^{x_0}  w(x,x_0) \ \mathrm{d} x , \qquad w(x,x_0) =  x \, (x_0 - x)^2 \, \sqrt{x^2 - 1}, \\
    f_0 &=\frac{2x_0^4-9x_0^2-8}{60} \sqrt{x_0^2-1}+ \frac{x_0}{4} \ln(x_0+\sqrt{x_0^2-1}),
\end{align}
with $x_0 = E_0/m_e$ and $E_0 = 1.292581~\mathrm{MeV}$, and takes the value $f_0 ( x_0 ) =  1.62989$. Following standard practice~\cite{Wilkinson:1982hu,Czarnecki:2004cw}, in Eq.~\eqref{eq:taun} we have lumped the Coulomb ($F_{NR}$) and recoil terms into an effective phase-space correction $\Delta_f$, separating the remaining radiative corrections into $\Delta_R$. In this factorization scheme, the various  corrections to the decay rate are defined by 
\begin{align}
    f_0 \left(1 + \Delta_f \right) &=    \int_{1}^{x_0}  w(x,x_0)  \,  F_{NR} \left( \beta (x) \right) \,      \left( 1 + \delta_{\rm recoil} \left( x m_e \right) \right) \ \mathrm{d} x, \label{eq:DeltaF1}     \\
    1 + \Delta_R  &= [g_V(\mu_\chi)]^2 \left( 1 + \frac{  \int_{1}^{x_0}  w(x,x_0)  \,  F_{NR} \left( \beta(x) \right) \left( 1 + \delta_{\rm recoil} \left( x m_e \right) \right)  \  \delta_{\rm RC} \left( x m_e, \mu_\chi \right) \ \mathrm{d} x}{f_0 (1 + \Delta_f )}     \right), \label{eq:DeltaR1}
\end{align}
where $\beta(x) = \sqrt{1-1/x^2}$. A few remarks are in order: 
\begin{itemize}
\item The decay rate in Eq.~\eqref{eq:taun} corresponds to the usual definition adopted in the literature~\cite{Czarnecki:2004cw}, upon identifying $f \equiv f_0 (1 + \Delta_f)$. Therefore, the total shift in the decay rate 
\begin{equation}
    \Delta_{\rm TOT} = - 1 + (1 + \Delta_f) (1 + \Delta_R), 
\end{equation}
which impacts the extraction of $V_{ud}$, requires specifying both $\Delta_f$ and $\Delta_R$. The expressions and numerical values of $\Delta_f$ and $\Delta_R$ in our EFT approach differ from the results found in the literature (see Ref.~\cite{Czarnecki:2004cw} and most recent calculations of $\Delta_R$~\cite{Cirigliano:2022yyo,Seng:2018yzq,Seng:2018qru,Czarnecki:2019mwq,Shiells:2020fqp,Hayen:2020cxh,Seng:2020wjq}). In what follows, when necessary we will discuss the origin of the differences. 

\item For $\Delta_f$, which encodes Coulomb and recoil corrections, we find 
\begin{equation}
    \Delta_f = 3.573 (5)  \%,
\end{equation}
where we estimated the uncertainty to be of the size of Coulomb corrections times the recoil cross term. The difference from the standard result $\Delta_f = 3.608 \times 10^{-2}$~\cite{Czarnecki:2004cw} is mainly due to the fact that we use the nonrelativistic Fermi function, for the reasons discussed above, while Ref.~\cite{Czarnecki:2004cw} uses the relativistic Fermi function. We also do not include the corrections induced by modeling the proton as a uniformly charged sphere of radius $R_p \simeq 1$~fm~\cite{Wilkinson:1982hu}: this is a small effect shifting $\Delta_f$ by $0.005\%$.

\item Up to the  accuracy of our calculation, the remaining radiative correction $\Delta_R$ in our framework is given by
\begin{equation}
    \Delta_R = \left[g_V (\mu_\chi)  \right]^2  \left( 1 + \frac{\alpha \left(\mu_\chi \right)}{2\pi} \left( \frac{3}{2} \ln \frac{\mu_\chi^2}{m_e^2} + \frac{5}{4} + \stackrel{\underline{\ }}{\hat{g}} \left(E_0 \right)  \right) \right) -1,  \label{eq:DeltaRf}
\end{equation}
where $\mu_\chi \sim m_e$ and $\stackrel{\underline{\ }}{\hat{g}} \left(E_0 \right) = -9.58766$ is obtained by averaging the subtracted Sirlin function $\hat g (E_e, E_0)$ over the phase space, according to Eq.~\eqref{eq:DeltaR1}. At leading order in $\alpha$, the $\mu_\chi$-scale dependence in Eq~\eqref{eq:DeltaRf} cancels between the coupling constant $g_V \left( \mu_\chi \right)$ and virtual one-loop contributions, while higher-order perturbative logarithms from virtual diagrams at scales $\mu_\chi \sim m_e$ are small. 

\item 
To separate hadronic and electroweak contributions to $g_V \left( \mu_\chi \right)$, and to make contact with some of the previous literature,  we provide the fixed-order result 
\begin{equation}
    \Delta_R = 2 \overline \Box_\mathrm{Had}^V (\mu_0) + \frac{\alpha}{2 \pi} \left[ 2 \left(1 - \frac{\alpha_s}{4 \pi} \right)  \ln \frac{M_Z^2}{\mu_0^2} + \frac{3}{2} \ln \frac{\mu_0^2}{m_e^2} +  \stackrel{\underline{\ }}{\hat{g}} \left(E_0 \right) \right].\label{eq:one_loop_delta_R}
\end{equation}
In the above relations, the explicit dependence on $\mu_0$ is canceled by the implicit dependence in $\overline \Box_\mathrm{Had}^V (\mu_0)$. The hadronic physics is included in $\overline \Box_\mathrm{Had}^V$, while the two logarithms in Eq.~\eqref{eq:one_loop_delta_R}, which are proportional to the anomalous dimensions, correspond to the ratios between electroweak vs hadronic and hadronic vs beta-decay scales.

\item  Our numerical result for $\Delta_R$ is 
\begin{equation}
    \Delta_\mathrm{R} = 4.044(27)\%,
\end{equation}
which, apart from the uncertainty coming from $g_V$ discussed in Sect. \ref{sec:numerics}, includes a perturbative uncertainty of $0.005\%$ obtained by varying the scale of the calculation $\mu_\chi$ in the range $ m^2_e/2 \le \mu^2_\chi \le 2 m^2_e$. Our result for $\Delta_R$ is $0.061\%$ above the most recent evaluation~\cite{Cirigliano:2022yyo} based on Refs.~\cite{Seng:2018yzq,Seng:2018qru,Czarnecki:2019mwq,Seng:2020wjq,Hayen:2020cxh,Shiells:2020fqp}. The sources of this difference are discussed in Section~\ref{sec2}. Combining $\Delta_f$ and $\Delta_R$ in the factorization scheme of Eq.~\eqref{eq:taun} we obtain
\begin{equation}
    \Delta_{\rm TOT} = 7.761 (27) \%.  \
\end{equation}
Using the results from Refs.~\cite{Cirigliano:2022yyo,Seng:2018yzq,Seng:2018qru,Czarnecki:2019mwq,Seng:2020wjq,Hayen:2020cxh,Shiells:2020fqp}, one gets $\Delta_{\rm TOT} = 7.735 (27)\%$, about one $\sigma$ below our result. The difference is due to two competing factors in our analysis: a positive shift of $+0.061\%$ in $\Delta_R$ and a negative shift of $-0.035\%$ in $\Delta_f$.

\item As a consistency check on the accuracy of the calculation and the size of cross terms (such as recoil $\times$ electromagnetic corrections), we have performed the phase-space integration in a different scheme that does not assume factorization of $F_{NR}$ and $\delta_{\rm recoil}$, defined by 
\begin{equation}
    \Gamma_n   \to   \frac{G_F^2  |V_{ud}|^2  m_e^5}{2 \pi^3}  \left(1 + 3 {\lambda}^2\right) \cdot f_0 \cdot \left( 1 + \Delta_{g_V} \right) \cdot \bigg(1 + \Delta_{\rm recoil} + \Delta_C + \Delta_{\rm RC} \bigg),
\end{equation}
with 
\begin{align}
    \Delta_{g_V} &= [g_V(\mu_\chi)]^2 - 1, \\
    \Delta_{\rm C} &= \frac{1}{f_0} \,  \int_{1}^{x_0}  w(x,x_0)  \, \left[  F_{NR} \left( \beta \left( x \right) \right) -   \left( 11 -  \frac{\left( E_0 - m_e \right)^2}{3 m^2_e} \right)  \frac{\alpha^2}{4 \beta \left( x \right)} - 1 \right] \ \mathrm{d} x, \\
    \Delta_{\rm RC} &=  \frac{1}{f_0} \,  \int_{1}^{x_0}  w(x,x_0)  \,    \delta_{\rm RC} \left( x m_e,\mu_\chi \right) \ \mathrm{d} x, \\
    \Delta_{\rm recoil} &= \frac{1}{f_0} \,  \int_{1}^{x_0}  w(x,x_0)  \,    \delta_{\rm recoil} \left( x m_e \right) \ \mathrm{d} x.
\end{align}
For the numerical values in this scheme, we find $\Delta_{g_V} = 5.060(27)\%$, $\Delta_\mathrm{C} = 3.375 \%$, $\Delta_\mathrm{RC} = -0.969\%$, and $\Delta_{\rm recoil} = 0.173\%$, leading to $\Delta_{\rm TOT} = 7.770\%$. The latter differs from the factorized result by 0.009\%, consistent with its expected size of ${\cal O}(\alpha^2)$ and the uncertainties quoted above.
\end{itemize} 

Finally, we extract the CKM matrix element $V_{ud}$ from precise measurements of the neutron lifetime with our updated calculation of radiative corrections and present the results in Section~\ref{sec2}.

\subsection{Comments on radiative corrections to nuclear decays}

Finally, we comment on the connection to the standard framework for the analysis of superallowed $0^+ \to 0^+$ transitions, described for example in Ref.~\cite{Hardy:2020qwl}. The corrections to nuclear beta decays are combined into the quantity ${\mathcal F t}$, related to the experimental $ft$ values as
\begin{equation}
    {\mathcal F t} = ft (1 + \delta_R^\prime) (1 + \delta_{NS} - \delta_C) = \frac{K}{2 G_F^2 |V_{ud}|^2 (1 + \Delta^V_R)},
\end{equation}
where $K$ is a constant and $\delta_C $ is the isospin-symmetry breaking contribution. The correction $\delta_{NS}$ corresponds to the transition-dependent nuclear structure correction. $\Delta_R^V$ is the transition-independent part of the radiative correction, which is related to the correction to neutron decay via
\begin{equation}
    \Delta_R^V  =  \Delta_R -  \frac{\alpha}{2 \pi} \bar{g} (E_0) = \left[ g_V \left( \mu_\chi \right) \right]^2 \left( 1 + \frac{\alpha \left( \mu_\chi \right)}{2 \pi}  \left(  \frac{3}{2} \ln \frac{\mu_\chi^2}{m_N^2} + \frac{5}{4}  \right) \right) - 1.\label{eq:DeltaR2}
\end{equation}
$\delta_R^\prime$ contains the so-called ``outer corrections'', which depend on the transition but not on the  nuclear structure, and corresponds to soft photon emissions from point-like nuclei. $\delta_R^\prime$ reduces to the Sirlin function at $\mathcal O(\alpha)$ and includes a set of $\mathcal O(Z \alpha^2)$ and $\mathcal O(Z^2 \alpha^3)$ corrections  \cite{Jaus:1970tah,Jaus:1972hua,Jaus:1986te,Sirlin:1986cc,Sirlin:1986hpu}. In addition, it contains the leading-logarithm renormalization group evolution from $m_N$ to $m_e$, using the RGE kernel derived in Ref.~\cite{Czarnecki:2004cw} and discussed in Ref.~\cite{Towner:2007np}. Therefore, the standard breakdown of radiative corrections corresponds to evaluating the coupling $g_V$ at a scale $\mu_\chi \sim \Lambda_\chi \sim m_N$ in Eq.~\eqref{eq:DeltaR2}, and then lumping the leading-logarithm RG evolution and the matrix element in $\delta_R^\prime$, namely 
\begin{equation}
    \left. \Delta^V_R \right|_{\rm Traditional} = \left[ g_V \left( m_N \right) \right]^2 \left( 1 +  \frac{5 \alpha (m_N)}{8 \pi} \right) -1 = 2.471(25) \%, 
\end{equation}
which agrees with the result compiled in Ref.~\cite{Cirigliano:2022yyo}.

From an EFT point of view aiming at describing nuclei starting from nucleon degrees of freedom, it is more natural to evolve the single-nucleon (and possibly two-nucleon, three-nucleon $\ldots$) coupling $g_V$ all the way down the scale $\mu_\chi =   m_e$, and only leave the evaluation of the fixed-order matrix element in $\delta_R^\prime$. This can be achieved by defining the universal correction $\Delta^V_R |_{\rm EFT}$:
\begin{equation}
    \left. \Delta^V_R \right|_{\rm EFT} = \left[ g_V \left( m_e \right) \right]^2 \left( 1 + \frac{5 \alpha \left( m_e \right)}{8 \pi} \right) -1,
\end{equation}
and appropriately redefining the Fermi function, the outer correction $\delta_R^\prime$, and the nuclear correction $\delta_{NS}$ in such a way that they do not contain large logarithms. This requires a new EFT analysis of both $\delta_R^\prime$ and $\delta_{NS}$.

\section{Conclusions and Outlook}
\label{sec8}

In this paper, we developed a systematically-improvable top-down effective field theory framework for radiative corrections to the neutron $\beta$ decay and low-energy (anti)neutrino-nucleon scattering. As a first step, we perform the matching at the electroweak scale of the Standard Model to the LEFT with effective four-fermion operators. We resum leading and next-to-leading large electroweak logarithms to all orders by evolving the semileptonic coupling constant in the LEFT from electroweak to GeV scale according to the LEFT RGEs. Next, we perform the matching to the heavy-baryon chiral perturbation theory at the hadronic scale and express the HBChPT low-energy constants in terms of non-perturbative correlation functions of quark currents. To avoid large logarithms in the evaluation of the matrix elements, we resum leading and subleading logarithms between the hadronic scale and electron-mass scale according to the RGEs in HBChPT. In this framework, all contributions from physics above the scale of the electron mass play the role of short-distance effects, which are captured by the Wilson coefficient $g_V$. We compare our framework to the traditional current-algebra approach and find an agreement at the one-loop level. Contrary to the traditional approach, we employ dimensional regularization with minimal subtraction ($\overline{\mathrm{MS}}$) and specify the scale and scheme dependence in all steps of the calculation allowing us to consistently include the next-to-leading logarithms and their resummation. In our approach, the so-called DIS region of the $\gamma W$ box is mapped onto a contribution to the Wilson coefficient in LEFT.

In our new EFT framework, we determined the low-energy vector coupling constant $g_V(m_e) - 1 = (2.499\pm0.012)\%$, which controls the neutron decay rate as well as low-energy (anti)neutrino-nucleon scattering, and provides the basis for one-body contributions to nuclear decays. We also extracted the CKM matrix element $V_{ud}$ from neutron decay measurements with our new values for the radiative corrections. An updated value $|V_{ud}| = 0.97402(42)$ based on the most precise determinations of the neutron lifetime and axial-vector to vector coupling constants ratio is smaller than previous results. The difference with respect to the previous analyses originates from the consistent inclusion of the next-to-leading logarithms and Coulomb corrections within our framework.

The effective field theory approach to radiative corrections in weak processes advocated in this paper can be extended to the analysis of the axial-vector coupling constant $g_A$, which is a natural next step that will be presented in future work. The developed EFT approach can straightforwardly be applied to precise first-principles cross-section calculations in low-energy (anti)neutrino-nucleon scattering and can be extended to describe neutral-current processes with nucleons at low energies. The EFT framework can also be generalized to address radiative corrections to nuclear decays. In fact, one of the advantages of EFT is that the effective couplings $g_V$ and $g_A$ already determine the one-body inner corrections to nuclear decays at $\mathcal O(G_F \alpha)$. Consequently, matrix elements of the weak Lagrangian of Eq.~\eqref{eq:Lagrangian_at_leading_order} should be computed to $\mathcal{O} \left( \alpha \right)$ in the low-energy nuclear many body theory. In this approach, the so-called ``nuclear $\gamma W$ box" arises from contributions at scales smaller than or equal to the Fermi momentum $k_F$, which are calculable in the nuclear EFT. Short-range physics is captured by $g_V$, and potentially, by two-nucleon and/or few-nucleon weak operators in HBChPT, whose contributions at a given order in $\epsilon_\chi$ can be be estimated by the power counting in chiral EFT. The full analysis of radiative corrections to nuclear beta decay will require the development of the EFT framework for few-nucleon systems to $\mathcal{O}(G_F \alpha \epsilon_\chi^n)$.

\section*{Acknowledgments}
We thank Chien-Yeah Seng for providing us with hadronic inputs in the evaluation of the $\gamma W$ box diagram and for explaining the details of his calculations, Andrzej Czarnecki and Bastian Kubis for useful correspondence, Ryan Plestid and Martin Hoferichter for useful discussions at the INT Program INT-23-1B, and Richard Hill for useful discussions and validations. We thank Jordy de Vries, Mikhail Gorchtein, Leendert Hayen, Duff Neill, Alessandro Vicini, Andreas von Manteuffel, Andr\'e Walker-Loud for useful discussions and comments on the manuscript. V.C. and W.D. acknowledge support by the U.S. DOE under Grant No. DE-FG02-00ER41132. This work is supported by the US Department of Energy through the Los Alamos National Laboratory and by LANL’s Laboratory Directed Research and Development (LDRD/PRD) program under projects 20210968PRD4 and 20210190ER. Los Alamos National Laboratory is operated by Triad National Security, LLC, for the National Nuclear Security Administration of U.S. Department of Energy (Contract No. 89233218CNA000001). We acknowledge support from the DOE Topical Collaboration ``Nuclear Theory for New Physics", award No. DE-SC0023663. FeynCalc~\cite{Mertig:1990an,Shtabovenko:2016sxi}, LoopTools~\cite{Hahn:1998yk}, and Mathematica~\cite{Mathematica} were extremely useful in this work.

\appendix

\section{Electromagnetic fine-structure constant in LEFT and $\chi$PT}
\label{sect:alpha}

In this appendix, we discuss the definition we adopt for the running fine-structure constant $\alpha (\mu)$ used in the LEFT and $\alpha_\chi (\mu_\chi)$ in $\chi$PT.

In any theory, including LEFT and $\chi$PT, charge renormalization is studied in connection with the photon self-energy tensor $\Pi^{\mu \nu} (q^2)$ and the vacuum polarization function $\Pi (q^2)$ defined by~\cite{Peskin:1995ev}
\begin{equation}
    \Pi^{\mu \nu} (q) = \left( g^{\mu \nu} q^2 - q^\mu q^\nu \right) \Pi (q^2).
\end{equation}
Including resummed self-energy corrections, the amplitude for scattering of two charged particles is proportional to the physical (renormalization scale and scheme-independent) combination 
\begin{equation}
    \kappa_{phys}  (q^2) =   \frac{\alpha_R}{1 - \Pi_R (q^2)}, \label{eq:kappa}
\end{equation}
where the subscript ``$R$" labels the renormalization scheme, $\alpha_R$ denotes the renormalized fine-structure constant and $\Pi_R$ is the corresponding subtracted, UV-finite vacuum polarization function. For example, in the ``on-shell" ($OS$) renormalization scheme the renormalized vacuum polarization function is defined by $\Pi_{OS} (q^2) = \Pi(q^2 ) - \Pi(0) \equiv \Delta \alpha (q^2)$. In this scheme, using observables at $q^2 \to 0$, one extracts $\alpha_{OS} = 1/137.036$. This scheme can be implemented in any field theory, including $\chi$PT, LEFT, and the full Standard Model.

Importantly, Eq.~\eqref{eq:kappa} allows one to relate $\alpha_R$ (in any scheme and at any renormalization scale) to $\alpha_{OS}$,  in terms of $\Pi_R (q^2=0)$. Eq.~\eqref{eq:kappa} can also be used to relate the electromagnetic couplings defined in {\it any} two renormalization schemes and even in two different EFTs.

\subsection{Charge renormalization in LEFT} 

Throughout this paper, we use the notation $\alpha (\mu)$ to indicate the electromagnetic coupling in LEFT, defined in the modified minimal subtraction scheme ($\overline{\rm MS}$).

At any value of $\mu < M_{W,Z}$, where the LEFT is applicable, $\alpha (\mu)$ can be defined by its relation to $\alpha_{OS}$ via Eq.~\eqref{eq:kappa}, leading to
\begin{subequations} \label{eq:aLEFT1}
\begin{equation}
    \alpha_{OS} = \frac{\alpha (\mu)}{1 - \Pi_{\overline{\rm MS}} (0)} = \frac{\alpha (\mu)}{1 - \Pi_{\overline{\rm MS}} (\tilde{\mu}^2) + \Pi_{OS} ( \tilde{\mu}^2) },
\end{equation} 
where in the second equality we have expressed $\Pi_{\overline{\rm MS}} (0)$ in terms of $\Pi_{\overline{\rm MS}} (\tilde{\mu}^2)$ and $\Pi_{OS} ( \tilde{\mu}^2)$, at the arbitrary scale $\tilde{\mu} \gg \Lambda_{QCD}$. In LEFT, the vacuum polarization receives contributions from charged fermions only. The contribution of charged leptons to both $ \Pi_{\overline{\rm MS}} (\tilde{\mu}^2)$ and $\Pi_{OS} ( \tilde{\mu}^2) = \Pi (\tilde{\mu}^2) - \Pi (0)$ can be computed in perturbation theory. For quarks, the calculation of $\Pi_{\overline{\rm MS}} (\tilde{\mu}^2)$ can be carried out in perturbation theory because $\tilde{\mu}^2 \gg \Lambda_{\rm QCD}^2$, with each quark flavor of charge $Q_q$ contributing (to zeroth order in the QCD coupling $\alpha_s$)
\begin{equation}
    \Pi^{(q)}_{\overline{\rm MS}} (\tilde{\mu}^2 \gg \Lambda_{\rm QCD}^2)= \frac{4}{3} N_C Q_q^2 \frac{\alpha}{4 \pi} \left[\ln \frac{\tilde{\mu}^2}{\mu^2} - \frac{5}{3} \right].
\end{equation}
The non-perturbative contributions are encoded in $\Pi_{OS} (\tilde{\mu}^2)$ and can be evaluated via a dispersion relation
\begin{equation}
    \Pi_{OS} (\tilde{\mu}^2) = \Delta \alpha (\tilde{\mu}^2) = - \frac{\alpha}{3 \pi} \,  \tilde{\mu}^2 \, {\rm Re}  \int_{4 m_\pi^2}^\infty \, \mathrm{d} s \frac{R(s)}{s ( s - \tilde{\mu}^2 + i 0^+)},
\end{equation}
\end{subequations}
where $R(s) = \sigma_{e^+ e^- \to {\rm hadrons}} (s) / \sigma_{e^+ e^- \to \mu^+ \mu^-} (s)$, see for example Ref.~\cite{Davier:2019can}.

The scheme matching conditions between the on-shell and $\overline{\rm MS}$ couplings given in Eqs.~\eqref{eq:aLEFT1} are conceptually clean and could be implemented at any value of $\mu < M_{W,Z}$. They are, however, not extremely practical, because the dispersive integral $\Delta \alpha (\tilde{\mu}^2)$ is usually given in the literature only at $\tilde{\mu} = M_Z$~\cite{ParticleDataGroup:2022pth,Davier:2019can}. To circumvent this issue, we define the LEFT $\overline{\rm MS}$ coupling $\alpha (\mu)$ by relating it to the $\overline{\rm MS}$ fine-structure constant in the full Standard Model with five quark flavors, denoted by $\hat \alpha^{(5)} (\mu)$ in the PDG review on the electroweak theory~\cite{ParticleDataGroup:2022pth}. $\hat \alpha^{(5)} (M_Z)$ is related to $\alpha_{OS}$ by an expression which is analogous to~\eqref{eq:aLEFT1}, with $\mu = \tilde{\mu} = M_Z$, up to an additional contribution to charge renormalization due to the $W$ boson~\cite{Fanchiotti:1992tu}. Taking this into account, we find at $\mu = \mu_{SM} \sim M_{W}$,
\begin{equation}
    \frac{1}{ \alpha (\mu_{SM}) } = \frac{1}{\hat \alpha^{(5)} (\mu_{SM})}   - \frac{1}{6 \pi}  + \frac{7}{2\pi} \ln \frac{M_W}{\mu_{SM}}.
\end{equation} 
 
The numerical value $\hat \alpha^{(5)} (M_Z)^{-1} = 127.951 (9)$~\cite{ParticleDataGroup:2022pth} implies $\hat \alpha^{(5)} (M_W)^{-1} = 127.989 (9)$ and $\alpha (M_W)^{-1} = 127.936 (9)$. We use the latter value as the initial condition for the RGE
\begin{align}
    \mu \frac{\mathrm{d} \alpha (\mu)}{\mathrm{d} \mu} &= - \frac{\beta_0 (\mu)}{2 \pi}   \alpha^2 (\mu) + \mathcal{O}(\alpha^3) ,\\  
    \beta_0 (\mu) &= - \frac{4}{3} \tilde n (\mu), \qquad  \qquad \tilde{n} (\mu) = \sum_f Q_f^2 n_f  \theta(\mu - m_f), 
\end{align}
where $Q_f$ denotes the fermion charge and $n_f$ is the multiplicity ($n_f=1$ for charged leptons, $n_f= N_C$ for quarks). To the accuracy we work, $\alpha (\mu)$ can be treated as continuous across heavy-fermion thresholds in LEFT.

\subsection{Charge renormalization in $\chi$PT}
\label{appa2}

Below the QCD scale $\Lambda_\chi \sim m_N$, we work with (baryon) $\chi$PT extended to include dynamical photons and leptons. For illustrative purposes, we consider $SU(2)$ $\chi$PT with both  muon and electron as dynamical fields. This is the field content relevant for matching to the LEFT at the scale $\Lambda_\chi$. We add the gauge-kinetic terms for photons and leptons and couple them to mesons and baryons by appropriate shifts to the external field sources that appear in chiral covariant derivatives, see Eq.~(\ref{eq:sources}). Upon redefining the photon field $A_\mu \to (1/e) A_\mu$, the electromagnetic coupling $e$ only appears in the photon kinetic term and the relevant renormalized Lagrangian, written in terms of renormalized coupling $\hat e_\chi$ and counterterms $Z_{A,\chi} -1$, reads
\begin{subequations} \label{eq:chpts}
\begin{eqnarray}
    {\cal L}_{A,\chi} &=& - \frac{1}{4 \hat e_\chi^2} F_{\mu \nu} F^{\mu \nu} -  \frac{1}{4 \hat e_\chi^2} F_{\mu \nu} F^{\mu \nu} \left( Z_{A,\chi} - 1 \right) + ...,\\ 
    Z_{A,\chi} &=& 1 +8\hat e_\chi^2 h_2 - 4 \hat  e_\chi^2 X_8. \label{eq:ZA}
\end{eqnarray}
\end{subequations}
The low-energy constant (LEC)  $h_2$  was introduced in Ref.~\cite{Gasser:1983yg} in the context of $SU(2)$ meson $\chi$PT. $X_8$ was introduced in Ref.~\cite{Knecht:1999ag},  which extended $\chi$PT to include dynamical leptons, for the study of semileptonic processes. In  $\chi$PT, the LECs contain a pure counterterm that subtracts the UV divergences of meson and lepton loops, and a finite, renormalized coupling that encodes contributions from heavy states not included in the EFT. Adopting dimensional regularization in $d=4 - 2 \varepsilon$ and following Refs.~\cite{Gasser:1983yg,Gasser:1984gg}, the generic LEC $ {\cal C}_i$ is  written as
\begin{eqnarray}
    {\cal C}_i  = {\cal C}_i^r (\mu_\chi)  - \frac{\gamma_i}{2} \frac{1}{(4 \pi)^2}  \left(  \frac{1}{\hat \varepsilon}  + 1 \right) , 
\end{eqnarray} 
where  $\gamma_{h_2} = 1/12$, $\gamma_{X_{8}} = -4/3$~\cite{Knecht:1999ag}, and $1/\hat \varepsilon =   1/ \varepsilon - \gamma_E + \ln \left( 4 \pi \right)$. The renormalized couplings ${\cal C}_i^r (\mu_\chi)$ depend on the scale in such a way that, after including loops, the physical amplitudes are $\mu_\chi$-independent. $h_2$ cancels divergences induced by pseudoscalar meson loops, while $X_8$ cancels divergences produced by loops with the electron and muon.

In the standard $\chi$PT scheme defined by  Eqs.~\eqref{eq:chpts}, we can study charge renormalization and vacuum polarization. The renormalized electromagnetic coupling $\hat \alpha_\chi \equiv \hat e_\chi^2/(4 \pi)$ and $\Pi_{\hat \chi} (q^2)$ are separately scale-independent. In fact, the subtracted vacuum polarization is given by
\begin{equation}
    \Pi_{\hat \chi} (q^2) = \Pi_\pi (q^2, m_\pi^2, \mu_\chi^2) + \Pi_\ell (q^2, m_e^2, \mu_\chi^2)+\Pi_\ell (q^2, m_\mu^2, \mu_\chi^2) + \Pi_{LECs} (\mu_\chi^2),
\end{equation}  
in terms of the pion loop, the lepton loop, and the counterterm contributions
\begin{subequations}
\begin{eqnarray} \label{eq:PiS}
    \Pi_\pi (q^2, m_\pi^2, \mu_\chi^2) &=& \frac{\hat \alpha_\chi}{4 \pi} \left[- \frac{1}{3} \frac{1}{\hat{\varepsilon}} + \int_0^1 \mathrm{d}x  \, (1 - 2 x)^2 \, \ln \left( \frac{m_\pi^2 - q^2 x (1-x)  -i 0^+}{\mu_\chi^2} \right)\right],\\ \label{eq:PiF}
    \Pi_\ell (q^2, m_\ell^2, \mu_\chi^2) &=& \frac{\hat \alpha_\chi}{4 \pi} \left[- \frac{4}{3} \frac{1}{\hat{\varepsilon}} +  8 \int_0^1 \mathrm{d}x  \, x (1-x) \, \ln \left( \frac{m_\ell^2 - q^2 x (1-x)  -i 0^+}{\mu_\chi^2} \right) \right],\\
    \Pi_{LECs}(\mu_\chi^2) &=& -8\hat e_\chi^2  h_2^r(\mu_\chi)  + 4 \hat e_\chi^2 X_8^r (\mu_\chi) \nonumber \\
    &+& \frac{\hat \alpha_\chi}{4 \pi} \left(\frac{1}{\hat \varepsilon} + 1 \right) \left[ \left(\frac{1}{3}\right)_{h_2} + \left(\frac{8}{3} \right)_{X_8} \right].
\end{eqnarray}
\end{subequations}
The divergent parts of the LECs, independently presented in Refs.~\cite{Gasser:1983yg,Knecht:1999ag}, cancel the loop contributions from light charged particles, as expected. The $\mu_\chi$ dependence cancels in the sum of loop and LECs contributions, so that  $\Pi_{\hat \chi} (q^2)$ does not depend on $\mu_\chi$. Moreover,  the relation between the known $\alpha_{OS}$ and $\hat \alpha_\chi$ depends on unknown LECs through $\Pi_{\hat \chi} (q^2=0)$ in the scheme-matching relation:
\begin{equation}
    \alpha_{OS}  = \frac{ \hat \alpha_\chi}{1 - \Pi_{\hat \chi} (0)}. 
\end{equation}

A more convenient renormalization scheme, that more closely resembles the standard minimal subtraction, is achieved by rewriting $Z_{A,\chi}$  in Eq.~\eqref{eq:ZA} as
\begin{subequations} \label{eq:chpts2}
\begin{eqnarray}
    Z_{A,\chi} &=& 1 + \bar{z}_{A,\chi} (\mu_\chi) + \delta z_{A,\chi},   \\ 
    \bar{z}_{A,\chi}  (\mu_\chi) &=&  8 e_\chi^2h_2^r(\mu_\chi)  - 4 \hat  e_\chi^2 X_8 (\mu_\chi),
\end{eqnarray}
where  $\delta z_{A,\chi}$ contains the divergent parts of the LECs, proportional to $(1/{\hat \varepsilon} + 1)$, and the fine-structure constant is redefined as 
\begin{equation}
    \alpha_\chi (\mu_\chi) \equiv  \frac{ \hat \alpha_\chi}{1 + \bar z_{A, \chi} (\mu_\chi)}. 
\end{equation} 
Such a redefinition corresponds to a different choice (scheme) in separating the counterterm Lagrangian. The first of Eqs.~\eqref{eq:chpts} now reads (up to higher-order terms in $\alpha_\chi$)
\begin{equation}
    {\cal L}_{A,\chi} = - \frac{1}{4 e_\chi^2} F_{\mu \nu} F^{\mu \nu} -  \frac{1}{4  e_\chi^2} F_{\mu \nu} F^{\mu \nu} \ \delta z_{A,\chi}   +  ....
\end{equation}
The corresponding finite vacuum polarization is given by
\begin{equation}
    \Pi_{\chi} (q^2) = \ \tilde \Pi_\pi (q^2, m_\pi^2, \mu_\chi^2) + \tilde \Pi_\ell (q^2, m_e^2, \mu_\chi^2)+\tilde \Pi_\ell (q^2, m_\mu^2, \mu_\chi^2) , \label{eq:Pichi}
\end{equation}  
where $\tilde \Pi_{\pi,\ell}$ are obtained from $\Pi_{\pi,\ell}$  by replacing $\hat \alpha_\chi \to \alpha_\chi$ and $1/\hat{\varepsilon} \to -1$ in \eqref{eq:PiS} and \eqref{eq:PiF}.
\end{subequations}

In this new scheme, the beta function for the renormalized coupling $\alpha_\chi (\mu_\chi)$ is similar to the standard minimal subtraction scheme. The running of $\alpha_\chi (\mu_\chi)$ is controlled by
\begin{eqnarray}
    \mu_\chi \frac{\mathrm{d} \alpha_\chi (\mu_\chi)}{\mathrm{d} \mu_\chi} &=& - \frac{\beta_0 (\mu_\chi)}{2 \pi}   \alpha^2 (\mu_\chi) + \mathcal{O}(\alpha_\chi^3) ,\\  
    \beta_0 (\mu_\chi) &=&  - \frac{4}{3} \tilde n_\ell (\mu_\chi) -\frac{1}{3} \tilde n_\pi (\mu_\chi) , \qquad\tilde{n}_{\ell,\pi} (\mu_\chi) = \sum_{\ell,\pi} Q_{\ell,\pi}^2 n_{\ell,\pi}  \, \theta(\mu_\chi - m_{\ell,\pi}), 
\end{eqnarray}
where the sum over charged leptons ($\ell$) includes the electron and the muon.

Another benefit of this renormalization scheme is that the relation to the on-shell coupling does not involve any unknown LECs:
\begin{equation}
    \alpha_{OS}  = \frac{ \alpha_\chi}{1 -  \Pi_{\chi} (0)},
\end{equation}
with $\Pi_\chi(0)$ obtained, at one loop, from Eq.~\eqref{eq:Pichi}. This implies
\begin{equation}
    \frac{1}{ \alpha_\chi  (\mu_\chi)} =   \frac{1}{\alpha_{OS}}  +  \frac{1}{3 \pi} \sum_{\ell=e,\mu} Q_\ell^2   \left(1 + \ln \frac{m_\ell^2}{\mu_\chi^2} \right)   \theta (\mu_\chi - m_\ell)  +  \frac{1}{12 \pi}  Q_\pi^2   \left(1 + \ln \frac{m_\pi^2}{\mu_\chi^2} \right) \theta (\mu_\chi - m_\pi).
\end{equation} 
The above formula accounts for the discontinuity of $\alpha_\chi (\mu_\chi)$ at the particle mass threshold, due to the fact that conventionally in $\chi$PT one subtracts $1/\hat \varepsilon + 1$ rather than $1/\hat \varepsilon$. In the main text of this manuscript, we imply $\alpha_\chi$ as soon as $\alpha$ has an argument $\mu_\chi$. Numerically, using the boundary conditions described above and running $\alpha (\mu)$ down from $\mu=M_Z$ and $\alpha_\chi (\mu_\chi)$ up from $\mu_\chi=m_e$, we find $1/\alpha_\chi (m_p) = 135.112$ while $1/\alpha (m_p) = 134.302$.

\section{Factorization of the decay rate in the nonrelativistic limit}\label{app:nrqed}

In this appendix, we provide a justification of the factorized form for the electromagnetic corrections provided in Eq. \eqref{eq:factorization}. This form can be rigorously derived for the nonrelativistic electron, but, even for the relativistic electron, it captures the leading series in $(\pi \alpha/\beta)^n $ and $\alpha/\pi (\pi \alpha/\beta)^n$. As these terms are enhanced by a factor of $\pi^2$ with respect to naive two-loop corrections, they are relevant at the level of $\sim 10^{-4}$, and their estimate requires the evaluation of the diagrams in Fig. \ref{fig:HBChPT_digarams} and of next-to-next-to-leading order real-virtual and real emission diagrams.
As we argue below, the first corrections to Eq. \eqref{eq:factorization} are of order $\mathcal O(\alpha^2)$. 

For $E_0 - E_e \ll m_e$, which does not apply to neutron decay but would apply, for example, to triton decay, we could further integrate out the scale of the electron mass, and match $\slashpi$EFT onto a theory with nonrelativistic electrons (NRQED). In this theory, the charged-current vector operator, with coupling constant $g_V$ in front, matches onto
\begin{equation}
    \mathcal L_{\rm NRQED} = - \sqrt{2} G_F V_{ud} \, g_V C_{\rm NQRED}  \, \bar{\psi}_{e} \gamma_0 \mathrm{P}_\mathrm{L} \nu_e  \, \bar{N}_v  \tau^+  N_v  + \mathcal O(\beta^2),
\end{equation}
where $\psi_e$ denotes a nonrelativistic electron field. The matching coefficient at one loop can be extracted from the matching of heavy-light onto heavy-heavy currents performed in Ref.  \cite{Grozin:2004yc}, and, in the $\overline{\rm MS}_\chi$ scheme, it is given by   
\begin{equation}
    C_{\rm NRQED}(\mu) = 1 + \frac{\alpha}{2\pi} \left( \frac{3}{4}\ln\frac{\mu^2}{m_e^2} - \frac{11}{4}   \right).
\end{equation}
The NRQED Lagrangian still contains various photon modes: soft ($k_0 \sim |\vec k| \sim m_e \beta$), ultrasoft ($k_0 \sim |\vec k| \sim m_e \beta^2$), and potential ($k_0 \sim  m_e \beta^2 \ll |\vec k| \sim m_e \beta$). We can integrate out soft and potential modes by matching onto potential NRQED (pNRQED) \cite{Pineda:1997ie,Pineda:1998kn}. In this EFT, the proton and electron interact via non-local potentials, and via the exchange of ultrasoft photons. In momentum space, the potential is given at leading order by the Coulomb potential,
\begin{equation}
    V_{\rm LO}(\vec q) =  - \frac{4 \pi \alpha}{\vec q^{\,2}}.
\end{equation}
In principle, the above potential should be evaluated using $\alpha$ in the Thomson limit, as the electron no longer contributes to the running of $\alpha$ within pNRQED. However, as we discuss below, the difference with the coupling in HBChPT, $\alpha(\mu_\chi)$, leads to very small effects numerically. Corrections to the potential are organized in powers of $\alpha$ and $\beta$. NLO ($\mathcal O(\alpha)$ and $\mathcal O(\beta)$) and N$^2$LO ($\mathcal O(\alpha^2)$, $\mathcal O( \alpha \beta)$, and $\mathcal O(\beta^2)$) corrections to the potential have been computed, both for systems of equal mass ($t\bar t$, quarkonium, and positronium), and for systems with different masses (such as the hydrogen atom, or the $B_c$ meson). The results can be found in Refs. \cite{Pineda:1997ie,Pineda:1998kn,Beneke:1999qg,Peset:2015vvi}. No corrections to the potential appear at NLO. At N$^2$LO, we can adapt the result of Refs. \cite{Pineda:1998kn,Peset:2015vvi} to the case of QED, $C_F = 1$, $C_A=0$, and $n_f = 0$. We find
\begin{equation}
    V_{\rm N^2LO} = + \frac{4 \pi \alpha}{m_e^2} \frac{c_D}{8},
\end{equation}
where $c_D$ is the Darwin term, and $c_D = 1$ at the order we are working. Finally, the interactions of ultrasoft photons with heavy quarks do not contribute up to N$^{3}$LO \cite{Beneke:1999qg,Pineda:1997ie}. This can for example be seen by the fact that one-loop real emission diagrams only contribute at $\mathcal O(\alpha \beta^2)$, three orders smaller than the leading order.

In the hypothetical case in which the electron emitted in neutron decay would be nonrelativistic, the decay rate could thus be expressed as
\begin{equation}
    \frac{\mathrm{d}\Gamma_n}{\mathrm{d}E_e} = \frac{G_F^2  \,  |V_{ud}|^2 }{(2\pi )^5}   \left(  1 + 3  \lambda^2 \right) \  p_e E_e (E_0 - E_e)^2  \left[ g_V (\mu_\chi) \right]^2 \ \left|C_{\rm NRQED} (\mu) \right|^2  \left|  \mathcal M_{\rm pNRQED}  \right|^2,
\end{equation}
with the pNRQED matrix element organized as
\begin{equation}
    \left|  \mathcal M_{\rm pNRQED}  \right|^2 = \sum_n \left(\frac{\alpha \pi}{\beta} \right)^n  \left\{1, \alpha, \beta, \alpha^2 , \alpha \beta, \beta^2, \ldots \right\}.
\end{equation}
From the previous discussion, the LO is provided by the iteration of the Coulomb potential, which leads to the nonrelativistic Fermi function
\begin{equation}
    \left| \mathcal M^{\rm LO}_{\rm pNRQED} \right|^2 = F_{NR}(\beta).
\end{equation}
Since there are no NLO potentials and the ultrasoft modes only contribute at N$^3$LO, $\mathcal M^{\rm LO}_{\rm pNRQED}$ does not receive any $\mathcal O(\alpha)$ corrections. $\mathcal O(\alpha)$ corrections are entirely contained in the matching coefficient $C_{\rm NRQED}$. Therefore, the leading $(\pi \alpha / \beta)^n$ and the subleading $\alpha/\pi (\pi \alpha / \beta)^n$ terms are captured by the factorized expression
\begin{equation}
    \frac{\mathrm{d}\Gamma_n}{\mathrm{d}E_e} = \frac{G_F^2  \,  |V_{ud}|^2 }{(2\pi )^5}   \left(  1 + 3  \lambda^2 \right) \  p_e E_e (E_0 - E_e)^2  \left[ g_V (\mu_\chi) \right]^2  \left|C_{\rm NRQED} (\mu)  \right|^2  F_{NR}(\beta).
\end{equation}
This result was proved in the case of $t \bar t$  production at threshold in Refs.~\cite{Czarnecki:1997vz,Hoang:1997sj,Hoang:1997ui,Beneke:1997jm}.

To make a   connection with the relativistic expressions, we now notice that
\begin{equation}
    (1 + \delta_{\rm RC}(E_e,\mu_\chi)) \xrightarrow{\beta \rightarrow 0} \left| C_{\rm NRQED}(\mu_\chi)\right|^2,
\end{equation}
where we neglected real emission diagrams which give a contribution that goes as $(E_0 - E_e)^2/E_e^2 \sim \mathcal O(\beta^4)$.

By expressing the decay rate as
\begin{equation}
    \frac{\mathrm{d}\Gamma_n}{\mathrm{d}E_e} = \frac{G_F^2  \,  |V_{ud}|^2 }{(2\pi )^5}   \left(  1 + 3  \lambda^2 \right) \  p_e E_e (E_0 - E_e)^2  \   \left[ g_V (\mu_\chi) \right]^2 \ F_{NR} (\beta) \bigg(  1  +   \delta_{\rm RC} (E_e, \mu_\chi) \bigg),
\end{equation}
our expression correctly reproduces the relativistic one-loop result, and, in addition, captures all subleading terms of $\alpha/\pi (\pi \alpha / \beta)^n$. Numerically, the various contributions to $\Delta_f$ (see Eq.~\eqref{eq:DeltaF1}) average to be as follows: $\left( \frac{\alpha^2}{\beta}, \alpha^2, \alpha^2 \ln \beta \right) = \left( 8, 5, 2 \right) \times 10^{-5}$. To obtain a diagrammatic expansion of the Fermi function $F_{NR}$ in the electromagnetic coupling constant at the level of decay rate, we evaluate it with $\chi$PT value $\alpha (\mu_\chi)$. We also test the ${\cal O}(\alpha^2)$ difference induced by computing $F_{NR}$ using $\alpha(\mu_\chi)$ versus $\alpha$ in the Thomson limit and obtain a result for $\Delta_f$ that is only $0.002\%$ higher. Therefore, theoretical improvements (i.e., knowledge of the terms $\mathcal{O}(\alpha^2 \ln \beta$)) would not induce much change in our results and uncertainty estimates.

\section{Details on the two-loop anomalous dimensions}
\label{sect:gammas}

As discussed in Sections\ \ref{sec:subsec31} and \ref{sec:RGE2}, we obtain the two-loop ${\cal O}( \alpha^2)$ anomalous dimensions, $\gamma_1$ and $\tilde \gamma_1$, in the LEFT and $\chi$PT by adapting calculations in the literature. For $\gamma_1$ in the LEFT, we use Refs.\ \cite{Buras:1989xd,Buras:1992zv} which consider the two-loop QCD anomalous dimension of a four-quark operator. As each diagram is given separately, their results can be modified to obtain the two-loop QED diagrams for the operator in Eq.~\eqref{eq:LLEFT}. This involves replacing the QCD couplings, multiplicities, and color factors with their QED counterparts. The same procedure also allows us to reproduce the ${\cal O}(\alpha\alpha_s)$ anomalous dimension $\gamma_{se}$. Although Refs.\ \cite{Buras:1989xd,Buras:1992zv} provide results using one particular scheme for the evanescent operators, $a=-1$, the full $a$ dependence can be recovered due to the fact that both the $1/\epsilon^2$ and $1/\epsilon$ coefficients of the diagrams are given, leading to the result in Eq.\ \eqref{eq:RGE}.

To obtain $\tilde \gamma_1$, relevant for the RGE in $\chi$PT, we use calculations for the heavy-light currents in heavy-quark effective theory,  \cite{Gimenez:1991bf}, see also Refs.~\cite{Ji:1991pr,Broadhurst:1991fy,Broadhurst:1991fz}. These calculations compute the two-loop anomalous dimension in QCD for a heavy-light current, $\bar Q \Gamma q$, where $Q$ denotes a nonrelativistic field, $q$ is a relativistic particle, and $\Gamma$ is an arbitrary Lorentz structure. As we argue below, the graphs in Ref.\ \cite{Gimenez:1991bf}  can be adapted to obtain the relevant diagrams for the  anomalous dimension of $g_V$, see Fig.\ \ref{fig:HBChPT_digarams}.

One way to see that the two cases are related to each other is by rewriting the weak operator in Eq.\ \eqref{eq:Lagrangian_at_leading_order} using Fierz identities,
\begin{equation}
(\bar p v^\mu n)(\bar e_L \gamma_\mu \nu_L) = \sum_{i,j}c_{ij}(\bar \nu_L^c \bar \Gamma_i n)(\bar p \bar \Gamma_j e^c)+\mathrm{E}\,,\label{eq:fierz}
\end{equation}
where $i,j$ run over the Dirac structures, $c_{ij}$ are coefficients determined by the Fierz relation, and $\mathrm{E}$ is an evanescent operator. The last bilinear on the right-hand side of Eq.\ \eqref{eq:fierz} now takes the same form as the heavy-light current, where the proton and the charge-conjugated electron, $e^c$,  play the role of the $Q$ and $q$ fields. This means the diagrams will take the same form as for the heavy-quark calculation, with the heavy-light vertex $\Gamma$ replaced by $\bar\Gamma_j$, while the neutral neutrino and neutron fields are irrelevant to the calculation.

The final ingredient to show a correspondence is the fact that the loop diagrams do not depend on the Dirac structure of the vertex \cite{Broadhurst:1991fz}. Since QED vertices, $\sim v^\mu$, and propagators, $\sim i/(v\cdot k)$, on the heavy-quark/proton line do not involve any Dirac structure, they cannot modify the original vertex. This is less obvious in the part of the diagrams involving the light-quark/electron line as it consists of a string of QED vertices, each of which comes with a propagator. However, one can show that, after performing the loop integrals, all gamma matrices will be contracted either with each other or with factors of $v^\mu$. This implies that the string of gamma matrices on the electron side also becomes proportional to the identity and leaves the original vertex unchanged. The independence of the gamma structure then allows us to adapt the results of Ref.\ \cite{Gimenez:1991bf} to obtain the diagrams with insertions of the right-hand side of Eq.\ \eqref{eq:fierz}.\footnote{Note that the independence of the gamma structure also means that the evanescent operator cannot contribute to matrix elements in $d=4$. In addition, one can use the same fact to directly show that insertions of the left-hand side of Eq.\ \eqref{eq:fierz} are related to the diagrams involving the heavy-light current, without the need for the Fierz rearrangement in Eq.\ \eqref{eq:fierz}.} The relevant replacements again involve replacing the QCD couplings, multiplicities, and color factors with their QED counterparts, where the appearing QED charges are now $Q_p$ and $Q_{e^+}=-Q_{e^-}$. The result of this procedure is given in Eq.\ \eqref{eq:RGE_LE}. Consequently, the anomalous dimensions can be obtained by a simple substitution $C_F =1,~C_A=0,~T_F=1$ in the QCD calculation of Ref.~\cite{Chetyrkin:2003vi}. This also allows us to obtain $\tilde\gamma_2$ as
\begin{equation}
    \tilde\gamma_2 = \frac{1}{64} \left[ \left( -80 \zeta_4 - 36 \zeta_3 + 64 \zeta_2 - \frac{37}{3}\right) + \tilde n \left(-\frac{176}{3}\zeta_3 + \frac{448}{9}\zeta_2 + \frac{470}{9}\right) + \frac{140}{27} \tilde n^2\right].
\end{equation}
Even though this expression contains terms enhanced by $\pi^4$, the numerical value of $\tilde\gamma_2$ is of natural size, $\tilde\gamma_2 \lesssim 2$ in $\slashpi$EFT, and its contribution is beyond the required accuracy for neutron $\beta$ decay.

\bibliography{beta_decay_EFT}{}

\begin{thebibliography}{143}%
\makeatletter
\providecommand \@ifxundefined [1]{%
 \@ifx{#1\undefined}
}%
\providecommand \@ifnum [1]{%
 \ifnum #1\expandafter \@firstoftwo
 \else \expandafter \@secondoftwo
 \fi
}%
\providecommand \@ifx [1]{%
 \ifx #1\expandafter \@firstoftwo
 \else \expandafter \@secondoftwo
 \fi
}%
\providecommand \natexlab [1]{#1}%
\providecommand \enquote  [1]{``#1''}%
\providecommand \bibnamefont  [1]{#1}%
\providecommand \bibfnamefont [1]{#1}%
\providecommand \citenamefont [1]{#1}%
\providecommand \href@noop [0]{\@secondoftwo}%
\providecommand \href [0]{\begingroup \@sanitize@url \@href}%
\providecommand \@href[1]{\@@startlink{#1}\@@href}%
\providecommand \@@href[1]{\endgroup#1\@@endlink}%
\providecommand \@sanitize@url [0]{\catcode `\\12\catcode `\$12\catcode
  `\&12\catcode `\#12\catcode `\^12\catcode `\_12\catcode `\%12\relax}%
\providecommand \@@startlink[1]{}%
\providecommand \@@endlink[0]{}%
\providecommand \url  [0]{\begingroup\@sanitize@url \@url }%
\providecommand \@url [1]{\endgroup\@href {#1}{\urlprefix }}%
\providecommand \urlprefix  [0]{URL }%
\providecommand \Eprint [0]{\href }%
\providecommand \doibase [0]{http://dx.doi.org/}%
\providecommand \selectlanguage [0]{\@gobble}%
\providecommand \bibinfo  [0]{\@secondoftwo}%
\providecommand \bibfield  [0]{\@secondoftwo}%
\providecommand \translation [1]{[#1]}%
\providecommand \BibitemOpen [0]{}%
\providecommand \bibitemStop [0]{}%
\providecommand \bibitemNoStop [0]{.\EOS\space}%
\providecommand \EOS [0]{\spacefactor3000\relax}%
\providecommand \BibitemShut  [1]{\csname bibitem#1\endcsname}%
\let\auto@bib@innerbib\@empty
\bibitem [{\citenamefont {Seng}\ \emph {et~al.}(2018)\citenamefont {Seng},
  \citenamefont {Gorchtein}, \citenamefont {Patel},\ and\ \citenamefont
  {Ramsey-Musolf}}]{Seng:2018yzq}%
  \BibitemOpen
  \bibfield  {author} {\bibinfo {author} {\bibfnamefont {C.-Y.}\ \bibnamefont
  {Seng}}, \bibinfo {author} {\bibfnamefont {M.}~\bibnamefont {Gorchtein}},
  \bibinfo {author} {\bibfnamefont {H.~H.}\ \bibnamefont {Patel}}, \ and\
  \bibinfo {author} {\bibfnamefont {M.~J.}\ \bibnamefont {Ramsey-Musolf}},\
  }\href {\doibase 10.1103/PhysRevLett.121.241804} {\bibfield  {journal}
  {\bibinfo  {journal} {Phys. Rev. Lett.}\ }\textbf {\bibinfo {volume} {121}},\
  \bibinfo {pages} {241804} (\bibinfo {year} {2018})},\ \Eprint
  {http://arxiv.org/abs/1807.10197}{arXiv:1807.10197 [hep-ph]}\BibitemShut
  {NoStop}%
\bibitem [{\citenamefont {Seng}\ \emph {et~al.}(2019)\citenamefont {Seng},
  \citenamefont {Gorchtein},\ and\ \citenamefont
  {Ramsey-Musolf}}]{Seng:2018qru}%
  \BibitemOpen
  \bibfield  {author} {\bibinfo {author} {\bibfnamefont {C.~Y.}\ \bibnamefont
  {Seng}}, \bibinfo {author} {\bibfnamefont {M.}~\bibnamefont {Gorchtein}}, \
  and\ \bibinfo {author} {\bibfnamefont {M.~J.}\ \bibnamefont
  {Ramsey-Musolf}},\ }\href {\doibase 10.1103/PhysRevD.100.013001} {\bibfield
  {journal} {\bibinfo  {journal} {Phys. Rev. D}\ }\textbf {\bibinfo {volume}
  {100}},\ \bibinfo {pages} {013001} (\bibinfo {year} {2019})},\ \Eprint
  {http://arxiv.org/abs/1812.03352}{arXiv:1812.03352 [nucl-th]}\BibitemShut
  {NoStop}%
\bibitem [{\citenamefont {Czarnecki}\ \emph {et~al.}(2019)\citenamefont
  {Czarnecki}, \citenamefont {Marciano},\ and\ \citenamefont
  {Sirlin}}]{Czarnecki:2019mwq}%
  \BibitemOpen
  \bibfield  {author} {\bibinfo {author} {\bibfnamefont {A.}~\bibnamefont
  {Czarnecki}}, \bibinfo {author} {\bibfnamefont {W.~J.}\ \bibnamefont
  {Marciano}}, \ and\ \bibinfo {author} {\bibfnamefont {A.}~\bibnamefont
  {Sirlin}},\ }\href {\doibase 10.1103/PhysRevD.100.073008} {\bibfield
  {journal} {\bibinfo  {journal} {Phys. Rev. D}\ }\textbf {\bibinfo {volume}
  {100}},\ \bibinfo {pages} {073008} (\bibinfo {year} {2019})},\ \Eprint
  {http://arxiv.org/abs/1907.06737}{arXiv:1907.06737 [hep-ph]}\BibitemShut
  {NoStop}%
\bibitem [{\citenamefont {Shiells}\ \emph {et~al.}(2021)\citenamefont
  {Shiells}, \citenamefont {Blunden},\ and\ \citenamefont
  {Melnitchouk}}]{Shiells:2020fqp}%
  \BibitemOpen
  \bibfield  {author} {\bibinfo {author} {\bibfnamefont {K.}~\bibnamefont
  {Shiells}}, \bibinfo {author} {\bibfnamefont {P.~G.}\ \bibnamefont
  {Blunden}}, \ and\ \bibinfo {author} {\bibfnamefont {W.}~\bibnamefont
  {Melnitchouk}},\ }\href {\doibase 10.1103/PhysRevD.104.033003} {\bibfield
  {journal} {\bibinfo  {journal} {Phys. Rev. D}\ }\textbf {\bibinfo {volume}
  {104}},\ \bibinfo {pages} {033003} (\bibinfo {year} {2021})},\ \Eprint
  {http://arxiv.org/abs/2012.01580}{arXiv:2012.01580 [hep-ph]}\BibitemShut
  {NoStop}%
\bibitem [{\citenamefont {Hayen}(2021)}]{Hayen:2020cxh}%
  \BibitemOpen
  \bibfield  {author} {\bibinfo {author} {\bibfnamefont {L.}~\bibnamefont
  {Hayen}},\ }\href {\doibase 10.1103/PhysRevD.103.113001} {\bibfield
  {journal} {\bibinfo  {journal} {Phys. Rev. D}\ }\textbf {\bibinfo {volume}
  {103}},\ \bibinfo {pages} {113001} (\bibinfo {year} {2021})},\ \Eprint
  {http://arxiv.org/abs/2010.07262}{arXiv:2010.07262 [hep-ph]}\BibitemShut
  {NoStop}%
\bibitem [{\citenamefont {Seng}\ \emph
  {et~al.}(2020{\natexlab{a}})\citenamefont {Seng}, \citenamefont {Feng},
  \citenamefont {Gorchtein},\ and\ \citenamefont {Jin}}]{Seng:2020wjq}%
  \BibitemOpen
  \bibfield  {author} {\bibinfo {author} {\bibfnamefont {C.-Y.}\ \bibnamefont
  {Seng}}, \bibinfo {author} {\bibfnamefont {X.}~\bibnamefont {Feng}}, \bibinfo
  {author} {\bibfnamefont {M.}~\bibnamefont {Gorchtein}}, \ and\ \bibinfo
  {author} {\bibfnamefont {L.-C.}\ \bibnamefont {Jin}},\ }\href {\doibase
  10.1103/PhysRevD.101.111301} {\bibfield  {journal} {\bibinfo  {journal}
  {Phys. Rev. D}\ }\textbf {\bibinfo {volume} {101}},\ \bibinfo {pages}
  {111301} (\bibinfo {year} {2020}{\natexlab{a}})},\ \Eprint
  {http://arxiv.org/abs/2003.11264}{arXiv:2003.11264 [hep-ph]}\BibitemShut
  {NoStop}%
\bibitem [{\citenamefont {Hardy}\ and\ \citenamefont
  {Towner}(2020)}]{Hardy:2020qwl}%
  \BibitemOpen
  \bibfield  {author} {\bibinfo {author} {\bibfnamefont {J.~C.}\ \bibnamefont
  {Hardy}}\ and\ \bibinfo {author} {\bibfnamefont {I.~S.}\ \bibnamefont
  {Towner}},\ }\href {\doibase 10.1103/PhysRevC.102.045501} {\bibfield
  {journal} {\bibinfo  {journal} {Phys. Rev. C}\ }\textbf {\bibinfo {volume}
  {102}},\ \bibinfo {pages} {045501} (\bibinfo {year} {2020})}\BibitemShut
  {NoStop}%
\bibitem [{\citenamefont {Cirigliano}\ \emph {et~al.}(2023)\citenamefont
  {Cirigliano}, \citenamefont {Crivellin}, \citenamefont {Hoferichter},\ and\
  \citenamefont {Moulson}}]{Cirigliano:2022yyo}%
  \BibitemOpen
  \bibfield  {author} {\bibinfo {author} {\bibfnamefont {V.}~\bibnamefont
  {Cirigliano}}, \bibinfo {author} {\bibfnamefont {A.}~\bibnamefont
  {Crivellin}}, \bibinfo {author} {\bibfnamefont {M.}~\bibnamefont
  {Hoferichter}}, \ and\ \bibinfo {author} {\bibfnamefont {M.}~\bibnamefont
  {Moulson}},\ }\href {\doibase 10.1016/j.physletb.2023.137748} {\bibfield
  {journal} {\bibinfo  {journal} {Phys. Lett. B}\ }\textbf {\bibinfo {volume}
  {838}},\ \bibinfo {pages} {137748} (\bibinfo {year} {2023})},\ \Eprint
  {http://arxiv.org/abs/2208.11707}{arXiv:2208.11707 [hep-ph]}\BibitemShut
  {NoStop}%
\bibitem [{\citenamefont {Zyla}\ \emph {et~al.}(2020)\citenamefont {Zyla} \emph
  {et~al.}}]{ParticleDataGroup:2020ssz}%
  \BibitemOpen
  \bibfield  {author} {\bibinfo {author} {\bibfnamefont {P.~A.}\ \bibnamefont
  {Zyla}} \emph {et~al.} (\bibinfo {collaboration} {Particle Data Group}),\
  }\href {\doibase 10.1093/ptep/ptaa104} {\bibfield  {journal} {\bibinfo
  {journal} {PTEP}\ }\textbf {\bibinfo {volume} {2020}},\ \bibinfo {pages}
  {083C01} (\bibinfo {year} {2020})}\BibitemShut {NoStop}%
\bibitem [{\citenamefont {Falkowski}\ \emph {et~al.}(2020)\citenamefont
  {Falkowski}, \citenamefont {Gonz\'alez-Alonso},\ and\ \citenamefont
  {Naviliat-Cuncic}}]{Falkowski:2020pma}%
  \BibitemOpen
  \bibfield  {author} {\bibinfo {author} {\bibfnamefont {A.}~\bibnamefont
  {Falkowski}}, \bibinfo {author} {\bibfnamefont {M.}~\bibnamefont
  {Gonz\'alez-Alonso}}, \ and\ \bibinfo {author} {\bibfnamefont
  {O.}~\bibnamefont {Naviliat-Cuncic}},\ }\href@noop {} {\  (\bibinfo {year}
  {2020})},\ \Eprint {http://arxiv.org/abs/2010.13797}{arXiv:2010.13797
  [hep-ph]}\BibitemShut {NoStop}%
\bibitem [{\citenamefont {Gonz\'alez-Alonso}\ \emph {et~al.}(2019)\citenamefont
  {Gonz\'alez-Alonso}, \citenamefont {Naviliat-Cuncic},\ and\ \citenamefont
  {Severijns}}]{Gonzalez-Alonso:2018omy}%
  \BibitemOpen
  \bibfield  {author} {\bibinfo {author} {\bibfnamefont {M.}~\bibnamefont
  {Gonz\'alez-Alonso}}, \bibinfo {author} {\bibfnamefont {O.}~\bibnamefont
  {Naviliat-Cuncic}}, \ and\ \bibinfo {author} {\bibfnamefont {N.}~\bibnamefont
  {Severijns}},\ }\href {\doibase 10.1016/j.ppnp.2018.08.002} {\bibfield
  {journal} {\bibinfo  {journal} {Prog. Part. Nucl. Phys.}\ }\textbf {\bibinfo
  {volume} {104}},\ \bibinfo {pages} {165} (\bibinfo {year} {2019})},\ \Eprint
  {http://arxiv.org/abs/1803.08732}{arXiv:1803.08732 [hep-ph]}\BibitemShut
  {NoStop}%
\bibitem [{\citenamefont {Nunokawa}\ \emph {et~al.}(2008)\citenamefont
  {Nunokawa}, \citenamefont {Parke},\ and\ \citenamefont
  {Valle}}]{Nunokawa:2007qh}%
  \BibitemOpen
  \bibfield  {author} {\bibinfo {author} {\bibfnamefont {H.}~\bibnamefont
  {Nunokawa}}, \bibinfo {author} {\bibfnamefont {S.~J.}\ \bibnamefont {Parke}},
  \ and\ \bibinfo {author} {\bibfnamefont {J.~W.~F.}\ \bibnamefont {Valle}},\
  }\href {\doibase 10.1016/j.ppnp.2007.10.001} {\bibfield  {journal} {\bibinfo
  {journal} {Prog. Part. Nucl. Phys.}\ }\textbf {\bibinfo {volume} {60}},\
  \bibinfo {pages} {338} (\bibinfo {year} {2008})},\ \Eprint
  {http://arxiv.org/abs/0710.0554}{arXiv:0710.0554 [hep-ph]}\BibitemShut
  {NoStop}%
\bibitem [{\citenamefont {Ayres}\ \emph {et~al.}(2007)\citenamefont {Ayres}
  \emph {et~al.}}]{NOvA:2007rmc}%
  \BibitemOpen
  \bibfield  {author} {\bibinfo {author} {\bibfnamefont {D.~S.}\ \bibnamefont
  {Ayres}} \emph {et~al.} (\bibinfo {collaboration} {NOvA}),\ }\href {\doibase
  10.2172/935497} {\  (\bibinfo {year} {2007}),\ 10.2172/935497}\BibitemShut
  {NoStop}%
\bibitem [{\citenamefont {Abe}\ \emph {et~al.}(2011)\citenamefont {Abe} \emph
  {et~al.}}]{T2K:2011qtm}%
  \BibitemOpen
  \bibfield  {author} {\bibinfo {author} {\bibfnamefont {K.}~\bibnamefont
  {Abe}} \emph {et~al.} (\bibinfo {collaboration} {T2K}),\ }\href {\doibase
  10.1016/j.nima.2011.06.067} {\bibfield  {journal} {\bibinfo  {journal} {Nucl.
  Instrum. Meth. A}\ }\textbf {\bibinfo {volume} {659}},\ \bibinfo {pages}
  {106} (\bibinfo {year} {2011})},\ \Eprint
  {http://arxiv.org/abs/1106.1238}{arXiv:1106.1238
  [physics.ins-det]}\BibitemShut {NoStop}%
\bibitem [{\citenamefont {Gando}\ \emph {et~al.}(2013)\citenamefont {Gando}
  \emph {et~al.}}]{KamLAND:2013rgu}%
  \BibitemOpen
  \bibfield  {author} {\bibinfo {author} {\bibfnamefont {A.}~\bibnamefont
  {Gando}} \emph {et~al.} (\bibinfo {collaboration} {KamLAND}),\ }\href
  {\doibase 10.1103/PhysRevD.88.033001} {\bibfield  {journal} {\bibinfo
  {journal} {Phys. Rev. D}\ }\textbf {\bibinfo {volume} {88}},\ \bibinfo
  {pages} {033001} (\bibinfo {year} {2013})},\ \Eprint
  {http://arxiv.org/abs/1303.4667}{arXiv:1303.4667 [hep-ex]}\BibitemShut
  {NoStop}%
\bibitem [{\citenamefont {An}\ \emph {et~al.}(2016)\citenamefont {An} \emph
  {et~al.}}]{JUNO:2015zny}%
  \BibitemOpen
  \bibfield  {author} {\bibinfo {author} {\bibfnamefont {F.}~\bibnamefont {An}}
  \emph {et~al.} (\bibinfo {collaboration} {JUNO}),\ }\href {\doibase
  10.1088/0954-3899/43/3/030401} {\bibfield  {journal} {\bibinfo  {journal} {J.
  Phys. G}\ }\textbf {\bibinfo {volume} {43}},\ \bibinfo {pages} {030401}
  (\bibinfo {year} {2016})},\ \Eprint
  {http://arxiv.org/abs/1507.05613}{arXiv:1507.05613
  [physics.ins-det]}\BibitemShut {NoStop}%
\bibitem [{\citenamefont {Abe}\ \emph {et~al.}(2015)\citenamefont {Abe} \emph
  {et~al.}}]{Hyper-KamiokandeProto-:2015xww}%
  \BibitemOpen
  \bibfield  {author} {\bibinfo {author} {\bibfnamefont {K.}~\bibnamefont
  {Abe}} \emph {et~al.} (\bibinfo {collaboration} {Hyper-Kamiokande Proto-}),\
  }\href {\doibase 10.1093/ptep/ptv061} {\bibfield  {journal} {\bibinfo
  {journal} {PTEP}\ }\textbf {\bibinfo {volume} {2015}},\ \bibinfo {pages}
  {053C02} (\bibinfo {year} {2015})},\ \Eprint
  {http://arxiv.org/abs/1502.05199}{arXiv:1502.05199 [hep-ex]}\BibitemShut
  {NoStop}%
\bibitem [{\citenamefont {Bak}\ \emph {et~al.}(2018)\citenamefont {Bak} \emph
  {et~al.}}]{RENO:2018dro}%
  \BibitemOpen
  \bibfield  {author} {\bibinfo {author} {\bibfnamefont {G.}~\bibnamefont
  {Bak}} \emph {et~al.} (\bibinfo {collaboration} {RENO}),\ }\href {\doibase
  10.1103/PhysRevLett.121.201801} {\bibfield  {journal} {\bibinfo  {journal}
  {Phys. Rev. Lett.}\ }\textbf {\bibinfo {volume} {121}},\ \bibinfo {pages}
  {201801} (\bibinfo {year} {2018})},\ \Eprint
  {http://arxiv.org/abs/1806.00248}{arXiv:1806.00248 [hep-ex]}\BibitemShut
  {NoStop}%
\bibitem [{\citenamefont {Adey}\ \emph {et~al.}(2018)\citenamefont {Adey} \emph
  {et~al.}}]{DayaBay:2018yms}%
  \BibitemOpen
  \bibfield  {author} {\bibinfo {author} {\bibfnamefont {D.}~\bibnamefont
  {Adey}} \emph {et~al.} (\bibinfo {collaboration} {Daya Bay}),\ }\href
  {\doibase 10.1103/PhysRevLett.121.241805} {\bibfield  {journal} {\bibinfo
  {journal} {Phys. Rev. Lett.}\ }\textbf {\bibinfo {volume} {121}},\ \bibinfo
  {pages} {241805} (\bibinfo {year} {2018})},\ \Eprint
  {http://arxiv.org/abs/1809.02261}{arXiv:1809.02261 [hep-ex]}\BibitemShut
  {NoStop}%
\bibitem [{\citenamefont {de~Kerret}\ \emph {et~al.}(2020)\citenamefont
  {de~Kerret} \emph {et~al.}}]{DoubleChooz:2019qbj}%
  \BibitemOpen
  \bibfield  {author} {\bibinfo {author} {\bibfnamefont {H.}~\bibnamefont
  {de~Kerret}} \emph {et~al.} (\bibinfo {collaboration} {Double Chooz}),\
  }\href {\doibase 10.1038/s41567-020-0831-y} {\bibfield  {journal} {\bibinfo
  {journal} {Nature Phys.}\ }\textbf {\bibinfo {volume} {16}},\ \bibinfo
  {pages} {558} (\bibinfo {year} {2020})},\ \Eprint
  {http://arxiv.org/abs/1901.09445}{arXiv:1901.09445 [hep-ex]}\BibitemShut
  {NoStop}%
\bibitem [{\citenamefont {Abe}\ \emph {et~al.}(2020)\citenamefont {Abe} \emph
  {et~al.}}]{T2K:2019bcf}%
  \BibitemOpen
  \bibfield  {author} {\bibinfo {author} {\bibfnamefont {K.}~\bibnamefont
  {Abe}} \emph {et~al.} (\bibinfo {collaboration} {T2K}),\ }\href {\doibase
  10.1038/s41586-020-2177-0} {\bibfield  {journal} {\bibinfo  {journal}
  {Nature}\ }\textbf {\bibinfo {volume} {580}},\ \bibinfo {pages} {339}
  (\bibinfo {year} {2020})},\ \bibinfo {note} {[Erratum: Nature 583, E16
  (2020)]},\ \Eprint {http://arxiv.org/abs/1910.03887}{arXiv:1910.03887
  [hep-ex]}\BibitemShut {NoStop}%
\bibitem [{\citenamefont {Acero}\ \emph {et~al.}(2019)\citenamefont {Acero}
  \emph {et~al.}}]{NOvA:2019cyt}%
  \BibitemOpen
  \bibfield  {author} {\bibinfo {author} {\bibfnamefont {M.~A.}\ \bibnamefont
  {Acero}} \emph {et~al.} (\bibinfo {collaboration} {NOvA}),\ }\href {\doibase
  10.1103/PhysRevLett.123.151803} {\bibfield  {journal} {\bibinfo  {journal}
  {Phys. Rev. Lett.}\ }\textbf {\bibinfo {volume} {123}},\ \bibinfo {pages}
  {151803} (\bibinfo {year} {2019})},\ \Eprint
  {http://arxiv.org/abs/1906.04907}{arXiv:1906.04907 [hep-ex]}\BibitemShut
  {NoStop}%
\bibitem [{\citenamefont {Abi}\ \emph {et~al.}(2020)\citenamefont {Abi} \emph
  {et~al.}}]{DUNE:2020ypp}%
  \BibitemOpen
  \bibfield  {author} {\bibinfo {author} {\bibfnamefont {B.}~\bibnamefont
  {Abi}} \emph {et~al.} (\bibinfo {collaboration} {DUNE}),\ }\href@noop {} {\
  (\bibinfo {year} {2020})},\ \Eprint
  {http://arxiv.org/abs/2002.03005}{arXiv:2002.03005 [hep-ex]}\BibitemShut
  {NoStop}%
\bibitem [{\citenamefont {Abusleme}\ \emph {et~al.}(2022)\citenamefont
  {Abusleme} \emph {et~al.}}]{JUNO:2021vlw}%
  \BibitemOpen
  \bibfield  {author} {\bibinfo {author} {\bibfnamefont {A.}~\bibnamefont
  {Abusleme}} \emph {et~al.} (\bibinfo {collaboration} {JUNO}),\ }\href
  {\doibase 10.1016/j.ppnp.2021.103927} {\bibfield  {journal} {\bibinfo
  {journal} {Prog. Part. Nucl. Phys.}\ }\textbf {\bibinfo {volume} {123}},\
  \bibinfo {pages} {103927} (\bibinfo {year} {2022})},\ \Eprint
  {http://arxiv.org/abs/2104.02565}{arXiv:2104.02565 [hep-ex]}\BibitemShut
  {NoStop}%
\bibitem [{\citenamefont {Mueller}\ \emph {et~al.}(2011)\citenamefont {Mueller}
  \emph {et~al.}}]{Mueller:2011nm}%
  \BibitemOpen
  \bibfield  {author} {\bibinfo {author} {\bibfnamefont {T.~A.}\ \bibnamefont
  {Mueller}} \emph {et~al.},\ }\href {\doibase 10.1103/PhysRevC.83.054615}
  {\bibfield  {journal} {\bibinfo  {journal} {Phys. Rev. C}\ }\textbf {\bibinfo
  {volume} {83}},\ \bibinfo {pages} {054615} (\bibinfo {year} {2011})},\
  \Eprint {http://arxiv.org/abs/1101.2663}{arXiv:1101.2663
  [hep-ex]}\BibitemShut {NoStop}%
\bibitem [{\citenamefont {Huber}(2011)}]{Huber:2011wv}%
  \BibitemOpen
  \bibfield  {author} {\bibinfo {author} {\bibfnamefont {P.}~\bibnamefont
  {Huber}},\ }\href {\doibase 10.1103/PhysRevC.85.029901} {\bibfield  {journal}
  {\bibinfo  {journal} {Phys. Rev. C}\ }\textbf {\bibinfo {volume} {84}},\
  \bibinfo {pages} {024617} (\bibinfo {year} {2011})},\ \bibinfo {note}
  {[Erratum: Phys.Rev.C 85, 029901 (2012)]},\ \Eprint
  {http://arxiv.org/abs/1106.0687}{arXiv:1106.0687 [hep-ph]}\BibitemShut
  {NoStop}%
\bibitem [{\citenamefont {Aliaga}\ \emph {et~al.}(2016)\citenamefont {Aliaga}
  \emph {et~al.}}]{MINERvA:2016iqn}%
  \BibitemOpen
  \bibfield  {author} {\bibinfo {author} {\bibfnamefont {L.}~\bibnamefont
  {Aliaga}} \emph {et~al.} (\bibinfo {collaboration} {MINERvA}),\ }\href
  {\doibase 10.1103/PhysRevD.94.092005} {\bibfield  {journal} {\bibinfo
  {journal} {Phys. Rev. D}\ }\textbf {\bibinfo {volume} {94}},\ \bibinfo
  {pages} {092005} (\bibinfo {year} {2016})},\ \bibinfo {note} {[Addendum:
  Phys.Rev.D 95, 039903 (2017)]},\ \Eprint
  {http://arxiv.org/abs/1607.00704}{arXiv:1607.00704 [hep-ex]}\BibitemShut
  {NoStop}%
\bibitem [{\citenamefont {Alvarez-Ruso}\ \emph {et~al.}(2018)\citenamefont
  {Alvarez-Ruso} \emph {et~al.}}]{NuSTEC:2017hzk}%
  \BibitemOpen
  \bibfield  {author} {\bibinfo {author} {\bibfnamefont {L.}~\bibnamefont
  {Alvarez-Ruso}} \emph {et~al.} (\bibinfo {collaboration} {NuSTEC}),\ }\href
  {\doibase 10.1016/j.ppnp.2018.01.006} {\bibfield  {journal} {\bibinfo
  {journal} {Prog. Part. Nucl. Phys.}\ }\textbf {\bibinfo {volume} {100}},\
  \bibinfo {pages} {1} (\bibinfo {year} {2018})},\ \Eprint
  {http://arxiv.org/abs/1706.03621}{arXiv:1706.03621 [hep-ph]}\BibitemShut
  {NoStop}%
\bibitem [{\citenamefont {Abusleme}\ \emph {et~al.}(2020)\citenamefont
  {Abusleme} \emph {et~al.}}]{JUNO:2020ijm}%
  \BibitemOpen
  \bibfield  {author} {\bibinfo {author} {\bibfnamefont {A.}~\bibnamefont
  {Abusleme}} \emph {et~al.} (\bibinfo {collaboration} {JUNO}),\ }\href@noop {}
  {\  (\bibinfo {year} {2020})},\ \Eprint
  {http://arxiv.org/abs/2005.08745}{arXiv:2005.08745
  [physics.ins-det]}\BibitemShut {NoStop}%
\bibitem [{\citenamefont {Khachatryan}\ \emph {et~al.}(2021)\citenamefont
  {Khachatryan} \emph {et~al.}}]{CLAS:2021neh}%
  \BibitemOpen
  \bibfield  {author} {\bibinfo {author} {\bibfnamefont {M.}~\bibnamefont
  {Khachatryan}} \emph {et~al.} (\bibinfo {collaboration} {CLAS, e4v}),\ }\href
  {\doibase 10.1038/s41586-021-04046-5} {\bibfield  {journal} {\bibinfo
  {journal} {Nature}\ }\textbf {\bibinfo {volume} {599}},\ \bibinfo {pages}
  {565} (\bibinfo {year} {2021})}\BibitemShut {NoStop}%
\bibitem [{\citenamefont {Behrends}\ \emph {et~al.}(1956)\citenamefont
  {Behrends}, \citenamefont {Finkelstein},\ and\ \citenamefont
  {Sirlin}}]{Behrends:1955mb}%
  \BibitemOpen
  \bibfield  {author} {\bibinfo {author} {\bibfnamefont {R.~E.}\ \bibnamefont
  {Behrends}}, \bibinfo {author} {\bibfnamefont {R.~J.}\ \bibnamefont
  {Finkelstein}}, \ and\ \bibinfo {author} {\bibfnamefont {A.}~\bibnamefont
  {Sirlin}},\ }\href {\doibase 10.1103/PhysRev.101.866} {\bibfield  {journal}
  {\bibinfo  {journal} {Phys. Rev.}\ }\textbf {\bibinfo {volume} {101}},\
  \bibinfo {pages} {866} (\bibinfo {year} {1956})}\BibitemShut {NoStop}%
\bibitem [{\citenamefont {Kinoshita}\ and\ \citenamefont
  {Sirlin}(1959)}]{Kinoshita:1958ru}%
  \BibitemOpen
  \bibfield  {author} {\bibinfo {author} {\bibfnamefont {T.}~\bibnamefont
  {Kinoshita}}\ and\ \bibinfo {author} {\bibfnamefont {A.}~\bibnamefont
  {Sirlin}},\ }\href {\doibase 10.1103/PhysRev.113.1652} {\bibfield  {journal}
  {\bibinfo  {journal} {Phys. Rev.}\ }\textbf {\bibinfo {volume} {113}},\
  \bibinfo {pages} {1652} (\bibinfo {year} {1959})}\BibitemShut {NoStop}%
\bibitem [{\citenamefont {Sirlin}(1967)}]{Sirlin:1967zza}%
  \BibitemOpen
  \bibfield  {author} {\bibinfo {author} {\bibfnamefont {A.}~\bibnamefont
  {Sirlin}},\ }\href {\doibase 10.1103/PhysRev.164.1767} {\bibfield  {journal}
  {\bibinfo  {journal} {Phys. Rev.}\ }\textbf {\bibinfo {volume} {164}},\
  \bibinfo {pages} {1767} (\bibinfo {year} {1967})}\BibitemShut {NoStop}%
\bibitem [{\citenamefont {Abers}\ \emph {et~al.}(1968)\citenamefont {Abers},
  \citenamefont {Dicus}, \citenamefont {Norton},\ and\ \citenamefont
  {Quinn}}]{Abers:1968zz}%
  \BibitemOpen
  \bibfield  {author} {\bibinfo {author} {\bibfnamefont {E.~S.}\ \bibnamefont
  {Abers}}, \bibinfo {author} {\bibfnamefont {D.~A.}\ \bibnamefont {Dicus}},
  \bibinfo {author} {\bibfnamefont {R.~E.}\ \bibnamefont {Norton}}, \ and\
  \bibinfo {author} {\bibfnamefont {H.~R.}\ \bibnamefont {Quinn}},\ }\href
  {\doibase 10.1103/PhysRev.167.1461} {\bibfield  {journal} {\bibinfo
  {journal} {Phys. Rev.}\ }\textbf {\bibinfo {volume} {167}},\ \bibinfo {pages}
  {1461} (\bibinfo {year} {1968})}\BibitemShut {NoStop}%
\bibitem [{\citenamefont {Sirlin}(1978)}]{Sirlin:1977sv}%
  \BibitemOpen
  \bibfield  {author} {\bibinfo {author} {\bibfnamefont {A.}~\bibnamefont
  {Sirlin}},\ }\href {\doibase 10.1103/RevModPhys.50.573} {\bibfield  {journal}
  {\bibinfo  {journal} {Rev. Mod. Phys.}\ }\textbf {\bibinfo {volume} {50}},\
  \bibinfo {pages} {573} (\bibinfo {year} {1978})},\ \bibinfo {note} {[Erratum:
  Rev.Mod.Phys. 50, 905 (1978)]}\BibitemShut {NoStop}%
\bibitem [{\citenamefont {Sirlin}(1982)}]{Sirlin:1981ie}%
  \BibitemOpen
  \bibfield  {author} {\bibinfo {author} {\bibfnamefont {A.}~\bibnamefont
  {Sirlin}},\ }\href {\doibase 10.1016/0550-3213(82)90303-0} {\bibfield
  {journal} {\bibinfo  {journal} {Nucl. Phys. B}\ }\textbf {\bibinfo {volume}
  {196}},\ \bibinfo {pages} {83} (\bibinfo {year} {1982})}\BibitemShut
  {NoStop}%
\bibitem [{\citenamefont {Marciano}\ and\ \citenamefont
  {Sirlin}(2006)}]{Marciano:2005ec}%
  \BibitemOpen
  \bibfield  {author} {\bibinfo {author} {\bibfnamefont {W.~J.}\ \bibnamefont
  {Marciano}}\ and\ \bibinfo {author} {\bibfnamefont {A.}~\bibnamefont
  {Sirlin}},\ }\href {\doibase 10.1103/PhysRevLett.96.032002} {\bibfield
  {journal} {\bibinfo  {journal} {Phys. Rev. Lett.}\ }\textbf {\bibinfo
  {volume} {96}},\ \bibinfo {pages} {032002} (\bibinfo {year} {2006})},\
  \Eprint
  {http://arxiv.org/abs/hep-ph/0510099}{arXiv:hep-ph/0510099}\BibitemShut
  {NoStop}%
\bibitem [{\citenamefont {Czarnecki}\ \emph {et~al.}(2004)\citenamefont
  {Czarnecki}, \citenamefont {Marciano},\ and\ \citenamefont
  {Sirlin}}]{Czarnecki:2004cw}%
  \BibitemOpen
  \bibfield  {author} {\bibinfo {author} {\bibfnamefont {A.}~\bibnamefont
  {Czarnecki}}, \bibinfo {author} {\bibfnamefont {W.~J.}\ \bibnamefont
  {Marciano}}, \ and\ \bibinfo {author} {\bibfnamefont {A.}~\bibnamefont
  {Sirlin}},\ }\href {\doibase 10.1103/PhysRevD.70.093006} {\bibfield
  {journal} {\bibinfo  {journal} {Phys. Rev. D}\ }\textbf {\bibinfo {volume}
  {70}},\ \bibinfo {pages} {093006} (\bibinfo {year} {2004})},\ \Eprint
  {http://arxiv.org/abs/hep-ph/0406324}{arXiv:hep-ph/0406324}\BibitemShut
  {NoStop}%
\bibitem [{\citenamefont {Seng}\ \emph
  {et~al.}(2020{\natexlab{b}})\citenamefont {Seng}, \citenamefont {Feng},
  \citenamefont {Gorchtein}, \citenamefont {Jin},\ and\ \citenamefont
  {Mei\ss{}ner}}]{Seng:2020jtz}%
  \BibitemOpen
  \bibfield  {author} {\bibinfo {author} {\bibfnamefont {C.-Y.}\ \bibnamefont
  {Seng}}, \bibinfo {author} {\bibfnamefont {X.}~\bibnamefont {Feng}}, \bibinfo
  {author} {\bibfnamefont {M.}~\bibnamefont {Gorchtein}}, \bibinfo {author}
  {\bibfnamefont {L.-C.}\ \bibnamefont {Jin}}, \ and\ \bibinfo {author}
  {\bibfnamefont {U.-G.}\ \bibnamefont {Mei\ss{}ner}},\ }\href {\doibase
  10.1007/JHEP10(2020)179} {\bibfield  {journal} {\bibinfo  {journal} {JHEP}\
  }\textbf {\bibinfo {volume} {10}},\ \bibinfo {pages} {179} (\bibinfo {year}
  {2020}{\natexlab{b}})},\ \Eprint
  {http://arxiv.org/abs/2009.00459}{arXiv:2009.00459 [hep-lat]}\BibitemShut
  {NoStop}%
\bibitem [{\citenamefont {Feng}\ \emph {et~al.}(2020)\citenamefont {Feng},
  \citenamefont {Gorchtein}, \citenamefont {Jin}, \citenamefont {Ma},\ and\
  \citenamefont {Seng}}]{Feng:2020zdc}%
  \BibitemOpen
  \bibfield  {author} {\bibinfo {author} {\bibfnamefont {X.}~\bibnamefont
  {Feng}}, \bibinfo {author} {\bibfnamefont {M.}~\bibnamefont {Gorchtein}},
  \bibinfo {author} {\bibfnamefont {L.-C.}\ \bibnamefont {Jin}}, \bibinfo
  {author} {\bibfnamefont {P.-X.}\ \bibnamefont {Ma}}, \ and\ \bibinfo {author}
  {\bibfnamefont {C.-Y.}\ \bibnamefont {Seng}},\ }\href {\doibase
  10.1103/PhysRevLett.124.192002} {\bibfield  {journal} {\bibinfo  {journal}
  {Phys. Rev. Lett.}\ }\textbf {\bibinfo {volume} {124}},\ \bibinfo {pages}
  {192002} (\bibinfo {year} {2020})},\ \Eprint
  {http://arxiv.org/abs/2003.09798}{arXiv:2003.09798 [hep-lat]}\BibitemShut
  {NoStop}%
\bibitem [{\citenamefont {Ma}\ \emph {et~al.}(2021)\citenamefont {Ma},
  \citenamefont {Feng}, \citenamefont {Gorchtein}, \citenamefont {Jin},\ and\
  \citenamefont {Seng}}]{Ma:2021azh}%
  \BibitemOpen
  \bibfield  {author} {\bibinfo {author} {\bibfnamefont {P.-X.}\ \bibnamefont
  {Ma}}, \bibinfo {author} {\bibfnamefont {X.}~\bibnamefont {Feng}}, \bibinfo
  {author} {\bibfnamefont {M.}~\bibnamefont {Gorchtein}}, \bibinfo {author}
  {\bibfnamefont {L.-C.}\ \bibnamefont {Jin}}, \ and\ \bibinfo {author}
  {\bibfnamefont {C.-Y.}\ \bibnamefont {Seng}},\ }\href {\doibase
  10.1103/PhysRevD.103.114503} {\bibfield  {journal} {\bibinfo  {journal}
  {Phys. Rev. D}\ }\textbf {\bibinfo {volume} {103}},\ \bibinfo {pages}
  {114503} (\bibinfo {year} {2021})},\ \Eprint
  {http://arxiv.org/abs/2102.12048}{arXiv:2102.12048 [hep-lat]}\BibitemShut
  {NoStop}%
\bibitem [{\citenamefont {Yoo}\ \emph {et~al.}(2022)\citenamefont {Yoo},
  \citenamefont {Bhattacharya}, \citenamefont {Gupta}, \citenamefont {Mondal},\
  and\ \citenamefont {Yoon}}]{Yoo:2022lmt}%
  \BibitemOpen
  \bibfield  {author} {\bibinfo {author} {\bibfnamefont {J.-S.}\ \bibnamefont
  {Yoo}}, \bibinfo {author} {\bibfnamefont {T.}~\bibnamefont {Bhattacharya}},
  \bibinfo {author} {\bibfnamefont {R.}~\bibnamefont {Gupta}}, \bibinfo
  {author} {\bibfnamefont {S.}~\bibnamefont {Mondal}}, \ and\ \bibinfo {author}
  {\bibfnamefont {B.}~\bibnamefont {Yoon}},\ }\href@noop {} {\  (\bibinfo
  {year} {2022})},\ \Eprint {http://arxiv.org/abs/2212.12830}{arXiv:2212.12830
  [hep-lat]}\BibitemShut {NoStop}%
\bibitem [{\citenamefont {Yoo}\ \emph {et~al.}(2023)\citenamefont {Yoo},
  \citenamefont {Bhattacharya}, \citenamefont {Gupta}, \citenamefont {Mondal},\
  and\ \citenamefont {Yoon}}]{Yoo:2023gln}%
  \BibitemOpen
  \bibfield  {author} {\bibinfo {author} {\bibfnamefont {J.-S.}\ \bibnamefont
  {Yoo}}, \bibinfo {author} {\bibfnamefont {T.}~\bibnamefont {Bhattacharya}},
  \bibinfo {author} {\bibfnamefont {R.}~\bibnamefont {Gupta}}, \bibinfo
  {author} {\bibfnamefont {S.}~\bibnamefont {Mondal}}, \ and\ \bibinfo {author}
  {\bibfnamefont {B.}~\bibnamefont {Yoon}},\ }\href@noop {} {\  (\bibinfo
  {year} {2023})},\ \Eprint {http://arxiv.org/abs/2305.03198}{arXiv:2305.03198
  [hep-lat]}\BibitemShut {NoStop}%
\bibitem [{\citenamefont {Cirigliano}\ \emph {et~al.}(2022)\citenamefont
  {Cirigliano}, \citenamefont {de~Vries}, \citenamefont {Hayen}, \citenamefont
  {Mereghetti},\ and\ \citenamefont {Walker-Loud}}]{Cirigliano:2022hob}%
  \BibitemOpen
  \bibfield  {author} {\bibinfo {author} {\bibfnamefont {V.}~\bibnamefont
  {Cirigliano}}, \bibinfo {author} {\bibfnamefont {J.}~\bibnamefont
  {de~Vries}}, \bibinfo {author} {\bibfnamefont {L.}~\bibnamefont {Hayen}},
  \bibinfo {author} {\bibfnamefont {E.}~\bibnamefont {Mereghetti}}, \ and\
  \bibinfo {author} {\bibfnamefont {A.}~\bibnamefont {Walker-Loud}},\ }\href
  {\doibase 10.1103/PhysRevLett.129.121801} {\bibfield  {journal} {\bibinfo
  {journal} {Phys. Rev. Lett.}\ }\textbf {\bibinfo {volume} {129}},\ \bibinfo
  {pages} {121801} (\bibinfo {year} {2022})},\ \Eprint
  {http://arxiv.org/abs/2202.10439}{arXiv:2202.10439 [nucl-th]}\BibitemShut
  {NoStop}%
\bibitem [{\citenamefont {Gorchtein}\ and\ \citenamefont
  {Seng}(2021)}]{Gorchtein:2021fce}%
  \BibitemOpen
  \bibfield  {author} {\bibinfo {author} {\bibfnamefont {M.}~\bibnamefont
  {Gorchtein}}\ and\ \bibinfo {author} {\bibfnamefont {C.-Y.}\ \bibnamefont
  {Seng}},\ }\href {\doibase 10.1007/JHEP10(2021)053} {\bibfield  {journal}
  {\bibinfo  {journal} {JHEP}\ }\textbf {\bibinfo {volume} {10}},\ \bibinfo
  {pages} {053} (\bibinfo {year} {2021})},\ \Eprint
  {http://arxiv.org/abs/2106.09185}{arXiv:2106.09185 [hep-ph]}\BibitemShut
  {NoStop}%
\bibitem [{\citenamefont {Ando}\ \emph {et~al.}(2004)\citenamefont {Ando},
  \citenamefont {Fearing}, \citenamefont {Gudkov}, \citenamefont {Kubodera},
  \citenamefont {Myhrer}, \citenamefont {Nakamura},\ and\ \citenamefont
  {Sato}}]{Ando:2004rk}%
  \BibitemOpen
  \bibfield  {author} {\bibinfo {author} {\bibfnamefont {S.}~\bibnamefont
  {Ando}}, \bibinfo {author} {\bibfnamefont {H.~W.}\ \bibnamefont {Fearing}},
  \bibinfo {author} {\bibfnamefont {V.~P.}\ \bibnamefont {Gudkov}}, \bibinfo
  {author} {\bibfnamefont {K.}~\bibnamefont {Kubodera}}, \bibinfo {author}
  {\bibfnamefont {F.}~\bibnamefont {Myhrer}}, \bibinfo {author} {\bibfnamefont
  {S.}~\bibnamefont {Nakamura}}, \ and\ \bibinfo {author} {\bibfnamefont
  {T.}~\bibnamefont {Sato}},\ }\href {\doibase 10.1016/j.physletb.2004.06.037}
  {\bibfield  {journal} {\bibinfo  {journal} {Phys. Lett. B}\ }\textbf
  {\bibinfo {volume} {595}},\ \bibinfo {pages} {250} (\bibinfo {year}
  {2004})},\ \Eprint
  {http://arxiv.org/abs/nucl-th/0402100}{arXiv:nucl-th/0402100}\BibitemShut
  {NoStop}%
\bibitem [{\citenamefont {Descotes-Genon}\ and\ \citenamefont
  {Moussallam}(2005)}]{Descotes-Genon:2005wrq}%
  \BibitemOpen
  \bibfield  {author} {\bibinfo {author} {\bibfnamefont {S.}~\bibnamefont
  {Descotes-Genon}}\ and\ \bibinfo {author} {\bibfnamefont {B.}~\bibnamefont
  {Moussallam}},\ }\href {\doibase 10.1140/epjc/s2005-02316-8} {\bibfield
  {journal} {\bibinfo  {journal} {Eur. Phys. J. C}\ }\textbf {\bibinfo {volume}
  {42}},\ \bibinfo {pages} {403} (\bibinfo {year} {2005})},\ \Eprint
  {http://arxiv.org/abs/hep-ph/0505077}{arXiv:hep-ph/0505077}\BibitemShut
  {NoStop}%
\bibitem [{\citenamefont {Bardeen}\ \emph {et~al.}(1978)\citenamefont
  {Bardeen}, \citenamefont {Buras}, \citenamefont {Duke},\ and\ \citenamefont
  {Muta}}]{Bardeen:1978yd}%
  \BibitemOpen
  \bibfield  {author} {\bibinfo {author} {\bibfnamefont {W.~A.}\ \bibnamefont
  {Bardeen}}, \bibinfo {author} {\bibfnamefont {A.~J.}\ \bibnamefont {Buras}},
  \bibinfo {author} {\bibfnamefont {D.~W.}\ \bibnamefont {Duke}}, \ and\
  \bibinfo {author} {\bibfnamefont {T.}~\bibnamefont {Muta}},\ }\href {\doibase
  10.1103/PhysRevD.18.3998} {\bibfield  {journal} {\bibinfo  {journal} {Phys.
  Rev. D}\ }\textbf {\bibinfo {volume} {18}},\ \bibinfo {pages} {3998}
  (\bibinfo {year} {1978})}\BibitemShut {NoStop}%
\bibitem [{\citenamefont {Gasser}\ and\ \citenamefont
  {Leutwyler}(1984)}]{Gasser:1983yg}%
  \BibitemOpen
  \bibfield  {author} {\bibinfo {author} {\bibfnamefont {J.}~\bibnamefont
  {Gasser}}\ and\ \bibinfo {author} {\bibfnamefont {H.}~\bibnamefont
  {Leutwyler}},\ }\href {\doibase 10.1016/0003-4916(84)90242-2} {\bibfield
  {journal} {\bibinfo  {journal} {Annals Phys.}\ }\textbf {\bibinfo {volume}
  {158}},\ \bibinfo {pages} {142} (\bibinfo {year} {1984})}\BibitemShut
  {NoStop}%
\bibitem [{\citenamefont {Jenkins}\ and\ \citenamefont
  {Manohar}(1991)}]{Jenkins:1990jv}%
  \BibitemOpen
  \bibfield  {author} {\bibinfo {author} {\bibfnamefont {E.~E.}\ \bibnamefont
  {Jenkins}}\ and\ \bibinfo {author} {\bibfnamefont {A.~V.}\ \bibnamefont
  {Manohar}},\ }\href {\doibase 10.1016/0370-2693(91)90266-S} {\bibfield
  {journal} {\bibinfo  {journal} {Phys. Lett. B}\ }\textbf {\bibinfo {volume}
  {255}},\ \bibinfo {pages} {558} (\bibinfo {year} {1991})}\BibitemShut
  {NoStop}%
\bibitem [{\citenamefont {Raha}\ \emph {et~al.}(2012)\citenamefont {Raha},
  \citenamefont {Myhrer},\ and\ \citenamefont {Kubodera}}]{Raha:2011aa}%
  \BibitemOpen
  \bibfield  {author} {\bibinfo {author} {\bibfnamefont {U.}~\bibnamefont
  {Raha}}, \bibinfo {author} {\bibfnamefont {F.}~\bibnamefont {Myhrer}}, \ and\
  \bibinfo {author} {\bibfnamefont {K.}~\bibnamefont {Kubodera}},\ }\href
  {\doibase 10.1103/PhysRevC.85.045502} {\bibfield  {journal} {\bibinfo
  {journal} {Phys. Rev. C}\ }\textbf {\bibinfo {volume} {85}},\ \bibinfo
  {pages} {045502} (\bibinfo {year} {2012})},\ \bibinfo {note} {[Erratum:
  Phys.Rev.C 86, 039903 (2012)]},\ \Eprint
  {http://arxiv.org/abs/1112.2007}{arXiv:1112.2007 [hep-ph]}\BibitemShut
  {NoStop}%
\bibitem [{\citenamefont {Falkowski}\ \emph {et~al.}(2021)\citenamefont
  {Falkowski}, \citenamefont {Gonz\'alez-Alonso}, \citenamefont {Palavri\'c},\
  and\ \citenamefont {Rodr\'\i{}guez-S\'anchez}}]{Falkowski:2021vdg}%
  \BibitemOpen
  \bibfield  {author} {\bibinfo {author} {\bibfnamefont {A.}~\bibnamefont
  {Falkowski}}, \bibinfo {author} {\bibfnamefont {M.}~\bibnamefont
  {Gonz\'alez-Alonso}}, \bibinfo {author} {\bibfnamefont {A.}~\bibnamefont
  {Palavri\'c}}, \ and\ \bibinfo {author} {\bibfnamefont {A.}~\bibnamefont
  {Rodr\'\i{}guez-S\'anchez}},\ }\href@noop {} {\  (\bibinfo {year} {2021})},\
  \Eprint {http://arxiv.org/abs/2112.07688}{arXiv:2112.07688
  [hep-ph]}\BibitemShut {NoStop}%
\bibitem [{\citenamefont {Wilkinson}(1982)}]{Wilkinson:1982hu}%
  \BibitemOpen
  \bibfield  {author} {\bibinfo {author} {\bibfnamefont {D.~H.}\ \bibnamefont
  {Wilkinson}},\ }\href {\doibase 10.1016/0375-9474(82)90051-3} {\bibfield
  {journal} {\bibinfo  {journal} {Nucl. Phys. A}\ }\textbf {\bibinfo {volume}
  {377}},\ \bibinfo {pages} {474} (\bibinfo {year} {1982})}\BibitemShut
  {NoStop}%
\bibitem [{\citenamefont {Sirlin}\ and\ \citenamefont
  {Zucchini}(1986)}]{Sirlin:1986cc}%
  \BibitemOpen
  \bibfield  {author} {\bibinfo {author} {\bibfnamefont {A.}~\bibnamefont
  {Sirlin}}\ and\ \bibinfo {author} {\bibfnamefont {R.}~\bibnamefont
  {Zucchini}},\ }\href {\doibase 10.1103/PhysRevLett.57.1994} {\bibfield
  {journal} {\bibinfo  {journal} {Phys. Rev. Lett.}\ }\textbf {\bibinfo
  {volume} {57}},\ \bibinfo {pages} {1994} (\bibinfo {year}
  {1986})}\BibitemShut {NoStop}%
\bibitem [{\citenamefont {Jaus}\ and\ \citenamefont
  {Rasche}(1970)}]{Jaus:1970tah}%
  \BibitemOpen
  \bibfield  {author} {\bibinfo {author} {\bibfnamefont {W.}~\bibnamefont
  {Jaus}}\ and\ \bibinfo {author} {\bibfnamefont {G.}~\bibnamefont {Rasche}},\
  }\href {\doibase 10.1016/0375-9474(70)90690-1} {\bibfield  {journal}
  {\bibinfo  {journal} {Nucl. Phys. A}\ }\textbf {\bibinfo {volume} {143}},\
  \bibinfo {pages} {202} (\bibinfo {year} {1970})}\BibitemShut {NoStop}%
\bibitem [{\citenamefont {Workman}\ \emph {et~al.}(2022)\citenamefont {Workman}
  \emph {et~al.}}]{ParticleDataGroup:2022pth}%
  \BibitemOpen
  \bibfield  {author} {\bibinfo {author} {\bibfnamefont {R.~L.}\ \bibnamefont
  {Workman}} \emph {et~al.} (\bibinfo {collaboration} {Particle Data Group}),\
  }\href {\doibase 10.1093/ptep/ptac097} {\bibfield  {journal} {\bibinfo
  {journal} {PTEP}\ }\textbf {\bibinfo {volume} {2022}},\ \bibinfo {pages}
  {083C01} (\bibinfo {year} {2022})}\BibitemShut {NoStop}%
\bibitem [{\citenamefont {Workman}\ and\ \citenamefont
  {Others}(2022)}]{Workman:2022ynf}%
  \BibitemOpen
  \bibfield  {author} {\bibinfo {author} {\bibfnamefont {R.~L.}\ \bibnamefont
  {Workman}}\ and\ \bibinfo {author} {\bibnamefont {Others}} (\bibinfo
  {collaboration} {Particle Data Group}),\ }\href {\doibase
  10.1093/ptep/ptac097} {\bibfield  {journal} {\bibinfo  {journal} {PTEP}\
  }\textbf {\bibinfo {volume} {2022}},\ \bibinfo {pages} {083C01} (\bibinfo
  {year} {2022})}\BibitemShut {NoStop}%
\bibitem [{\citenamefont {Gonzalez}\ \emph {et~al.}(2021)\citenamefont
  {Gonzalez} \emph {et~al.}}]{UCNt:2021pcg}%
  \BibitemOpen
  \bibfield  {author} {\bibinfo {author} {\bibfnamefont {F.~M.}\ \bibnamefont
  {Gonzalez}} \emph {et~al.} (\bibinfo {collaboration}
  {UCN\ensuremath{\tau}}),\ }\href {\doibase 10.1103/PhysRevLett.127.162501}
  {\bibfield  {journal} {\bibinfo  {journal} {Phys. Rev. Lett.}\ }\textbf
  {\bibinfo {volume} {127}},\ \bibinfo {pages} {162501} (\bibinfo {year}
  {2021})},\ \Eprint {http://arxiv.org/abs/2106.10375}{arXiv:2106.10375
  [nucl-ex]}\BibitemShut {NoStop}%
\bibitem [{\citenamefont {M\"arkisch}\ \emph {et~al.}(2019)\citenamefont
  {M\"arkisch} \emph {et~al.}}]{Markisch:2018ndu}%
  \BibitemOpen
  \bibfield  {author} {\bibinfo {author} {\bibfnamefont {B.}~\bibnamefont
  {M\"arkisch}} \emph {et~al.},\ }\href {\doibase
  10.1103/PhysRevLett.122.242501} {\bibfield  {journal} {\bibinfo  {journal}
  {Phys. Rev. Lett.}\ }\textbf {\bibinfo {volume} {122}},\ \bibinfo {pages}
  {242501} (\bibinfo {year} {2019})},\ \Eprint
  {http://arxiv.org/abs/1812.04666}{arXiv:1812.04666 [nucl-ex]}\BibitemShut
  {NoStop}%
\bibitem [{\citenamefont {Dubbers}\ \emph {et~al.}(2019)\citenamefont
  {Dubbers}, \citenamefont {Saul}, \citenamefont {M\"arkisch}, \citenamefont
  {Soldner},\ and\ \citenamefont {Abele}}]{Dubbers:2018kgh}%
  \BibitemOpen
  \bibfield  {author} {\bibinfo {author} {\bibfnamefont {D.}~\bibnamefont
  {Dubbers}}, \bibinfo {author} {\bibfnamefont {H.}~\bibnamefont {Saul}},
  \bibinfo {author} {\bibfnamefont {B.}~\bibnamefont {M\"arkisch}}, \bibinfo
  {author} {\bibfnamefont {T.}~\bibnamefont {Soldner}}, \ and\ \bibinfo
  {author} {\bibfnamefont {H.}~\bibnamefont {Abele}},\ }\href {\doibase
  10.1016/j.physletb.2019.02.013} {\bibfield  {journal} {\bibinfo  {journal}
  {Phys. Lett. B}\ }\textbf {\bibinfo {volume} {791}},\ \bibinfo {pages} {6}
  (\bibinfo {year} {2019})},\ \Eprint
  {http://arxiv.org/abs/1812.00626}{arXiv:1812.00626 [nucl-ex]}\BibitemShut
  {NoStop}%
\bibitem [{\citenamefont {Knecht}\ \emph {et~al.}(2000)\citenamefont {Knecht},
  \citenamefont {Neufeld}, \citenamefont {Rupertsberger},\ and\ \citenamefont
  {Talavera}}]{Knecht:1999ag}%
  \BibitemOpen
  \bibfield  {author} {\bibinfo {author} {\bibfnamefont {M.}~\bibnamefont
  {Knecht}}, \bibinfo {author} {\bibfnamefont {H.}~\bibnamefont {Neufeld}},
  \bibinfo {author} {\bibfnamefont {H.}~\bibnamefont {Rupertsberger}}, \ and\
  \bibinfo {author} {\bibfnamefont {P.}~\bibnamefont {Talavera}},\ }\href
  {\doibase 10.1007/s100529900265} {\bibfield  {journal} {\bibinfo  {journal}
  {Eur. Phys. J. C}\ }\textbf {\bibinfo {volume} {12}},\ \bibinfo {pages} {469}
  (\bibinfo {year} {2000})},\ \Eprint
  {http://arxiv.org/abs/hep-ph/9909284}{arXiv:hep-ph/9909284}\BibitemShut
  {NoStop}%
\bibitem [{\citenamefont {van Ritbergen}\ and\ \citenamefont
  {Stuart}(2000)}]{vanRitbergen:1999fi}%
  \BibitemOpen
  \bibfield  {author} {\bibinfo {author} {\bibfnamefont {T.}~\bibnamefont {van
  Ritbergen}}\ and\ \bibinfo {author} {\bibfnamefont {R.~G.}\ \bibnamefont
  {Stuart}},\ }\href {\doibase 10.1016/S0550-3213(99)00572-6} {\bibfield
  {journal} {\bibinfo  {journal} {Nucl. Phys. B}\ }\textbf {\bibinfo {volume}
  {564}},\ \bibinfo {pages} {343} (\bibinfo {year} {2000})},\ \Eprint
  {http://arxiv.org/abs/hep-ph/9904240}{arXiv:hep-ph/9904240}\BibitemShut
  {NoStop}%
\bibitem [{\citenamefont {Barczyk}\ \emph {et~al.}(2008)\citenamefont {Barczyk}
  \emph {et~al.}}]{FAST:2007rsc}%
  \BibitemOpen
  \bibfield  {author} {\bibinfo {author} {\bibfnamefont {A.}~\bibnamefont
  {Barczyk}} \emph {et~al.} (\bibinfo {collaboration} {FAST}),\ }\href
  {\doibase 10.1016/j.physletb.2008.04.006} {\bibfield  {journal} {\bibinfo
  {journal} {Phys. Lett. B}\ }\textbf {\bibinfo {volume} {663}},\ \bibinfo
  {pages} {172} (\bibinfo {year} {2008})},\ \Eprint
  {http://arxiv.org/abs/0707.3904}{arXiv:0707.3904 [hep-ex]}\BibitemShut
  {NoStop}%
\bibitem [{\citenamefont {Casella}\ \emph {et~al.}(2013)\citenamefont {Casella}
  \emph {et~al.}}]{Casella:2013bla}%
  \BibitemOpen
  \bibfield  {author} {\bibinfo {author} {\bibfnamefont {C.}~\bibnamefont
  {Casella}} \emph {et~al.},\ }\href {\doibase 10.1016/j.nima.2012.10.032}
  {\bibfield  {journal} {\bibinfo  {journal} {Nucl. Instrum. Meth. A}\ }\textbf
  {\bibinfo {volume} {700}},\ \bibinfo {pages} {1} (\bibinfo {year}
  {2013})}\BibitemShut {NoStop}%
\bibitem [{\citenamefont {Tishchenko}\ \emph {et~al.}(2013)\citenamefont
  {Tishchenko} \emph {et~al.}}]{MuLan:2012sih}%
  \BibitemOpen
  \bibfield  {author} {\bibinfo {author} {\bibfnamefont {V.}~\bibnamefont
  {Tishchenko}} \emph {et~al.} (\bibinfo {collaboration} {MuLan}),\ }\href
  {\doibase 10.1103/PhysRevD.87.052003} {\bibfield  {journal} {\bibinfo
  {journal} {Phys. Rev. D}\ }\textbf {\bibinfo {volume} {87}},\ \bibinfo
  {pages} {052003} (\bibinfo {year} {2013})},\ \Eprint
  {http://arxiv.org/abs/1211.0960}{arXiv:1211.0960 [hep-ex]}\BibitemShut
  {NoStop}%
\bibitem [{\citenamefont {Hill}\ and\ \citenamefont
  {Tomalak}(2020)}]{Hill:2019xqk}%
  \BibitemOpen
  \bibfield  {author} {\bibinfo {author} {\bibfnamefont {R.~J.}\ \bibnamefont
  {Hill}}\ and\ \bibinfo {author} {\bibfnamefont {O.}~\bibnamefont {Tomalak}},\
  }\href {\doibase 10.1016/j.physletb.2020.135466} {\bibfield  {journal}
  {\bibinfo  {journal} {Phys. Lett. B}\ }\textbf {\bibinfo {volume} {805}},\
  \bibinfo {pages} {135466} (\bibinfo {year} {2020})},\ \Eprint
  {http://arxiv.org/abs/1911.01493}{arXiv:1911.01493 [hep-ph]}\BibitemShut
  {NoStop}%
\bibitem [{\citenamefont {Dekens}\ and\ \citenamefont
  {Stoffer}(2019)}]{Dekens:2019ept}%
  \BibitemOpen
  \bibfield  {author} {\bibinfo {author} {\bibfnamefont {W.}~\bibnamefont
  {Dekens}}\ and\ \bibinfo {author} {\bibfnamefont {P.}~\bibnamefont
  {Stoffer}},\ }\href {\doibase 10.1007/JHEP10(2019)197} {\bibfield  {journal}
  {\bibinfo  {journal} {JHEP}\ }\textbf {\bibinfo {volume} {10}},\ \bibinfo
  {pages} {197} (\bibinfo {year} {2019})},\ \bibinfo {note} {[Erratum: JHEP 11,
  148 (2022)]},\ \Eprint {http://arxiv.org/abs/1908.05295}{arXiv:1908.05295
  [hep-ph]}\BibitemShut {NoStop}%
\bibitem [{\citenamefont {Buras}\ and\ \citenamefont
  {Weisz}(1990)}]{Buras:1989xd}%
  \BibitemOpen
  \bibfield  {author} {\bibinfo {author} {\bibfnamefont {A.~J.}\ \bibnamefont
  {Buras}}\ and\ \bibinfo {author} {\bibfnamefont {P.~H.}\ \bibnamefont
  {Weisz}},\ }\href {\doibase 10.1016/0550-3213(90)90223-Z} {\bibfield
  {journal} {\bibinfo  {journal} {Nucl. Phys. B}\ }\textbf {\bibinfo {volume}
  {333}},\ \bibinfo {pages} {66} (\bibinfo {year} {1990})}\BibitemShut
  {NoStop}%
\bibitem [{\citenamefont {Dugan}\ and\ \citenamefont
  {Grinstein}(1991)}]{Dugan:1990df}%
  \BibitemOpen
  \bibfield  {author} {\bibinfo {author} {\bibfnamefont {M.~J.}\ \bibnamefont
  {Dugan}}\ and\ \bibinfo {author} {\bibfnamefont {B.}~\bibnamefont
  {Grinstein}},\ }\href {\doibase 10.1016/0370-2693(91)90680-O} {\bibfield
  {journal} {\bibinfo  {journal} {Phys. Lett. B}\ }\textbf {\bibinfo {volume}
  {256}},\ \bibinfo {pages} {239} (\bibinfo {year} {1991})}\BibitemShut
  {NoStop}%
\bibitem [{\citenamefont {Herrlich}\ and\ \citenamefont
  {Nierste}(1995)}]{Herrlich:1994kh}%
  \BibitemOpen
  \bibfield  {author} {\bibinfo {author} {\bibfnamefont {S.}~\bibnamefont
  {Herrlich}}\ and\ \bibinfo {author} {\bibfnamefont {U.}~\bibnamefont
  {Nierste}},\ }\href {\doibase 10.1016/0550-3213(95)00474-7} {\bibfield
  {journal} {\bibinfo  {journal} {Nucl. Phys. B}\ }\textbf {\bibinfo {volume}
  {455}},\ \bibinfo {pages} {39} (\bibinfo {year} {1995})},\ \Eprint
  {http://arxiv.org/abs/hep-ph/9412375}{arXiv:hep-ph/9412375}\BibitemShut
  {NoStop}%
\bibitem [{\citenamefont {Erler}(2004)}]{Erler:2002mv}%
  \BibitemOpen
  \bibfield  {author} {\bibinfo {author} {\bibfnamefont {J.}~\bibnamefont
  {Erler}},\ }\href@noop {} {\bibfield  {journal} {\bibinfo  {journal} {Rev.
  Mex. Fis.}\ }\textbf {\bibinfo {volume} {50}},\ \bibinfo {pages} {200}
  (\bibinfo {year} {2004})},\ \Eprint
  {http://arxiv.org/abs/hep-ph/0211345}{arXiv:hep-ph/0211345}\BibitemShut
  {NoStop}%
\bibitem [{\citenamefont {Buras}\ \emph {et~al.}(1993)\citenamefont {Buras},
  \citenamefont {Jamin},\ and\ \citenamefont {Lautenbacher}}]{Buras:1992zv}%
  \BibitemOpen
  \bibfield  {author} {\bibinfo {author} {\bibfnamefont {A.~J.}\ \bibnamefont
  {Buras}}, \bibinfo {author} {\bibfnamefont {M.}~\bibnamefont {Jamin}}, \ and\
  \bibinfo {author} {\bibfnamefont {M.~E.}\ \bibnamefont {Lautenbacher}},\
  }\href {\doibase 10.1016/0550-3213(93)90398-9} {\bibfield  {journal}
  {\bibinfo  {journal} {Nucl. Phys. B}\ }\textbf {\bibinfo {volume} {400}},\
  \bibinfo {pages} {75} (\bibinfo {year} {1993})},\ \Eprint
  {http://arxiv.org/abs/hep-ph/9211321}{arXiv:hep-ph/9211321}\BibitemShut
  {NoStop}%
\bibitem [{\citenamefont {Buras}\ \emph {et~al.}(1990)\citenamefont {Buras},
  \citenamefont {Jamin},\ and\ \citenamefont {Weisz}}]{Buras:1990fn}%
  \BibitemOpen
  \bibfield  {author} {\bibinfo {author} {\bibfnamefont {A.~J.}\ \bibnamefont
  {Buras}}, \bibinfo {author} {\bibfnamefont {M.}~\bibnamefont {Jamin}}, \ and\
  \bibinfo {author} {\bibfnamefont {P.~H.}\ \bibnamefont {Weisz}},\ }\href
  {\doibase 10.1016/0550-3213(90)90373-L} {\bibfield  {journal} {\bibinfo
  {journal} {Nucl. Phys. B}\ }\textbf {\bibinfo {volume} {347}},\ \bibinfo
  {pages} {491} (\bibinfo {year} {1990})}\BibitemShut {NoStop}%
\bibitem [{\citenamefont {Buchalla}\ \emph {et~al.}(1996)\citenamefont
  {Buchalla}, \citenamefont {Buras},\ and\ \citenamefont
  {Lautenbacher}}]{Buchalla:1995vs}%
  \BibitemOpen
  \bibfield  {author} {\bibinfo {author} {\bibfnamefont {G.}~\bibnamefont
  {Buchalla}}, \bibinfo {author} {\bibfnamefont {A.~J.}\ \bibnamefont {Buras}},
  \ and\ \bibinfo {author} {\bibfnamefont {M.~E.}\ \bibnamefont
  {Lautenbacher}},\ }\href {\doibase 10.1103/RevModPhys.68.1125} {\bibfield
  {journal} {\bibinfo  {journal} {Rev. Mod. Phys.}\ }\textbf {\bibinfo {volume}
  {68}},\ \bibinfo {pages} {1125} (\bibinfo {year} {1996})},\ \Eprint
  {http://arxiv.org/abs/hep-ph/9512380}{arXiv:hep-ph/9512380}\BibitemShut
  {NoStop}%
\bibitem [{\citenamefont {Neubert}(2000)}]{Neubert:2000ag}%
  \BibitemOpen
  \bibfield  {author} {\bibinfo {author} {\bibfnamefont {M.}~\bibnamefont
  {Neubert}},\ }in\ \href {\doibase 10.1142/9789812799982_0002} {\emph
  {\bibinfo {booktitle} {{Theoretical Advanced Study Institute in Elementary
  Particle Physics (TASI 2000): Flavor Physics for the Millennium}}}}\
  (\bibinfo {year} {2000})\ pp.\ \bibinfo {pages} {483--535},\ \Eprint
  {http://arxiv.org/abs/hep-ph/0012204}{arXiv:hep-ph/0012204}\BibitemShut
  {NoStop}%
\bibitem [{\citenamefont {Urech}(1995)}]{Urech:1994hd}%
  \BibitemOpen
  \bibfield  {author} {\bibinfo {author} {\bibfnamefont {R.}~\bibnamefont
  {Urech}},\ }\href {\doibase 10.1016/0550-3213(95)90707-N} {\bibfield
  {journal} {\bibinfo  {journal} {Nucl. Phys. B}\ }\textbf {\bibinfo {volume}
  {433}},\ \bibinfo {pages} {234} (\bibinfo {year} {1995})},\ \Eprint
  {http://arxiv.org/abs/hep-ph/9405341}{arXiv:hep-ph/9405341}\BibitemShut
  {NoStop}%
\bibitem [{\citenamefont {Moussallam}(1997)}]{Moussallam:1997xx}%
  \BibitemOpen
  \bibfield  {author} {\bibinfo {author} {\bibfnamefont {B.}~\bibnamefont
  {Moussallam}},\ }\href {\doibase 10.1016/S0550-3213(97)00464-1} {\bibfield
  {journal} {\bibinfo  {journal} {Nucl. Phys. B}\ }\textbf {\bibinfo {volume}
  {504}},\ \bibinfo {pages} {381} (\bibinfo {year} {1997})},\ \Eprint
  {http://arxiv.org/abs/hep-ph/9701400}{arXiv:hep-ph/9701400}\BibitemShut
  {NoStop}%
\bibitem [{\citenamefont {Meissner}\ and\ \citenamefont
  {Steininger}(1998)}]{Meissner:1997ii}%
  \BibitemOpen
  \bibfield  {author} {\bibinfo {author} {\bibfnamefont {U.~G.}\ \bibnamefont
  {Meissner}}\ and\ \bibinfo {author} {\bibfnamefont {S.}~\bibnamefont
  {Steininger}},\ }\href {\doibase 10.1016/S0370-2693(97)01418-4} {\bibfield
  {journal} {\bibinfo  {journal} {Phys. Lett. B}\ }\textbf {\bibinfo {volume}
  {419}},\ \bibinfo {pages} {403} (\bibinfo {year} {1998})},\ \Eprint
  {http://arxiv.org/abs/hep-ph/9709453}{arXiv:hep-ph/9709453}\BibitemShut
  {NoStop}%
\bibitem [{\citenamefont {Muller}\ and\ \citenamefont
  {Meissner}(1999)}]{Muller:1999ww}%
  \BibitemOpen
  \bibfield  {author} {\bibinfo {author} {\bibfnamefont {G.}~\bibnamefont
  {Muller}}\ and\ \bibinfo {author} {\bibfnamefont {U.-G.}\ \bibnamefont
  {Meissner}},\ }\href {\doibase 10.1016/S0550-3213(99)00339-9} {\bibfield
  {journal} {\bibinfo  {journal} {Nucl. Phys. B}\ }\textbf {\bibinfo {volume}
  {556}},\ \bibinfo {pages} {265} (\bibinfo {year} {1999})},\ \Eprint
  {http://arxiv.org/abs/hep-ph/9903375}{arXiv:hep-ph/9903375}\BibitemShut
  {NoStop}%
\bibitem [{\citenamefont {Gasser}\ \emph {et~al.}(2002)\citenamefont {Gasser},
  \citenamefont {Ivanov}, \citenamefont {Lipartia}, \citenamefont {Mojzis},\
  and\ \citenamefont {Rusetsky}}]{Gasser:2002am}%
  \BibitemOpen
  \bibfield  {author} {\bibinfo {author} {\bibfnamefont {J.}~\bibnamefont
  {Gasser}}, \bibinfo {author} {\bibfnamefont {M.~A.}\ \bibnamefont {Ivanov}},
  \bibinfo {author} {\bibfnamefont {E.}~\bibnamefont {Lipartia}}, \bibinfo
  {author} {\bibfnamefont {M.}~\bibnamefont {Mojzis}}, \ and\ \bibinfo {author}
  {\bibfnamefont {A.}~\bibnamefont {Rusetsky}},\ }\href {\doibase
  10.1007/s10052-002-1013-z} {\bibfield  {journal} {\bibinfo  {journal} {Eur.
  Phys. J. C}\ }\textbf {\bibinfo {volume} {26}},\ \bibinfo {pages} {13}
  (\bibinfo {year} {2002})},\ \Eprint
  {http://arxiv.org/abs/hep-ph/0206068}{arXiv:hep-ph/0206068}\BibitemShut
  {NoStop}%
\bibitem [{\citenamefont {Tomalak}(2023)}]{Tomalak:2023xgm}%
  \BibitemOpen
  \bibfield  {author} {\bibinfo {author} {\bibfnamefont {O.}~\bibnamefont
  {Tomalak}},\ }\href {\doibase 10.1007/s00601-023-01802-3} {\bibfield
  {journal} {\bibinfo  {journal} {Few Body Syst.}\ }\textbf {\bibinfo {volume}
  {64}},\ \bibinfo {pages} {23} (\bibinfo {year} {2023})},\ \Eprint
  {http://arxiv.org/abs/2302.00642}{arXiv:2302.00642 [hep-ph]}\BibitemShut
  {NoStop}%
\bibitem [{\citenamefont {Bernard}\ \emph {et~al.}(1995)\citenamefont
  {Bernard}, \citenamefont {Kaiser},\ and\ \citenamefont
  {Meissner}}]{Bernard:1995dp}%
  \BibitemOpen
  \bibfield  {author} {\bibinfo {author} {\bibfnamefont {V.}~\bibnamefont
  {Bernard}}, \bibinfo {author} {\bibfnamefont {N.}~\bibnamefont {Kaiser}}, \
  and\ \bibinfo {author} {\bibfnamefont {U.-G.}\ \bibnamefont {Meissner}},\
  }\href {\doibase 10.1142/S0218301395000092} {\bibfield  {journal} {\bibinfo
  {journal} {Int. J. Mod. Phys. E}\ }\textbf {\bibinfo {volume} {4}},\ \bibinfo
  {pages} {193} (\bibinfo {year} {1995})},\ \Eprint
  {http://arxiv.org/abs/hep-ph/9501384}{arXiv:hep-ph/9501384}\BibitemShut
  {NoStop}%
\bibitem [{\citenamefont {Adler}\ and\ \citenamefont
  {Tung}(1969)}]{Adler:1969ei}%
  \BibitemOpen
  \bibfield  {author} {\bibinfo {author} {\bibfnamefont {S.~L.}\ \bibnamefont
  {Adler}}\ and\ \bibinfo {author} {\bibfnamefont {W.-K.}\ \bibnamefont
  {Tung}},\ }\href {\doibase 10.1103/PhysRevLett.22.978} {\bibfield  {journal}
  {\bibinfo  {journal} {Phys. Rev. Lett.}\ }\textbf {\bibinfo {volume} {22}},\
  \bibinfo {pages} {978} (\bibinfo {year} {1969})}\BibitemShut {NoStop}%
\bibitem [{\citenamefont {Adler}\ and\ \citenamefont
  {Tung}(1970)}]{Adler:1970hu}%
  \BibitemOpen
  \bibfield  {author} {\bibinfo {author} {\bibfnamefont {S.~L.}\ \bibnamefont
  {Adler}}\ and\ \bibinfo {author} {\bibfnamefont {W.-K.}\ \bibnamefont
  {Tung}},\ }\href {\doibase 10.1103/PhysRevD.1.2846} {\bibfield  {journal}
  {\bibinfo  {journal} {Phys. Rev. D}\ }\textbf {\bibinfo {volume} {1}},\
  \bibinfo {pages} {2846} (\bibinfo {year} {1970})},\ \bibinfo {note}
  {[Erratum: Phys.Rev.D 2, 2514--2514 (1970)]}\BibitemShut {NoStop}%
\bibitem [{\citenamefont {Ji}(1993)}]{Ji:1993ey}%
  \BibitemOpen
  \bibfield  {author} {\bibinfo {author} {\bibfnamefont {X.-D.}\ \bibnamefont
  {Ji}},\ }\href {\doibase 10.1016/0550-3213(93)90642-3} {\bibfield  {journal}
  {\bibinfo  {journal} {Nucl. Phys. B}\ }\textbf {\bibinfo {volume} {402}},\
  \bibinfo {pages} {217} (\bibinfo {year} {1993})}\BibitemShut {NoStop}%
\bibitem [{\citenamefont {Blumlein}\ and\ \citenamefont
  {Kochelev}(1996)}]{Blumlein:1996tp}%
  \BibitemOpen
  \bibfield  {author} {\bibinfo {author} {\bibfnamefont {J.}~\bibnamefont
  {Blumlein}}\ and\ \bibinfo {author} {\bibfnamefont {N.}~\bibnamefont
  {Kochelev}},\ }\href {\doibase 10.1016/0370-2693(96)00583-7} {\bibfield
  {journal} {\bibinfo  {journal} {Phys. Lett. B}\ }\textbf {\bibinfo {volume}
  {381}},\ \bibinfo {pages} {296} (\bibinfo {year} {1996})},\ \Eprint
  {http://arxiv.org/abs/hep-ph/9603397}{arXiv:hep-ph/9603397}\BibitemShut
  {NoStop}%
\bibitem [{\citenamefont {Maul}\ \emph {et~al.}(1997)\citenamefont {Maul},
  \citenamefont {Ehrnsperger}, \citenamefont {Stein},\ and\ \citenamefont
  {Schafer}}]{Maul:1996dx}%
  \BibitemOpen
  \bibfield  {author} {\bibinfo {author} {\bibfnamefont {M.}~\bibnamefont
  {Maul}}, \bibinfo {author} {\bibfnamefont {B.}~\bibnamefont {Ehrnsperger}},
  \bibinfo {author} {\bibfnamefont {E.}~\bibnamefont {Stein}}, \ and\ \bibinfo
  {author} {\bibfnamefont {A.}~\bibnamefont {Schafer}},\ }\href {\doibase
  10.48550/arXiv.hep-ph/9602377} {\bibfield  {journal} {\bibinfo  {journal} {Z.
  Phys. A}\ }\textbf {\bibinfo {volume} {356}},\ \bibinfo {pages} {443}
  (\bibinfo {year} {1997})},\ \Eprint
  {http://arxiv.org/abs/hep-ph/9602377}{arXiv:hep-ph/9602377}\BibitemShut
  {NoStop}%
\bibitem [{\citenamefont {Blumlein}(2013)}]{Blumlein:2012bf}%
  \BibitemOpen
  \bibfield  {author} {\bibinfo {author} {\bibfnamefont {J.}~\bibnamefont
  {Blumlein}},\ }\href {\doibase 10.1016/j.ppnp.2012.09.006} {\bibfield
  {journal} {\bibinfo  {journal} {Prog. Part. Nucl. Phys.}\ }\textbf {\bibinfo
  {volume} {69}},\ \bibinfo {pages} {28} (\bibinfo {year} {2013})},\ \Eprint
  {http://arxiv.org/abs/1208.6087}{arXiv:1208.6087 [hep-ph]}\BibitemShut
  {NoStop}%
\bibitem [{\citenamefont {Drell}\ and\ \citenamefont
  {Sullivan}(1967)}]{Drell:1966kk}%
  \BibitemOpen
  \bibfield  {author} {\bibinfo {author} {\bibfnamefont {S.~D.}\ \bibnamefont
  {Drell}}\ and\ \bibinfo {author} {\bibfnamefont {J.~D.}\ \bibnamefont
  {Sullivan}},\ }\href {\doibase 10.1103/PhysRev.154.1477} {\bibfield
  {journal} {\bibinfo  {journal} {Phys. Rev.}\ }\textbf {\bibinfo {volume}
  {154}},\ \bibinfo {pages} {1477} (\bibinfo {year} {1967})}\BibitemShut
  {NoStop}%
\bibitem [{\citenamefont {Bjorken}(1966)}]{Bjorken:1966jh}%
  \BibitemOpen
  \bibfield  {author} {\bibinfo {author} {\bibfnamefont {J.~D.}\ \bibnamefont
  {Bjorken}},\ }\href {\doibase 10.1103/PhysRev.148.1467} {\bibfield  {journal}
  {\bibinfo  {journal} {Phys. Rev.}\ }\textbf {\bibinfo {volume} {148}},\
  \bibinfo {pages} {1467} (\bibinfo {year} {1966})}\BibitemShut {NoStop}%
\bibitem [{\citenamefont {Larin}\ \emph {et~al.}(1991)\citenamefont {Larin},
  \citenamefont {Tkachov},\ and\ \citenamefont {Vermaseren}}]{Larin:1990zw}%
  \BibitemOpen
  \bibfield  {author} {\bibinfo {author} {\bibfnamefont {S.~A.}\ \bibnamefont
  {Larin}}, \bibinfo {author} {\bibfnamefont {F.~V.}\ \bibnamefont {Tkachov}},
  \ and\ \bibinfo {author} {\bibfnamefont {J.~A.~M.}\ \bibnamefont
  {Vermaseren}},\ }\href {\doibase 10.1103/PhysRevLett.66.862} {\bibfield
  {journal} {\bibinfo  {journal} {Phys. Rev. Lett.}\ }\textbf {\bibinfo
  {volume} {66}},\ \bibinfo {pages} {862} (\bibinfo {year} {1991})}\BibitemShut
  {NoStop}%
\bibitem [{\citenamefont {Larin}\ and\ \citenamefont
  {Vermaseren}(1991)}]{Larin:1991tj}%
  \BibitemOpen
  \bibfield  {author} {\bibinfo {author} {\bibfnamefont {S.~A.}\ \bibnamefont
  {Larin}}\ and\ \bibinfo {author} {\bibfnamefont {J.~A.~M.}\ \bibnamefont
  {Vermaseren}},\ }\href {\doibase 10.1016/0370-2693(91)90839-I} {\bibfield
  {journal} {\bibinfo  {journal} {Phys. Lett. B}\ }\textbf {\bibinfo {volume}
  {259}},\ \bibinfo {pages} {345} (\bibinfo {year} {1991})}\BibitemShut
  {NoStop}%
\bibitem [{\citenamefont {Baikov}\ \emph {et~al.}(2010)\citenamefont {Baikov},
  \citenamefont {Chetyrkin},\ and\ \citenamefont {Kuhn}}]{Baikov:2010je}%
  \BibitemOpen
  \bibfield  {author} {\bibinfo {author} {\bibfnamefont {P.~A.}\ \bibnamefont
  {Baikov}}, \bibinfo {author} {\bibfnamefont {K.~G.}\ \bibnamefont
  {Chetyrkin}}, \ and\ \bibinfo {author} {\bibfnamefont {J.~H.}\ \bibnamefont
  {Kuhn}},\ }\href {\doibase 10.1103/PhysRevLett.104.132004} {\bibfield
  {journal} {\bibinfo  {journal} {Phys. Rev. Lett.}\ }\textbf {\bibinfo
  {volume} {104}},\ \bibinfo {pages} {132004} (\bibinfo {year} {2010})},\
  \Eprint {http://arxiv.org/abs/1001.3606}{arXiv:1001.3606
  [hep-ph]}\BibitemShut {NoStop}%
\bibitem [{\citenamefont {Borah}\ \emph {et~al.}(2020)\citenamefont {Borah},
  \citenamefont {Hill}, \citenamefont {Lee},\ and\ \citenamefont
  {Tomalak}}]{Borah:2020gte}%
  \BibitemOpen
  \bibfield  {author} {\bibinfo {author} {\bibfnamefont {K.}~\bibnamefont
  {Borah}}, \bibinfo {author} {\bibfnamefont {R.~J.}\ \bibnamefont {Hill}},
  \bibinfo {author} {\bibfnamefont {G.}~\bibnamefont {Lee}}, \ and\ \bibinfo
  {author} {\bibfnamefont {O.}~\bibnamefont {Tomalak}},\ }\href {\doibase
  10.1103/PhysRevD.102.074012} {\bibfield  {journal} {\bibinfo  {journal}
  {Phys. Rev. D}\ }\textbf {\bibinfo {volume} {102}},\ \bibinfo {pages}
  {074012} (\bibinfo {year} {2020})},\ \Eprint
  {http://arxiv.org/abs/2003.13640}{arXiv:2003.13640 [hep-ph]}\BibitemShut
  {NoStop}%
\bibitem [{\citenamefont {Meyer}\ \emph {et~al.}(2016)\citenamefont {Meyer},
  \citenamefont {Betancourt}, \citenamefont {Gran},\ and\ \citenamefont
  {Hill}}]{Meyer:2016oeg}%
  \BibitemOpen
  \bibfield  {author} {\bibinfo {author} {\bibfnamefont {A.~S.}\ \bibnamefont
  {Meyer}}, \bibinfo {author} {\bibfnamefont {M.}~\bibnamefont {Betancourt}},
  \bibinfo {author} {\bibfnamefont {R.}~\bibnamefont {Gran}}, \ and\ \bibinfo
  {author} {\bibfnamefont {R.~J.}\ \bibnamefont {Hill}},\ }\href {\doibase
  10.1103/PhysRevD.93.113015} {\bibfield  {journal} {\bibinfo  {journal} {Phys.
  Rev. D}\ }\textbf {\bibinfo {volume} {93}},\ \bibinfo {pages} {113015}
  (\bibinfo {year} {2016})},\ \Eprint
  {http://arxiv.org/abs/1603.03048}{arXiv:1603.03048 [hep-ph]}\BibitemShut
  {NoStop}%
\bibitem [{\citenamefont {Fukugita}\ and\ \citenamefont
  {Kubota}(2004)}]{Fukugita:2004cq}%
  \BibitemOpen
  \bibfield  {author} {\bibinfo {author} {\bibfnamefont {M.}~\bibnamefont
  {Fukugita}}\ and\ \bibinfo {author} {\bibfnamefont {T.}~\bibnamefont
  {Kubota}},\ }\href@noop {} {\bibfield  {journal} {\bibinfo  {journal} {Acta
  Phys. Polon. B}\ }\textbf {\bibinfo {volume} {35}},\ \bibinfo {pages} {1687}
  (\bibinfo {year} {2004})},\ \Eprint
  {http://arxiv.org/abs/hep-ph/0403149}{arXiv:hep-ph/0403149}\BibitemShut
  {NoStop}%
\bibitem [{\citenamefont {Tomalak}\ \emph {et~al.}(2022)\citenamefont
  {Tomalak}, \citenamefont {Chen}, \citenamefont {Hill}, \citenamefont
  {McFarland},\ and\ \citenamefont {Wret}}]{Tomalak:2022xup}%
  \BibitemOpen
  \bibfield  {author} {\bibinfo {author} {\bibfnamefont {O.}~\bibnamefont
  {Tomalak}}, \bibinfo {author} {\bibfnamefont {Q.}~\bibnamefont {Chen}},
  \bibinfo {author} {\bibfnamefont {R.~J.}\ \bibnamefont {Hill}}, \bibinfo
  {author} {\bibfnamefont {K.~S.}\ \bibnamefont {McFarland}}, \ and\ \bibinfo
  {author} {\bibfnamefont {C.}~\bibnamefont {Wret}},\ }\href {\doibase
  10.1103/PhysRevD.106.093006} {\bibfield  {journal} {\bibinfo  {journal}
  {Phys. Rev. D}\ }\textbf {\bibinfo {volume} {106}},\ \bibinfo {pages}
  {093006} (\bibinfo {year} {2022})},\ \Eprint
  {http://arxiv.org/abs/2204.11379}{arXiv:2204.11379 [hep-ph]}\BibitemShut
  {NoStop}%
\bibitem [{\citenamefont {Voloshin}\ and\ \citenamefont
  {Shifman}(1987)}]{Voloshin:1986dir}%
  \BibitemOpen
  \bibfield  {author} {\bibinfo {author} {\bibfnamefont {M.~B.}\ \bibnamefont
  {Voloshin}}\ and\ \bibinfo {author} {\bibfnamefont {M.~A.}\ \bibnamefont
  {Shifman}},\ }\href@noop {} {\bibfield  {journal} {\bibinfo  {journal} {Sov.
  J. Nucl. Phys.}\ }\textbf {\bibinfo {volume} {45}},\ \bibinfo {pages} {292}
  (\bibinfo {year} {1987})}\BibitemShut {NoStop}%
\bibitem [{\citenamefont {Politzer}\ and\ \citenamefont
  {Wise}(1988)}]{Politzer:1988wp}%
  \BibitemOpen
  \bibfield  {author} {\bibinfo {author} {\bibfnamefont {H.~D.}\ \bibnamefont
  {Politzer}}\ and\ \bibinfo {author} {\bibfnamefont {M.~B.}\ \bibnamefont
  {Wise}},\ }\href {\doibase 10.1016/0370-2693(88)90718-6} {\bibfield
  {journal} {\bibinfo  {journal} {Phys. Lett. B}\ }\textbf {\bibinfo {volume}
  {206}},\ \bibinfo {pages} {681} (\bibinfo {year} {1988})}\BibitemShut
  {NoStop}%
\bibitem [{\citenamefont {Gimenez}(1992)}]{Gimenez:1991bf}%
  \BibitemOpen
  \bibfield  {author} {\bibinfo {author} {\bibfnamefont {V.}~\bibnamefont
  {Gimenez}},\ }\href {\doibase 10.1016/0550-3213(92)90112-O} {\bibfield
  {journal} {\bibinfo  {journal} {Nucl. Phys. B}\ }\textbf {\bibinfo {volume}
  {375}},\ \bibinfo {pages} {582} (\bibinfo {year} {1992})}\BibitemShut
  {NoStop}%
\bibitem [{\citenamefont {Ji}\ and\ \citenamefont {Musolf}(1991)}]{Ji:1991pr}%
  \BibitemOpen
  \bibfield  {author} {\bibinfo {author} {\bibfnamefont {X.-D.}\ \bibnamefont
  {Ji}}\ and\ \bibinfo {author} {\bibfnamefont {M.~J.}\ \bibnamefont
  {Musolf}},\ }\href {\doibase 10.1016/0370-2693(91)91916-J} {\bibfield
  {journal} {\bibinfo  {journal} {Phys. Lett. B}\ }\textbf {\bibinfo {volume}
  {257}},\ \bibinfo {pages} {409} (\bibinfo {year} {1991})}\BibitemShut
  {NoStop}%
\bibitem [{\citenamefont {Broadhurst}\ \emph {et~al.}(1991)\citenamefont
  {Broadhurst}, \citenamefont {Gray},\ and\ \citenamefont
  {Schilcher}}]{Broadhurst:1991fy}%
  \BibitemOpen
  \bibfield  {author} {\bibinfo {author} {\bibfnamefont {D.~J.}\ \bibnamefont
  {Broadhurst}}, \bibinfo {author} {\bibfnamefont {N.}~\bibnamefont {Gray}}, \
  and\ \bibinfo {author} {\bibfnamefont {K.}~\bibnamefont {Schilcher}},\ }\href
  {\doibase 10.1007/BF01412333} {\bibfield  {journal} {\bibinfo  {journal} {Z.
  Phys. C}\ }\textbf {\bibinfo {volume} {52}},\ \bibinfo {pages} {111}
  (\bibinfo {year} {1991})}\BibitemShut {NoStop}%
\bibitem [{\citenamefont {Broadhurst}\ and\ \citenamefont
  {Grozin}(1991)}]{Broadhurst:1991fz}%
  \BibitemOpen
  \bibfield  {author} {\bibinfo {author} {\bibfnamefont {D.~J.}\ \bibnamefont
  {Broadhurst}}\ and\ \bibinfo {author} {\bibfnamefont {A.~G.}\ \bibnamefont
  {Grozin}},\ }\href {\doibase 10.1016/0370-2693(91)90532-U} {\bibfield
  {journal} {\bibinfo  {journal} {Phys. Lett. B}\ }\textbf {\bibinfo {volume}
  {267}},\ \bibinfo {pages} {105} (\bibinfo {year} {1991})},\ \Eprint
  {http://arxiv.org/abs/hep-ph/9908362}{arXiv:hep-ph/9908362}\BibitemShut
  {NoStop}%
\bibitem [{\citenamefont {Hoang}(1997{\natexlab{a}})}]{Hoang:1997sj}%
  \BibitemOpen
  \bibfield  {author} {\bibinfo {author} {\bibfnamefont {A.~H.}\ \bibnamefont
  {Hoang}},\ }\href {\doibase 10.1103/PhysRevD.56.7276} {\bibfield  {journal}
  {\bibinfo  {journal} {Phys. Rev. D}\ }\textbf {\bibinfo {volume} {56}},\
  \bibinfo {pages} {7276} (\bibinfo {year} {1997}{\natexlab{a}})},\ \Eprint
  {http://arxiv.org/abs/hep-ph/9703404}{arXiv:hep-ph/9703404}\BibitemShut
  {NoStop}%
\bibitem [{\citenamefont {Hoang}(1997{\natexlab{b}})}]{Hoang:1997ui}%
  \BibitemOpen
  \bibfield  {author} {\bibinfo {author} {\bibfnamefont {A.~H.}\ \bibnamefont
  {Hoang}},\ }\href {\doibase 10.1103/PhysRevD.56.5851} {\bibfield  {journal}
  {\bibinfo  {journal} {Phys. Rev. D}\ }\textbf {\bibinfo {volume} {56}},\
  \bibinfo {pages} {5851} (\bibinfo {year} {1997}{\natexlab{b}})},\ \Eprint
  {http://arxiv.org/abs/hep-ph/9704325}{arXiv:hep-ph/9704325}\BibitemShut
  {NoStop}%
\bibitem [{\citenamefont {Czarnecki}\ and\ \citenamefont
  {Melnikov}(1998)}]{Czarnecki:1997vz}%
  \BibitemOpen
  \bibfield  {author} {\bibinfo {author} {\bibfnamefont {A.}~\bibnamefont
  {Czarnecki}}\ and\ \bibinfo {author} {\bibfnamefont {K.}~\bibnamefont
  {Melnikov}},\ }\href {\doibase 10.1103/PhysRevLett.80.2531} {\bibfield
  {journal} {\bibinfo  {journal} {Phys. Rev. Lett.}\ }\textbf {\bibinfo
  {volume} {80}},\ \bibinfo {pages} {2531} (\bibinfo {year} {1998})},\ \Eprint
  {http://arxiv.org/abs/hep-ph/9712222}{arXiv:hep-ph/9712222}\BibitemShut
  {NoStop}%
\bibitem [{\citenamefont {Beneke}\ \emph {et~al.}(1999)\citenamefont {Beneke},
  \citenamefont {Signer},\ and\ \citenamefont {Smirnov}}]{Beneke:1999qg}%
  \BibitemOpen
  \bibfield  {author} {\bibinfo {author} {\bibfnamefont {M.}~\bibnamefont
  {Beneke}}, \bibinfo {author} {\bibfnamefont {A.}~\bibnamefont {Signer}}, \
  and\ \bibinfo {author} {\bibfnamefont {V.~A.}\ \bibnamefont {Smirnov}},\
  }\href {\doibase 10.1016/S0370-2693(99)00343-3} {\bibfield  {journal}
  {\bibinfo  {journal} {Phys. Lett. B}\ }\textbf {\bibinfo {volume} {454}},\
  \bibinfo {pages} {137} (\bibinfo {year} {1999})},\ \Eprint
  {http://arxiv.org/abs/hep-ph/9903260}{arXiv:hep-ph/9903260}\BibitemShut
  {NoStop}%
\bibitem [{\citenamefont {Sommerfeld}(1931)}]{Sommerfeld:1931qaf}%
  \BibitemOpen
  \bibfield  {author} {\bibinfo {author} {\bibfnamefont {A.}~\bibnamefont
  {Sommerfeld}},\ }\href {\doibase 10.1002/andp.19314030302} {\bibfield
  {journal} {\bibinfo  {journal} {Annalen Phys.}\ }\textbf {\bibinfo {volume}
  {403}},\ \bibinfo {pages} {257} (\bibinfo {year} {1931})}\BibitemShut
  {NoStop}%
\bibitem [{\citenamefont {Fermi}(1934)}]{Fermi:1934hr}%
  \BibitemOpen
  \bibfield  {author} {\bibinfo {author} {\bibfnamefont {E.}~\bibnamefont
  {Fermi}},\ }\href {\doibase 10.1007/BF01351864} {\bibfield  {journal}
  {\bibinfo  {journal} {Z. Phys.}\ }\textbf {\bibinfo {volume} {88}},\ \bibinfo
  {pages} {161} (\bibinfo {year} {1934})}\BibitemShut {NoStop}%
\bibitem [{\citenamefont {Konopinski}\ and\ \citenamefont
  {Uhlenbeck}(1935)}]{Konopinski:1935zz}%
  \BibitemOpen
  \bibfield  {author} {\bibinfo {author} {\bibfnamefont {E.~J.}\ \bibnamefont
  {Konopinski}}\ and\ \bibinfo {author} {\bibfnamefont {G.~E.}\ \bibnamefont
  {Uhlenbeck}},\ }\href {\doibase 10.1103/PhysRev.48.7} {\bibfield  {journal}
  {\bibinfo  {journal} {Phys. Rev.}\ }\textbf {\bibinfo {volume} {48}},\
  \bibinfo {pages} {7} (\bibinfo {year} {1935})}\BibitemShut {NoStop}%
\bibitem [{\citenamefont {Morita}(1963)}]{Morita:1963zz}%
  \BibitemOpen
  \bibfield  {author} {\bibinfo {author} {\bibfnamefont {M.}~\bibnamefont
  {Morita}},\ }\href {\doibase 10.1143/PTPS.26.1} {\bibfield  {journal}
  {\bibinfo  {journal} {Prog. Theor. Phys. Suppl.}\ }\textbf {\bibinfo {volume}
  {26}},\ \bibinfo {pages} {1} (\bibinfo {year} {1963})}\BibitemShut {NoStop}%
\bibitem [{\citenamefont {Wilson}(1968)}]{Wilson:1968pwx}%
  \BibitemOpen
  \bibfield  {author} {\bibinfo {author} {\bibfnamefont {F.~L.}\ \bibnamefont
  {Wilson}},\ }\href {\doibase 10.1119/1.1974382} {\bibfield  {journal}
  {\bibinfo  {journal} {Am. J. Phys.}\ }\textbf {\bibinfo {volume} {36}},\
  \bibinfo {pages} {1150} (\bibinfo {year} {1968})}\BibitemShut {NoStop}%
\bibitem [{\citenamefont {Halpern}(1970)}]{Halpern:1970it}%
  \BibitemOpen
  \bibfield  {author} {\bibinfo {author} {\bibfnamefont {T.~A.}\ \bibnamefont
  {Halpern}},\ }\href {\doibase 10.1103/PhysRevC.1.1928} {\bibfield  {journal}
  {\bibinfo  {journal} {Phys. Rev. C}\ }\textbf {\bibinfo {volume} {1}},\
  \bibinfo {pages} {1928} (\bibinfo {year} {1970})}\BibitemShut {NoStop}%
\bibitem [{\citenamefont {Halpern}\ and\ \citenamefont
  {Chern}(1968)}]{Halpern:1968zz}%
  \BibitemOpen
  \bibfield  {author} {\bibinfo {author} {\bibfnamefont {T.~A.}\ \bibnamefont
  {Halpern}}\ and\ \bibinfo {author} {\bibfnamefont {B.}~\bibnamefont
  {Chern}},\ }\href {\doibase 10.1103/PhysRev.175.1314} {\bibfield  {journal}
  {\bibinfo  {journal} {Phys. Rev.}\ }\textbf {\bibinfo {volume} {175}},\
  \bibinfo {pages} {1314} (\bibinfo {year} {1968})}\BibitemShut {NoStop}%
\bibitem [{\citenamefont {Hoferichter}\ \emph {et~al.}(2010)\citenamefont
  {Hoferichter}, \citenamefont {Kubis},\ and\ \citenamefont
  {Meissner}}]{Hoferichter:2009gn}%
  \BibitemOpen
  \bibfield  {author} {\bibinfo {author} {\bibfnamefont {M.}~\bibnamefont
  {Hoferichter}}, \bibinfo {author} {\bibfnamefont {B.}~\bibnamefont {Kubis}},
  \ and\ \bibinfo {author} {\bibfnamefont {U.~G.}\ \bibnamefont {Meissner}},\
  }\href {\doibase 10.1016/j.nuclphysa.2009.11.012} {\bibfield  {journal}
  {\bibinfo  {journal} {Nucl. Phys. A}\ }\textbf {\bibinfo {volume} {833}},\
  \bibinfo {pages} {18} (\bibinfo {year} {2010})},\ \Eprint
  {http://arxiv.org/abs/0909.4390}{arXiv:0909.4390 [hep-ph]}\BibitemShut
  {NoStop}%
\bibitem [{\citenamefont {Matsuzaki}\ and\ \citenamefont
  {Tanaka}(2012)}]{Matsuzaki:2012qb}%
  \BibitemOpen
  \bibfield  {author} {\bibinfo {author} {\bibfnamefont {A.}~\bibnamefont
  {Matsuzaki}}\ and\ \bibinfo {author} {\bibfnamefont {H.}~\bibnamefont
  {Tanaka}},\ }\href {\doibase 10.1103/PhysRevC.86.065502} {\bibfield
  {journal} {\bibinfo  {journal} {Phys. Rev. C}\ }\textbf {\bibinfo {volume}
  {86}},\ \bibinfo {pages} {065502} (\bibinfo {year} {2012})},\ \Eprint
  {http://arxiv.org/abs/1203.6461}{arXiv:1203.6461 [hep-ph]}\BibitemShut
  {NoStop}%
\bibitem [{\citenamefont {Matsuzaki}\ and\ \citenamefont
  {Tanaka}(2013)}]{Matsuzaki:2013twa}%
  \BibitemOpen
  \bibfield  {author} {\bibinfo {author} {\bibfnamefont {A.}~\bibnamefont
  {Matsuzaki}}\ and\ \bibinfo {author} {\bibfnamefont {H.}~\bibnamefont
  {Tanaka}},\ }\href@noop {} {\  (\bibinfo {year} {2013})},\ \Eprint
  {http://arxiv.org/abs/1310.4082}{arXiv:1310.4082 [hep-th]}\BibitemShut
  {NoStop}%
\bibitem [{\citenamefont {Heller}\ \emph {et~al.}(2019)\citenamefont {Heller},
  \citenamefont {Tomalak}, \citenamefont {Vanderhaeghen},\ and\ \citenamefont
  {Wu}}]{Heller:2019dyv}%
  \BibitemOpen
  \bibfield  {author} {\bibinfo {author} {\bibfnamefont {M.}~\bibnamefont
  {Heller}}, \bibinfo {author} {\bibfnamefont {O.}~\bibnamefont {Tomalak}},
  \bibinfo {author} {\bibfnamefont {M.}~\bibnamefont {Vanderhaeghen}}, \ and\
  \bibinfo {author} {\bibfnamefont {S.}~\bibnamefont {Wu}},\ }\href {\doibase
  10.1103/PhysRevD.100.076013} {\bibfield  {journal} {\bibinfo  {journal}
  {Phys. Rev. D}\ }\textbf {\bibinfo {volume} {100}},\ \bibinfo {pages}
  {076013} (\bibinfo {year} {2019})},\ \Eprint
  {http://arxiv.org/abs/1906.02706}{arXiv:1906.02706 [hep-ph]}\BibitemShut
  {NoStop}%
\bibitem [{\citenamefont {Beneke}\ \emph {et~al.}(1998)\citenamefont {Beneke},
  \citenamefont {Signer},\ and\ \citenamefont {Smirnov}}]{Beneke:1997jm}%
  \BibitemOpen
  \bibfield  {author} {\bibinfo {author} {\bibfnamefont {M.}~\bibnamefont
  {Beneke}}, \bibinfo {author} {\bibfnamefont {A.}~\bibnamefont {Signer}}, \
  and\ \bibinfo {author} {\bibfnamefont {V.~A.}\ \bibnamefont {Smirnov}},\
  }\href {\doibase 10.1103/PhysRevLett.80.2535} {\bibfield  {journal} {\bibinfo
   {journal} {Phys. Rev. Lett.}\ }\textbf {\bibinfo {volume} {80}},\ \bibinfo
  {pages} {2535} (\bibinfo {year} {1998})},\ \Eprint
  {http://arxiv.org/abs/hep-ph/9712302}{arXiv:hep-ph/9712302}\BibitemShut
  {NoStop}%
\bibitem [{\citenamefont {Beneke}\ \emph {et~al.}(2010)\citenamefont {Beneke},
  \citenamefont {Czakon}, \citenamefont {Falgari}, \citenamefont {Mitov},\ and\
  \citenamefont {Schwinn}}]{Beneke:2009ye}%
  \BibitemOpen
  \bibfield  {author} {\bibinfo {author} {\bibfnamefont {M.}~\bibnamefont
  {Beneke}}, \bibinfo {author} {\bibfnamefont {M.}~\bibnamefont {Czakon}},
  \bibinfo {author} {\bibfnamefont {P.}~\bibnamefont {Falgari}}, \bibinfo
  {author} {\bibfnamefont {A.}~\bibnamefont {Mitov}}, \ and\ \bibinfo {author}
  {\bibfnamefont {C.}~\bibnamefont {Schwinn}},\ }\href {\doibase
  10.1016/j.physletb.2010.05.038} {\bibfield  {journal} {\bibinfo  {journal}
  {Phys. Lett. B}\ }\textbf {\bibinfo {volume} {690}},\ \bibinfo {pages} {483}
  (\bibinfo {year} {2010})},\ \bibinfo {note} {[Erratum: Phys.Lett.B 778,
  464--464 (2018)]},\ \Eprint {http://arxiv.org/abs/0911.5166}{arXiv:0911.5166
  [hep-ph]}\BibitemShut {NoStop}%
\bibitem [{\citenamefont {Kawamura}\ \emph {et~al.}(2012)\citenamefont
  {Kawamura}, \citenamefont {Lo~Presti}, \citenamefont {Moch},\ and\
  \citenamefont {Vogt}}]{Kawamura:2012cr}%
  \BibitemOpen
  \bibfield  {author} {\bibinfo {author} {\bibfnamefont {H.}~\bibnamefont
  {Kawamura}}, \bibinfo {author} {\bibfnamefont {N.~A.}\ \bibnamefont
  {Lo~Presti}}, \bibinfo {author} {\bibfnamefont {S.}~\bibnamefont {Moch}}, \
  and\ \bibinfo {author} {\bibfnamefont {A.}~\bibnamefont {Vogt}},\ }\href
  {\doibase 10.1016/j.nuclphysb.2012.07.001} {\bibfield  {journal} {\bibinfo
  {journal} {Nucl. Phys. B}\ }\textbf {\bibinfo {volume} {864}},\ \bibinfo
  {pages} {399} (\bibinfo {year} {2012})},\ \Eprint
  {http://arxiv.org/abs/1205.5727}{arXiv:1205.5727 [hep-ph]}\BibitemShut
  {NoStop}%
\bibitem [{\citenamefont {Piclum}\ and\ \citenamefont
  {Schwinn}(2018)}]{Piclum:2018ndt}%
  \BibitemOpen
  \bibfield  {author} {\bibinfo {author} {\bibfnamefont {J.}~\bibnamefont
  {Piclum}}\ and\ \bibinfo {author} {\bibfnamefont {C.}~\bibnamefont
  {Schwinn}},\ }\href {\doibase 10.1007/JHEP03(2018)164} {\bibfield  {journal}
  {\bibinfo  {journal} {JHEP}\ }\textbf {\bibinfo {volume} {03}},\ \bibinfo
  {pages} {164} (\bibinfo {year} {2018})},\ \Eprint
  {http://arxiv.org/abs/1801.05788}{arXiv:1801.05788 [hep-ph]}\BibitemShut
  {NoStop}%
\bibitem [{\citenamefont {Grozin}(2004)}]{Grozin:2004yc}%
  \BibitemOpen
  \bibfield  {author} {\bibinfo {author} {\bibfnamefont {A.~G.}\ \bibnamefont
  {Grozin}},\ }\href {\doibase 10.1007/b79301} {\bibfield  {journal} {\bibinfo
  {journal} {Springer Tracts Mod. Phys.}\ }\textbf {\bibinfo {volume} {201}},\
  \bibinfo {pages} {1} (\bibinfo {year} {2004})}\BibitemShut {NoStop}%
\bibitem [{\citenamefont {Behrens}\ and\ \citenamefont
  {J\"{a}necke}(1969)}]{Behrens:1969}%
  \BibitemOpen
  \bibfield  {author} {\bibinfo {author} {\bibfnamefont {H.}~\bibnamefont
  {Behrens}}\ and\ \bibinfo {author} {\bibfnamefont {H.}~\bibnamefont
  {J\"{a}necke}},\ }\href {\doibase 10.1007/b19939} {\bibfield  {journal}
  {\bibinfo  {journal} {Springer-Verlag, Berlin}\ }\textbf {\bibinfo {volume}
  {4}},\ \bibinfo {pages} {315} (\bibinfo {year} {1969})}\BibitemShut {NoStop}%
\bibitem [{\citenamefont {Behrens}\ and\ \citenamefont
  {B\"uhring}(1970)}]{Behrens:1970ncy}%
  \BibitemOpen
  \bibfield  {author} {\bibinfo {author} {\bibfnamefont {H.}~\bibnamefont
  {Behrens}}\ and\ \bibinfo {author} {\bibfnamefont {W.}~\bibnamefont
  {B\"uhring}},\ }\href {\doibase 10.1016/0375-9474(70)90413-6} {\bibfield
  {journal} {\bibinfo  {journal} {Nucl. Phys. A}\ }\textbf {\bibinfo {volume}
  {150}},\ \bibinfo {pages} {481} (\bibinfo {year} {1970})}\BibitemShut
  {NoStop}%
\bibitem [{\citenamefont {Behrens}\ and\ \citenamefont
  {B\"uhring}(1971)}]{Behrens:1971rbq}%
  \BibitemOpen
  \bibfield  {author} {\bibinfo {author} {\bibfnamefont {H.}~\bibnamefont
  {Behrens}}\ and\ \bibinfo {author} {\bibfnamefont {W.}~\bibnamefont
  {B\"uhring}},\ }\href {\doibase 10.1016/0375-9474(71)90489-1} {\bibfield
  {journal} {\bibinfo  {journal} {Nucl. Phys. A}\ }\textbf {\bibinfo {volume}
  {162}},\ \bibinfo {pages} {111} (\bibinfo {year} {1971})}\BibitemShut
  {NoStop}%
\bibitem [{\citenamefont {Behrens}\ and\ \citenamefont
  {B\"uhring}(1972)}]{Behrens:1972zya}%
  \BibitemOpen
  \bibfield  {author} {\bibinfo {author} {\bibfnamefont {H.}~\bibnamefont
  {Behrens}}\ and\ \bibinfo {author} {\bibfnamefont {W.}~\bibnamefont
  {B\"uhring}},\ }\href {\doibase 10.1016/0375-9474(72)90371-5} {\bibfield
  {journal} {\bibinfo  {journal} {Nucl. Phys. A}\ }\textbf {\bibinfo {volume}
  {179}},\ \bibinfo {pages} {297} (\bibinfo {year} {1972})}\BibitemShut
  {NoStop}%
\bibitem [{\citenamefont {Jaus}(1972)}]{Jaus:1972hua}%
  \BibitemOpen
  \bibfield  {author} {\bibinfo {author} {\bibfnamefont {W.}~\bibnamefont
  {Jaus}},\ }\href {\doibase 10.1016/0370-2693(72)90610-7} {\bibfield
  {journal} {\bibinfo  {journal} {Phys. Lett. B}\ }\textbf {\bibinfo {volume}
  {40}},\ \bibinfo {pages} {616} (\bibinfo {year} {1972})}\BibitemShut
  {NoStop}%
\bibitem [{\citenamefont {Jaus}\ and\ \citenamefont
  {Rasche}(1987)}]{Jaus:1986te}%
  \BibitemOpen
  \bibfield  {author} {\bibinfo {author} {\bibfnamefont {W.}~\bibnamefont
  {Jaus}}\ and\ \bibinfo {author} {\bibfnamefont {G.}~\bibnamefont {Rasche}},\
  }\href {\doibase 10.1103/PhysRevD.35.3420} {\bibfield  {journal} {\bibinfo
  {journal} {Phys. Rev. D}\ }\textbf {\bibinfo {volume} {35}},\ \bibinfo
  {pages} {3420} (\bibinfo {year} {1987})}\BibitemShut {NoStop}%
\bibitem [{\citenamefont {Sirlin}(1987)}]{Sirlin:1986hpu}%
  \BibitemOpen
  \bibfield  {author} {\bibinfo {author} {\bibfnamefont {A.}~\bibnamefont
  {Sirlin}},\ }\href {\doibase 10.1103/PhysRevD.35.3423} {\bibfield  {journal}
  {\bibinfo  {journal} {Phys. Rev. D}\ }\textbf {\bibinfo {volume} {35}},\
  \bibinfo {pages} {3423} (\bibinfo {year} {1987})}\BibitemShut {NoStop}%
\bibitem [{\citenamefont {Towner}\ and\ \citenamefont
  {Hardy}(2008)}]{Towner:2007np}%
  \BibitemOpen
  \bibfield  {author} {\bibinfo {author} {\bibfnamefont {I.~S.}\ \bibnamefont
  {Towner}}\ and\ \bibinfo {author} {\bibfnamefont {J.~C.}\ \bibnamefont
  {Hardy}},\ }\href {\doibase 10.1103/PhysRevC.77.025501} {\bibfield  {journal}
  {\bibinfo  {journal} {Phys. Rev. C}\ }\textbf {\bibinfo {volume} {77}},\
  \bibinfo {pages} {025501} (\bibinfo {year} {2008})},\ \Eprint
  {http://arxiv.org/abs/0710.3181}{arXiv:0710.3181 [nucl-th]}\BibitemShut
  {NoStop}%
\bibitem [{\citenamefont {Mertig}\ \emph {et~al.}(1991)\citenamefont {Mertig},
  \citenamefont {Bohm},\ and\ \citenamefont {Denner}}]{Mertig:1990an}%
  \BibitemOpen
  \bibfield  {author} {\bibinfo {author} {\bibfnamefont {R.}~\bibnamefont
  {Mertig}}, \bibinfo {author} {\bibfnamefont {M.}~\bibnamefont {Bohm}}, \ and\
  \bibinfo {author} {\bibfnamefont {A.}~\bibnamefont {Denner}},\ }\href
  {\doibase 10.1016/0010-4655(91)90130-D} {\bibfield  {journal} {\bibinfo
  {journal} {Comput. Phys. Commun.}\ }\textbf {\bibinfo {volume} {64}},\
  \bibinfo {pages} {345} (\bibinfo {year} {1991})}\BibitemShut {NoStop}%
\bibitem [{\citenamefont {Shtabovenko}\ \emph {et~al.}(2016)\citenamefont
  {Shtabovenko}, \citenamefont {Mertig},\ and\ \citenamefont
  {Orellana}}]{Shtabovenko:2016sxi}%
  \BibitemOpen
  \bibfield  {author} {\bibinfo {author} {\bibfnamefont {V.}~\bibnamefont
  {Shtabovenko}}, \bibinfo {author} {\bibfnamefont {R.}~\bibnamefont {Mertig}},
  \ and\ \bibinfo {author} {\bibfnamefont {F.}~\bibnamefont {Orellana}},\
  }\href {\doibase 10.1016/j.cpc.2016.06.008} {\bibfield  {journal} {\bibinfo
  {journal} {Comput. Phys. Commun.}\ }\textbf {\bibinfo {volume} {207}},\
  \bibinfo {pages} {432} (\bibinfo {year} {2016})},\ \Eprint
  {http://arxiv.org/abs/1601.01167}{arXiv:1601.01167 [hep-ph]}\BibitemShut
  {NoStop}%
\bibitem [{\citenamefont {Hahn}\ and\ \citenamefont
  {Perez-Victoria}(1999)}]{Hahn:1998yk}%
  \BibitemOpen
  \bibfield  {author} {\bibinfo {author} {\bibfnamefont {T.}~\bibnamefont
  {Hahn}}\ and\ \bibinfo {author} {\bibfnamefont {M.}~\bibnamefont
  {Perez-Victoria}},\ }\href {\doibase 10.1016/S0010-4655(98)00173-8}
  {\bibfield  {journal} {\bibinfo  {journal} {Comput. Phys. Commun.}\ }\textbf
  {\bibinfo {volume} {118}},\ \bibinfo {pages} {153} (\bibinfo {year}
  {1999})},\ \Eprint
  {http://arxiv.org/abs/hep-ph/9807565}{arXiv:hep-ph/9807565}\BibitemShut
  {NoStop}%
\bibitem [{\citenamefont {Inc.}(2022)}]{Mathematica}%
  \BibitemOpen
  \bibfield  {author} {\bibinfo {author} {\bibfnamefont {W.~R.}\ \bibnamefont
  {Inc.}},\ }\href@noop {} {\enquote {\bibinfo {title} {Mathematica, {V}ersion
  12.2.0.0},}\ } (\bibinfo {year} {2022}),\ \bibinfo {note} {champaign, IL,
  2022}\BibitemShut {NoStop}%
\bibitem [{\citenamefont {Peskin}\ and\ \citenamefont
  {Schroeder}(1995)}]{Peskin:1995ev}%
  \BibitemOpen
  \bibfield  {author} {\bibinfo {author} {\bibfnamefont {M.~E.}\ \bibnamefont
  {Peskin}}\ and\ \bibinfo {author} {\bibfnamefont {D.~V.}\ \bibnamefont
  {Schroeder}},\ }\href@noop {} {\emph {\bibinfo {title} {{An Introduction to
  quantum field theory}}}}\ (\bibinfo  {publisher} {Addison-Wesley},\ \bibinfo
  {address} {Reading, USA},\ \bibinfo {year} {1995})\BibitemShut {NoStop}%
\bibitem [{\citenamefont {Davier}\ \emph {et~al.}(2020)\citenamefont {Davier},
  \citenamefont {Hoecker}, \citenamefont {Malaescu},\ and\ \citenamefont
  {Zhang}}]{Davier:2019can}%
  \BibitemOpen
  \bibfield  {author} {\bibinfo {author} {\bibfnamefont {M.}~\bibnamefont
  {Davier}}, \bibinfo {author} {\bibfnamefont {A.}~\bibnamefont {Hoecker}},
  \bibinfo {author} {\bibfnamefont {B.}~\bibnamefont {Malaescu}}, \ and\
  \bibinfo {author} {\bibfnamefont {Z.}~\bibnamefont {Zhang}},\ }\href
  {\doibase 10.1140/epjc/s10052-020-7792-2} {\bibfield  {journal} {\bibinfo
  {journal} {Eur. Phys. J. C}\ }\textbf {\bibinfo {volume} {80}},\ \bibinfo
  {pages} {241} (\bibinfo {year} {2020})},\ \bibinfo {note} {[Erratum:
  Eur.Phys.J.C 80, 410 (2020)]},\ \Eprint
  {http://arxiv.org/abs/1908.00921}{arXiv:1908.00921 [hep-ph]}\BibitemShut
  {NoStop}%
\bibitem [{\citenamefont {Fanchiotti}\ \emph {et~al.}(1993)\citenamefont
  {Fanchiotti}, \citenamefont {Kniehl},\ and\ \citenamefont
  {Sirlin}}]{Fanchiotti:1992tu}%
  \BibitemOpen
  \bibfield  {author} {\bibinfo {author} {\bibfnamefont {S.}~\bibnamefont
  {Fanchiotti}}, \bibinfo {author} {\bibfnamefont {B.~A.}\ \bibnamefont
  {Kniehl}}, \ and\ \bibinfo {author} {\bibfnamefont {A.}~\bibnamefont
  {Sirlin}},\ }\href {\doibase 10.1103/PhysRevD.48.307} {\bibfield  {journal}
  {\bibinfo  {journal} {Phys. Rev. D}\ }\textbf {\bibinfo {volume} {48}},\
  \bibinfo {pages} {307} (\bibinfo {year} {1993})},\ \Eprint
  {http://arxiv.org/abs/hep-ph/9212285}{arXiv:hep-ph/9212285}\BibitemShut
  {NoStop}%
\bibitem [{\citenamefont {Gasser}\ and\ \citenamefont
  {Leutwyler}(1985)}]{Gasser:1984gg}%
  \BibitemOpen
  \bibfield  {author} {\bibinfo {author} {\bibfnamefont {J.}~\bibnamefont
  {Gasser}}\ and\ \bibinfo {author} {\bibfnamefont {H.}~\bibnamefont
  {Leutwyler}},\ }\href {\doibase 10.1016/0550-3213(85)90492-4} {\bibfield
  {journal} {\bibinfo  {journal} {Nucl. Phys. B}\ }\textbf {\bibinfo {volume}
  {250}},\ \bibinfo {pages} {465} (\bibinfo {year} {1985})}\BibitemShut
  {NoStop}%
\bibitem [{\citenamefont {Pineda}\ and\ \citenamefont
  {Soto}(1998)}]{Pineda:1997ie}%
  \BibitemOpen
  \bibfield  {author} {\bibinfo {author} {\bibfnamefont {A.}~\bibnamefont
  {Pineda}}\ and\ \bibinfo {author} {\bibfnamefont {J.}~\bibnamefont {Soto}},\
  }\href {\doibase 10.1016/S0370-2693(97)01537-2} {\bibfield  {journal}
  {\bibinfo  {journal} {Phys. Lett. B}\ }\textbf {\bibinfo {volume} {420}},\
  \bibinfo {pages} {391} (\bibinfo {year} {1998})},\ \Eprint
  {http://arxiv.org/abs/hep-ph/9711292}{arXiv:hep-ph/9711292}\BibitemShut
  {NoStop}%
\bibitem [{\citenamefont {Pineda}\ and\ \citenamefont
  {Soto}(1999)}]{Pineda:1998kn}%
  \BibitemOpen
  \bibfield  {author} {\bibinfo {author} {\bibfnamefont {A.}~\bibnamefont
  {Pineda}}\ and\ \bibinfo {author} {\bibfnamefont {J.}~\bibnamefont {Soto}},\
  }\href {\doibase 10.1103/PhysRevD.59.016005} {\bibfield  {journal} {\bibinfo
  {journal} {Phys. Rev. D}\ }\textbf {\bibinfo {volume} {59}},\ \bibinfo
  {pages} {016005} (\bibinfo {year} {1999})},\ \Eprint
  {http://arxiv.org/abs/hep-ph/9805424}{arXiv:hep-ph/9805424}\BibitemShut
  {NoStop}%
\bibitem [{\citenamefont {Peset}\ \emph {et~al.}(2016)\citenamefont {Peset},
  \citenamefont {Pineda},\ and\ \citenamefont {Stahlhofen}}]{Peset:2015vvi}%
  \BibitemOpen
  \bibfield  {author} {\bibinfo {author} {\bibfnamefont {C.}~\bibnamefont
  {Peset}}, \bibinfo {author} {\bibfnamefont {A.}~\bibnamefont {Pineda}}, \
  and\ \bibinfo {author} {\bibfnamefont {M.}~\bibnamefont {Stahlhofen}},\
  }\href {\doibase 10.1007/JHEP05(2016)017} {\bibfield  {journal} {\bibinfo
  {journal} {JHEP}\ }\textbf {\bibinfo {volume} {05}},\ \bibinfo {pages} {017}
  (\bibinfo {year} {2016})},\ \Eprint
  {http://arxiv.org/abs/1511.08210}{arXiv:1511.08210 [hep-ph]}\BibitemShut
  {NoStop}%
\bibitem [{\citenamefont {Chetyrkin}\ and\ \citenamefont
  {Grozin}(2003)}]{Chetyrkin:2003vi}%
  \BibitemOpen
  \bibfield  {author} {\bibinfo {author} {\bibfnamefont {K.~G.}\ \bibnamefont
  {Chetyrkin}}\ and\ \bibinfo {author} {\bibfnamefont {A.~G.}\ \bibnamefont
  {Grozin}},\ }\href {\doibase 10.1016/S0550-3213(03)00490-5} {\bibfield
  {journal} {\bibinfo  {journal} {Nucl. Phys. B}\ }\textbf {\bibinfo {volume}
  {666}},\ \bibinfo {pages} {289} (\bibinfo {year} {2003})},\ \Eprint
  {http://arxiv.org/abs/hep-ph/0303113}{arXiv:hep-ph/0303113}\BibitemShut
  {NoStop}%
\end{thebibliography}%

\end{document}